\documentclass[11pt]{article}
\usepackage{epsfig} 
\newcommand{\sect}[1]{ \section{#1} \setcounter{equation}{0} } 

\newcommand{\half}{\mbox{\small{$\frac{1}{2}$}}}

\newcommand{\MSbar}{\overline{\mbox{MS}}} 
 
\newcommand{\Nc}{N_{\!c}}
\newcommand{\Nf}{N_{\!f}}

\begin{document}
\title{Four loop renormalization of $\phi^3$ theory in six dimensions}
\author{J.A. Gracey, \\ Theoretical Physics Division, \\ 
Department of Mathematical Sciences, \\ University of Liverpool, \\ P.O. Box 
147, \\ Liverpool, \\ L69 3BX, \\ United Kingdom.} 
\date{}
\maketitle 

\vspace{5cm} 
\noindent 
{\bf Abstract.} We renormalize six dimensional $\phi^3$ theory in the
modified minimal subtraction ($\MSbar$) scheme at four loops. From the 
resulting $\beta$-function, anomalous dimension and mass anomalous dimension we
compute four loop critical exponents relevant to the Lee-Yang edge singularity 
and percolation problems. Using resummation methods and information on the 
exponents of the relevant two dimensional conformal field theory we obtain 
estimates for exponents in dimensions $3$, $4$ and $5$ which are in reasonable 
agreement with other techniques for these two problems. The renormalization 
group functions for the more general theory with an $O(N)$ symmetry are also 
computed in order to obtain estimates of exponents at various fixed points in 
five dimensions. Included in this $O(N)$ analysis is the full evaluation of the
mass operator mixing matrix of anomalous dimensions at four loops. We show that
its eigen-exponents are in agreement with the mass exponents computed at 
$O(1/N^2)$ in the non-perturbative large $N$ expansion.  

\vspace{-18cm}
\hspace{13cm}
{\bf LTH 1046}

\newpage 

\sect{Introduction.}

Recently there has been renewed interest in analysing the $\phi^3$ scalar
quantum field theory which is perturbatively renormalizable in six spacetime
dimensions. This interest in primarily due in the main to the modern
development of the original conformal bootstrap method, \cite{1,2,3,4,5,6,7}, 
to study the fixed point structure of field theories in a nonperturbative way, 
\cite{8,9,10,11}. One of the aims is, for example, to ascertain whether a 
conformal window exists and if so for what range of parameters of the symmetry 
group. Such studies are not limited to two dimensions where the structure of 
the conformal group being infinite dimensional, is fundamentally different to 
the finite dimensional conformal group in $d$~$>$~$2$ where $d$ is the 
spacetime dimension. Instead models exhibiting conformal symmetry in three, 
four and higher dimensions are of interest. One motivation for such studies 
rested in part on applications to dualities in higher spin AdS/CFT's 
\cite{12,13} as well as model building beyond the Standard Model. Indeed in 
terms of gauge theories the original study in quantum chromodynamics (QCD) of 
\cite{14} suggested that for a certain range of the number of quarks there 
could be a nontrivial fixed point in strictly {\em four} dimensions. Such a 
fixed point, which is known as the Banks-Zaks fixed point, may have a 
connection with the chiral symmetry phase transition, \cite{14}. Another 
motivation for studying conformal properties in higher dimensional theories 
rests in trying to generalize properties of theorems such as the two
dimensional $c$-theorem, \cite{15}, to analogues in three and four dimensions. 
Though to have some insight into such extensions one has to be aware of the 
fixed point structure of the underlying quantum field theory. In this respect 
there has been interest in tackling this problem in scalar $\phi^3$ theory in 
six and lower dimensions. See, for example, \cite{16,17,18,19} for recent in 
depth studies of the $O(N)$ symmetric $\phi^3$ theory. For instance, the fixed 
point structure has been comprehensively studied perturbatively to three loops 
in \cite{18,20}. This has also subsequently been extended to the theory with an
$Sp(N)$ symmetry in \cite{21}.

Those works exploit the renormalization of the theory from a generation ago,
\cite{22,23,24}, when modern multiloop computational techniques were not 
available. One highlight of \cite{18,20} was the estimate of the conformal 
window. It is possible to compute order by order in perturbation theory the 
value of $N$ as a function of the spacetime dimension for which the stable 
infrared fixed point ceases to exist. This critical value is denoted by 
$N_{cr}$. Evaluating the expression for $N_{cr}$ in five dimensions the one 
loop result of \cite{25,18} was that the leading order value is 
$N_{cr}$~$=$~$1038$. Indeed the first examination of the conformal window in
six dimensional $O(N)$ $\phi^3$ theory was given in \cite{25}. In \cite{20} the
three loop analysis reduced this to $N_{cr}$~$=$~$64$. Such a large reduction 
gives credence to the hope that the value can be reduced further. Indeed there 
appears to be support for this in conformal bootstrap analyses, 
\cite{26,27,28}, and moreover the series for $N_{cr}$ in \cite{20} appears to 
be convergent. However, partly in order to resolve this but mainly for 
applications to other problems, the primary aim of this article is to extend 
the work of \cite{22,23,24} to {\em four} loops. This is now possible given the
advances in methods to evaluate high loop massless Feynman diagrams and in 
particular $2$-point functions in the period since \cite{23,24}. Moreover, 
while this will involve a large amount of integration such a renormalization 
could not proceed in the absence of powerful symbolic manipulation languages as
well as powerful computing resources. Therefore, we will not only provide a 
comprehensive analysis of the four loop structure of the renormalization group 
functions of $\phi^3$ theory in the modified minimal subtraction ($\MSbar$) 
scheme, but we will also outline the computational algorithm used. In this 
respect we will provide the basic four loop massless $2$-point function Feynman
integrals which are central to the calculations to allow others to extend the 
programme to other models.

In indicating our analysis will be extensive this not only means that we will
focus on the $O(N)$ symmetric theory but also simpler models where there is no
or different symmetry. This is because of the connection of these theories to
condensed matter or statistical physics problems. The simplest such case is the
theory with one scalar field. It is of interest because of the relation to
the Lee-Yang edge singularity problem, \cite{29}. When the coupling constant
of the single field theory is purely imaginary, \cite{29}, then the critical 
exponent $\sigma$, which is determined at the $d$-dimensional Wilson-Fisher 
fixed point, is the main quantity of physical interest. It can be determined
via the renormalization group functions of the theory. The exponent $\sigma$
has been estimated by different techniques such as high temperature series and 
Monte Carlo methods for the discrete dimensions less than six. Several 
references to such work are \cite{30,31,32}, for instance. More recently the 
conformal bootstrap programme have been applied to it, \cite{33}. With these
evidently more powerful methods giving reasonable agreement it is therefore 
timely to extend the $\epsilon$-expansion analysis of critical exponents to 
$O(\epsilon^4)$ where we will use $d$~$=$~$6$~$-$~$2\epsilon$ throughout. Not 
only will this improve the estimates given in \cite{23,24} but we will follow 
the method of \cite{32} where constrained Pad\'{e} approximants were used. 
Central to this idea is the exploitation of the properties of the underlying 
{\em two} dimensional conformal field theory. There the corresponding critical 
exponents are known exactly. This information is used like a boundary condition
on the Pad\'{e} approximant and, as will be evident, will significantly improve 
exponent estimates for large values of $\epsilon$. It will transpire that this 
analytic perturbative approach will give exponent estimates which are not 
unreasonable in comparison with numerically intense alternatives. While these 
comments have been driven by the Lee-Yang edge singularity connection to 
$\phi^3$ theory, they will equally apply to a similar, in terms of the 
underlying quantum field theory, but different physical problem. This is the 
percolation problem which can be formulated in continuum $\phi^3$ theory by a 
multiplet of scalar fields with a group valued coupling constant, \cite{34}. In
the appropriate replica limit the critical exponents at the Wilson-Fisher fixed 
point equate to the $1$-state Potts model and hence percolation. We will 
provide estimates for a large set of critical exponents in the dimensions 
between two and six. Again it will be the case that the estimates are in 
keeping with other approaches.

That we draw attention to these applications in the same context as the recent
$O(N)$ $\phi^3$ theory studies is not unconnected. As outlined in \cite{20} the
$O(N)$ theory has to have a connectivity with an underlying two dimensional
conformal field theory in respect of the various critical points which emerge.
If ultimately the $d$-dimensional theory can be identified with such a two 
dimensional theory then it will open the possibility of extended constrained
Pad\'{e} approximants to the exponents of the various fixed points of the 
$O(N)$ theory. Though it is not currently clear what the application to a 
physical problem is at the moment. One other motivation in \cite{18} was to 
make the connection of $\phi^3$ theory with $O(N)$ $\phi^4$ theory in the 
dimension range $4$~$<$~$d$~$<$~$6$ at the Wilson-Fisher fixed point via the 
large $N$ expansion in $d$-dimensions. It is well known that in 
$2$~$<$~$d$~$<$~$4$ the $O(N)$ $\phi^4$ theory is in the same universality 
class as the $O(N)$ nonlinear $\sigma$ model and the three dimensional 
Heisenberg ferromagnet. Above four dimensions it turns out that at the 
Wilson-Fisher fixed point the $d$-dimensional theory is in the same 
universality class as one of the $O(N)$ $\phi^3$ fixed points. So using the 
large $N$ expansion we can evaluate the four loop $d$-dimensional critical 
exponents and compare them with those known from the explicit large $N$ 
expansion, \cite{35,36,37,38}. This will provide a nontrivial check on our 
perturbative results. Indeed we will compute the full mixing matrix for the 
mass operators in the $O(N)$ case and show the subtlety in connecting the mass 
eigen-critical exponents with the two separate mass critical epxonents in the 
$d$-dimensional $O(N)$ $\phi^4$ theory. In this context it is interesting to 
note that the determination of the exponent $\eta$ at $O(1/N^3)$, \cite{37}, 
was originally evaluated as a function of $d$ by using a $d$-dimensional 
conformal bootstrap method but in an analytic as opposed to the modern
numerical approach.

The article is organized as follows. We summarize the relevant background to
performing the four loop renormalization of the basic six dimensional $\phi^3$
theory in section $2$. The notation used throughout is given there as well as
the technical details of how the computation was organized. In particular we
emphasise that all the basic renormalization constants can be extracted purely
from an evaluation of the $2$-point function. Results of this method are 
summarized in the subsequent section where the renormalization group functions
are recorded for various symmetry configurations. Included in this are 
theories with $SU(\Nc)$ symmetry which were examined as early toy examples of
the strong interactions. Sections $4$ and $5$ provide details respectively of 
the results for the Lee-Yang edge singularity and percolation problems. The 
final main problem we analyse is the full four loop structure of the $O(N)$ 
version of $\phi^3$ theory which is provided in section $6$. Conclusions are 
given in section $7$. In addition two appendices are provided. The first gives 
the $\epsilon$ expansions of the relevant basic but nontrivial integrals we 
needed at four loops. The second extends the three loop results for various 
values of $N$ given in the appendix A of \cite{28} to four loops but also 
includes the mass eigen-critical exponents for each fixed point as an expansion
in $\epsilon$.
 
\sect{Background.}

In this section we outline the technical aspects of the four loop 
renormalization of six dimensional $\phi^3$ theory. We will consider the 
theory in different guises depending on the particular application required
to extract, for instance, critical exponents. These will depend essentially on
different decorations of the underlying Lagrangian, which is
\begin{equation}
L ~=~ \frac{1}{2} \left( \partial_\mu \phi \right)^2 ~+~ \frac{g}{6} \phi^3
\label{lagphi3}
\end{equation}
with various symmetry groups. Here $g$ is the coupling constant which is
dimensionless in six dimensions. For example, both the Lee-Yang edge 
singularity and percolation problems can be accommodated with the more general 
theory, \cite{23,24}, 
\begin{equation}
L ~=~ \frac{1}{2} \left( \partial_\mu \phi^i \right)^2 ~+~ 
\frac{g}{6} d^{ijk} \phi^i \phi^j \phi^k
\label{lagphi3d}
\end{equation}
where there are a multiplet of fields $\phi^i$ and a group theory tensor which 
is totally symmetric in its indices in addition to a coupling constant $g$. The
Lagrangians for theories with more than one set of fields will be discussed in 
later sections but the underlying calculational procedure is effectively the 
same as that introduced here. For (\ref{lagphi3d}) we use the same assumptions 
as \cite{23,24} for the renormalizability of the Lagrangian. Briefly this 
reduces to two observations. First, for $2$-point self-energy graphs, including
subgraphs, the product of two coupling tensors satisfies 
\begin{equation}
d^{i_1 i_3 i_4} d^{i_2 i_3 i_4} ~=~ T_2 \delta^{i_1 i_2} 
\end{equation}
where we use a different and more systematic notation here compared to 
\cite{23,24} for what corresponds to the group theory Casimirs. The second 
assumption in \cite{23,24} is that the product of coupling tensors is such that
if the product has three free indices then it is proportional to $d^{ijk}$ 
itself. In this way (\ref{lagphi3d}) will clearly be renormalizable. Given this
it might be thought that there is a sizeable number of group invariants which 
can appear at high loop order. This is not the case as it transpires that to 
four loops the following Casimirs suffice to write the renormalization group 
functions in a concise form. These are 
\begin{eqnarray}
&& d^{i i_1 i_2} d^{j i_1 i_3} d^{k i_2 i_3} ~=~ 
T_3 d^{i j k} ~~,~~ 
d^{i i_1 i_2} d^{j i_3 i_4} d^{k i_5 i_6} d^{i_1 i_3 i_5} 
d^{i_2 i_4 i_6} ~=~ T_5 d^{i j k} \nonumber \\
&& d^{i i_1 i_2} d^{j i_3 i_4} d^{k i_5 i_6} d^{i_1 i_3 i_7} d^{i_2 i_5 i_8} 
d^{i_4 i_6 i_9} d^{i_7 i_8 i_9} ~=~ T_{71} d^{i j k} \nonumber \\
&& d^{i i_1 i_2} d^{j i_3 i_4} d^{k i_5 i_6} d^{i_1 i_3 i_7} d^{i_2 i_5 i_8} 
d^{i_4 i_8 i_9} d^{i_6 i_7 i_9} ~=~ T_{72} d^{i j k} \nonumber \\
&& d^{i i_1 i_2} d^{j i_3 i_4} d^{k i_5 i_{12}} d^{i_1 i_5 i_6}
d^{i_2 i_7 i_8} d^{i_3 i_9 i_{12}} d^{i_4 i_7 i_{10}}
d^{i_6 i_8 i_{11}} d^{i_9 i_{10} i_{11}} ~=~ T_{91} d^{i j k} \nonumber \\
&& d^{i i_1 i_2} d^{j i_3 i_4} d^{k i_{11} i_{12}}
d^{i_1 i_5 i_6} d^{i_2 i_7 i_8} d^{i_3 i_5 i_9}
d^{i_4 i_7 i_{10}} d^{i_6 i_8 i_{11}} d^{i_9 i_{10} i_{12}} ~=~ T_{92} 
d^{i j k} \nonumber \\
&& d^{i i_1 i_2} d^{j i_3 i_4} d^{k i_6 i_{12}}
d^{i_1 i_5 i_6} d^{i_2 i_7 i_8} d^{i_3 i_5 i_9}
d^{i_4 i_7 i_{10}} d^{i_8 i_{11} i_{12}} d^{i_9 i_{10} i_{11}} ~=~ 
T_{93} d^{i j k} \nonumber \\
&& d^{i i_1 i_2} d^{j i_3 i_4} d^{k i_5 i_{12}}
d^{i_1 i_5 i_6} d^{i_2 i_7 i_8} d^{i_3 i_9 i_{12}}
d^{i_4 i_{10} i_{11}} d^{i_6 i_7 i_{10}} d^{i_8 i_9 i_{11}} ~=~ 
T_{94} d^{i j k} \nonumber \\
&& d^{i i_1 i_2} d^{j i_3 i_4} d^{k i_8 i_{12}}
d^{i_1 i_5 i_6} d^{i_2 i_7 i_8} d^{i_3 i_5 i_9}
d^{i_4 i_{10} i_{11}} d^{i_6 i_7 i_{10}} d^{i_9 i_{11} i_{12}} ~=~ 
T_{95} d^{i j k} \nonumber \\
&& d^{i i_1 i_2} d^{j i_3 i_4} d^{k i_{11} i_{12}}
d^{i_1 i_3 i_5} d^{i_2 i_6 i_7} d^{i_4 i_6 i_8}
d^{i_5 i_9 i_{10}} d^{i_7 i_9 i_{11}} d^{i_8 i_{10} i_{12}} ~=~ 
T_{96} d^{i j k} \nonumber \\
&& d^{i i_1 i_2} d^{j i_3 i_4} d^{k i_5 i_{12}}
d^{i_1 i_5 i_6} d^{i_2 i_7 i_8} d^{i_3 i_9 i_{12}}
d^{i_4 i_7 i_{10}} d^{i_6 i_{10} i_{11}} d^{i_8 i_9 i_{11}} ~=~ 
T_{97} d^{i j k} \nonumber \\
&& d^{i i_1 i_2} d^{j i_3 i_4} d^{k i_6 i_{12}}
d^{i_1 i_5 i_6} d^{i_2 i_7 i_8} d^{i_3 i_5 i_9}
d^{i_4 i_7 i_{10}} d^{i_8 i_9 i_{11}} d^{i_{10} i_{11} i_{12}} ~=~ 
T_{98} d^{i j k} \nonumber \\
&& d^{i i_1 i_2} d^{j i_3 i_4} d^{k i_{11} i_{12}}
d^{i_1 i_5 i_6} d^{i_2 i_7 i_8} d^{i_3 i_5 i_9}
d^{i_4 i_7 i_{10}} d^{i_6 i_{10} i_{11}} d^{i_8 i_9 i_{12}} ~=~ 
T_{99} d^{i j k} ~. 
\label{tensT}
\end{eqnarray}
where we have included the previous invariants up to three loops of 
\cite{23,24}. So at most there are nine new group invariants at four loops in 
principle. Though it will turn out that while these nine correspond to 
different graph topologies the four loop Casimirs, $T_{9m}$, do not all have 
distinct values. The syntax here is that the first label on $T_n$ corresponds 
to the number of tensors $d^{ijk}$ in the product and the second label 
distinguishes between different invariants. If the first label is even then it 
arises in a $2$-point function while an odd label indicates an invariant which 
will appear in the coupling constant renormalization. Labelling the actual
group indices in each product in (\ref{tensT}) with subscripts allows one to 
straightforwardly construct the underlying topologies graphically. Each paired 
index corresponds to a propagator and these will join at vertices defined by 
each coupling tensor. To make contact with the notation used in \cite{23,24} 
for the invariants up to three loops, we note that $T_2$~$=$~$\alpha$, 
$T_3$~$=$~$\beta$, $T_5$~$=$~$\gamma$, $T_{71}$~$=$~$\delta$ and 
$T_{72}$~$=$~$\lambda$ when one observes that the defining graphs given in 
Figure $1$ of \cite{23} appear in reversed mirror image. We prefer the notation
$T_n$ here to avoid confusion with the standard notation of various critical 
exponents which also use Greek letters.

Although the renormalization of both (\ref{lagphi3}) and (\ref{lagphi3d}) are
the same we will focus the technical discussion on the former as the extension 
to tensor couplings is not onerous. For (\ref{lagphi3}) we have to determine
the wave function and coupling constant renormalization constants. Once these
have been established then the determination of the mass anomalous dimension
follows as a corollary even though the Lagrangian is massless. Indeed we work
with massless fields throughout as this renders the four loop $\beta$-function
accessible with a relatively minimal amount of computation. The first stage is 
to compute the $2$-point and $3$-point functions of (\ref{lagphi3}). The 
Feynman graphs are generated using the {\sc Qgraf} package, \cite{39}. There 
are respectively $1$, $2$, $10$ and $64$ one, two, three and four loops 
diagrams to evaluate to the simple pole in $\epsilon$ in dimensional 
regularization which we use throughout. Indeed we note that the index labelling
in (\ref{tensT}) is based on the {\sc Qgraf} output. We note that throughout 
our convention is $d$~$=$~$6$~$-$~$2\epsilon$. For the $3$-point vertex the 
respective numbers are $1$, $7$, $56$ and $540$ which is an order of magnitude 
increase at four loops compared to the $2$-point case. However, we will compute
the coupling constant renormalization constant for (\ref{lagphi3}) purely from 
the $2$-point graphs by exploiting certain properties of the specific field 
theory. It is based on the following observation. If for the moment one 
considers the massive extension of (\ref{lagphi3}) then the propagator can be 
formally expanded in powers of $m^2$ where $m$ is the mass via  
\begin{equation}
\frac{1}{[k^2-m^2]} ~=~ \frac{1}{k^2} ~+~ \frac{m^2}{(k^2)^2} ~+~ O(m^4) ~.
\label{propexp}
\end{equation}
The first term on the right hand side of course corresponds to the massless 
theory. However, the second term represents the zero momentum insertion of the 
unit operator on a propagator. Diagrammatically for the self energy 
renormalization this corresponds to a $2$-point function with a zero momentum 
insertion but more importantly this term would correspond to a $3$-point 
function graph where one of the external legs has a nullified momentum. In 
other words this would be equivalent to a graph contributing to the coupling 
constant renormalization. As we are computing in dimensional regularization and
only interested in the $\MSbar$ renormalization scheme the coupling constant
renormalization constant can be correctly extracted from this nullified
external momentum configuration. Indeed in four dimensional gauge theories this
is the standard procedure for three loop renormalization \cite{40,41}. One 
concern is that nullifying an external momentum in (\ref{lagphi3}) could 
introduce unwanted infrared divergences which would be indistinguishable from 
ultraviolet ones in dimensional regularization. Indeed in four dimensions if a 
propagator was present in a massless graph which had the form of the second 
term of (\ref{propexp}) then it would be infrared singular. That it is not an 
issue for (\ref{lagphi3}) is because in six dimensions such a propagator is by 
contrast infrared safe even in scalar field theories unlike in four dimensions.
Therefore, this observation radically reduces the number of Feynman graphs to 
be computed. One downside is that there are more integrals to determine but 
this is surmountable as we will indicate later. One concern which may arise 
with the expansion approach is that symmetry factors in the $2$-point function 
may not be set in such a way that after expansion the term relating to the 
vertex has the incorrect factor. We have checked that it is the case to three 
loops. We illustrate this at one loop. There the basic one loop self-energy 
graph has a symmetry factor of $\half$ but the one loop $3$-point vertex has a 
unit symmetry factor. However, using (\ref{propexp}) there are two $O(m^2)$ 
terms which cancel the factor of $\half$ so that everything tallies. One final 
point resides in the determination of the renormalization of the mass operator 
$\half \phi^2$ in (\ref{lagphi3}). It transpires that in this theory the mass 
and coupling constant renormalization are equivalent. This is implicit in the 
algorithm we discussed for the determination of the latter and has already been
noted in, for instance, \cite{23,24}. So in (\ref{lagphi3}) no extra 
computation has to be carried out. However, for (\ref{lagphi3d}) the mass and 
coupling constant renormalization constants are different. This is evident in 
the algorithm we have introduced. Both can be determined using our method of 
only computing the graphs for the $2$-point function by extending 
(\ref{propexp}) and mapping each massless propagator with
\begin{equation}
\frac{\delta^{ij}}{k^2} ~\mapsto~ \frac{\delta^{ij}}{k^2} ~+~ 
\frac{m_1^2\delta^{ij}}{(k^2)^2} ~+~ \frac{m_2^2 g d^{ijk_e}}{(k^2)^2} 
\label{propexpd}
\end{equation}
after the graphs have been generated with {\sc Qgraf}. We have included the
group theory structure on the propagators and $m_i^2$ are parameters of the
dimension of a mass. These parameters are included as a label to distinguish 
which term is which when one comes to extract the various wave function, mass 
and coupling constant renormalization constants in the sum of all the 
contributions to our $2$-point function. That involving $m_1$ corresponds to 
the mass operator. The final term of (\ref{propexpd}) with label $m_2$ 
corresponds to the zero momentum insertion for the renormalization of the
coupling constant $g$. The index $k_e$ in (\ref{propexpd}) is the index
associated with the third leg of the $3$-point function and is a fixed free
external index.  

Having discussed the method to isolate the necessary graphs and integrals
contributing to the wave function, coupling constant and mass operator
renormalization constants we now discuss the method used to determine their
divergent part. The approach is to use the Laporta algorithm, \cite{42}. This
is a systematic application of integration by parts to establish towers of
relations between integrals which are then solved algebraically to express all
integrals in terms of a relatively small set of basic integrals which are
called masters. The values for these have to be determined directly as 
integration by parts can no longer be used. Once these are found then the
integration algorithm is complete and we run an automatic determination of
the respective Green's functions to find the underlying renormalization 
constants. Crucial to this is the symbolic manipulation language {\sc Form},
\cite{43}, and its threaded version {\sc Tform}, \cite{44}, which we use 
extensively to handle the large amounts of algebra. This arises since the 
reduction to masters produces relations where the coefficients are large 
rational polynomials in $d$. Not only have these to be expanded in powers of 
$\epsilon$ and the masters substituted but the denominators may contain factors
of $(d-6)$ which are termed spurious poles. Thus not only have the polynomial 
coefficients to be expanded to higher order in $\epsilon$ but some masters need
to be evaluated to a precision beyond the simple pole. Even for the simple 
scalar theory, (\ref{lagphi3}), where there are no tensor integrals certain 
graphs become tedious to evaluate. For the present four loop computation we 
have used the {\sc Reduze} package, \cite{45}. One useful feature is that the 
output relations between integrals can be readily converted to {\sc Form} input 
notation and thus included as a module within the automatic {\sc Form} 
evaluation.

The final ingredient in the algorithm is the explicit values for the master
integrals. This requires a method other than integration by parts and we have
adapted several approaches which have been used for similar problems in other
contexts such as \cite{45,46}. In addition we were able to exploit a feature of
{\sc Reduze} which means that we could do this systematically. By this we mean 
the following. When the relation between integrals are solved in {\sc Reduze} 
the algorithm has some internal criterion for determining which integrals are 
to be used as the masters in terms of which all other integrals are expressed. 
This is not always the best choice for the calculation of interest. For 
instance, some of the masters may be simple self-energy graphs where each 
subgraph is itself a lower loop self-energy graph or products of lower loop 
graphs. Such integrals are straightforward to determine and are retained in the
set of masters. However, more difficult integrals remain such as those which 
are primitively divergent and it is their simple and only pole which will be 
needed. The feature of {\sc Reduze} which one exploits is that one can specify 
a set of integrals which the package identifies as the masters. This is done 
after the initial reduction has been determined and a database constructed, 
\cite{46}. Therefore, we have chosen a set of basic master integrals which, 
aside from those which are simple to evaluate, are finite using Weinberg's 
theorem, \cite{47}. This technique has been elaborated on recently in 
\cite{48}. While this may seem to resolve the computational algorithm one has 
to be careful. This is because of the problem of spurious poles in $\epsilon$ 
which means that not only the leading term of these finite integrals are 
required, but sometimes also several terms in powers of $\epsilon$. However, 
one also has to have some information on a master choice to be able to solve 
for these and other required coefficients. 

To do this we use the known four loop self-energy master integrals given in 
\cite{46}. These were determined by application of the glue-and-cut method of 
{\em five} loop primitive massless vacuum diagrams to varying orders in the
$\epsilon$ expansion. The results are consistent with a subsequent independent 
numerical sector decomposition evaluation given in \cite{49}. Analytic 
evaluations were also developed thereafter in \cite{50} to even higher powers 
in the $\epsilon$ expansion compared to \cite{46}. The only problem is that 
those masters were computed near {\it four} and not six dimensions which is the
dimension we require them for to complete our master integral determination.
This is achieved by using the method of \cite{51,52}. It allows one to relate a
$d$-dimensional integral to a set of integrals in $(d+2)$-dimensions. The 
latter set will always be of the same topology as the original lower 
dimensional one but with increased propagator powers. Indeed if the 
$d$-dimensional integral has $P$ propagators then the $(d+2)$-dimensional 
integrals will have $(P+L)$ propagators by simple power counting where $L$ is
the number of loops. In our case if the four dimensional integral is a known 
master then it can be related to the as yet undetermined six dimensional 
integral. Though to do this one has to use the {\sc Reduze} database to effect 
the reduction to the unknown master and a set of previously evaluated masters. 
In other words if one builds the system of master evaluation from integrals 
with low numbers of propagators then one can move up the tower of unknown 
masters systematically until all the ones required for the renormalization have
been found. In the list of masters given in \cite{46} there is enough 
information in terms of the $\epsilon$ expansion there in $(4$~$-$~$2\epsilon)$
dimensions to ascertain all the Feynman graphs contributing to the full 
renormalization of (\ref{lagphi3d}) to four loops. To assist with future work 
we have recorded the values of certain integrals in $(6$~$-$~$2\epsilon)$ 
dimensions in Appendix A. Rather than present the ones used in our efforts, so 
that others are not restricted to our particular basis of masters, we have 
presented the values for the same topologies given in \cite{46}. These 
integrals were denoted by $M_i$ and we have given several to quite a high order
in $\epsilon$. This was partly because of the spurious pole problem. As a 
simple check on this evaluation of the four loop masters we have followed the
same procedure at three loops using the three loop four dimensional masters 
given in \cite{46}. When these were included in the automatic three loop 
renormalization we correctly reproduced the results of \cite{22,23,24}. While a
simple check we also had to do this exercise anyway as we needed higher terms 
in $\epsilon$ at three loops since each three loop integral will be multiplied 
by counterterms. Hence they contribute to the four loop renormalization
constants. Finally to effect the renormalization within the automatic 
computation we follow the method of \cite{41}. In this one computes the Green's
function in terms of the bare coupling constant. Then after all the graphs have
been summed the bare variable is rescaled by the coupling constant 
renormalization constant. This systematically introduces the appropriate 
counterterms automatically and the overall remaining divergence in the sum is 
fixed by the associated unknown renormalization constant. One advantage of this
approach is that it avoids the subtraction at the level of individual diagrams 
which is tedious and not possible to encapsulate easily in an automatic 
symbolic manipulation algorithm. Finally, with this we have provided all the 
computational pieces to fully renormalize six dimensional $\phi^3$ theory in 
the $\MSbar$ scheme at four loops.

\sect{Results.}

We are now in a position to formally record the renormalization group functions
at four loops for various formulations of six dimensional $\phi^3$ theory. 
Throughout all our results will be in the $\MSbar$ scheme and we have included 
an electronic file data with the main results of the article. For 
(\ref{lagphi3d}) we have 
\begin{eqnarray}
\beta(g) &=&
\left[ - T_2 + 4 T_3 \right] \frac{g^3}{8} ~+~ 
\left[ - 11 T_2^2 + 66 T_2 T_3 - 108 T_3^2 - 72 T_5 \right] \frac{g^5}{288}
\nonumber \\
&& +~ \left[ - 821 T_2^3 + 6078 T_2^2 T_3 - 12564 T_2 T_3^2
+ 2592 \zeta_3 T_2 T_5 - 9288 T_2 T_5 - 11664 T_3^3 
\right. \nonumber \\
&& \left. ~~~~ 
- 51840 \zeta_3 T_3 T_5 + 61344 T_3 T_5 + 20736 T_{71} + 62208 \zeta_3 T_{72} 
- 20736 T_{72} \right]
\frac{g^7}{41472} \nonumber \\
&& +~ \left[ - 20547 T_2^4 + 1728 \zeta_3 T_2^3 T_3 + 185774 T_2^3 T_3
- 31104 \zeta_3 T_2^2 T_3^2 - 510960 T_2^2 T_3^2 
\right. \nonumber \\
&& \left. ~~~~ 
+ 127008 \zeta_3T_2^2 T_5 
- 23328 \zeta_4 T_2^2 T_5 - 285336 T_2^2 T_5 + 373248 \zeta_3 T_2 T_3^3 
- 437472 T_2 T_3^3 
\right. \nonumber \\
&& \left. ~~~~ 
- 2716416 \zeta_3 T_2 T_3 T_5 + 559872 \zeta_4 T_2 T_3 T_5 
+ 2744064 T_2 T_3 T_5 + 124416 \zeta_3 T_2 T_{71} 
\right. \nonumber \\
&& \left. ~~~~ 
- 622080 \zeta_5 T_2 T_{71} 
+ 1005696 T_2 T_{71} + 2457216 \zeta_3 T_2 T_{72} - 559872 \zeta_4 T_2 T_{72} 
\right. \nonumber \\
&& \left. ~~~~ 
- 1005696 T_2 T_{72} - 1866240 \zeta_3 T_3^4 + 3005424 T_3^4
+ 8315136 \zeta_3 T_3^2 T_5 - 1866240 \zeta_4 T_3^2 T_5 
\right. \nonumber \\
&& \left. ~~~~ 
- 6763392 T_3^2 T_5 - 5474304 \zeta_3 T_3 T_{71} + 14929920 \zeta_5 T_3 T_{71} 
- 7755264 T_3 T_{71} 
\right. \nonumber \\
&& \left. ~~~~ 
- 6096384 \zeta_3 T_3 T_{72} + 2239488 \zeta_4 T_3 T_{72} 
+ 7755264 T_3 T_{72} + 1306368 \zeta_3 T_5^2 
\right. \nonumber \\
&& \left. ~~~~ 
- 1321920 T_5^2 - 7464960 \zeta_3 T_{91} + 7464960 \zeta_5 T_{91} 
- 746496 T_{91} - 7464960 \zeta_3 T_{98} 
\right. \nonumber \\
&& \left. ~~~~ 
+ 7464960 \zeta_5 T_{98} + 1492992 \zeta_3 T_{99} 
- 1866240 \zeta_5 T_{99} - 3732480 \zeta_3 T_{92} 
\right. \nonumber \\
&& \left. ~~~~ 
+ 3732480 \zeta_5 T_{92} + 10450944 \zeta_3 T_{93} - 14929920 \zeta_5 T_{93} 
- 2239488 \zeta_3 T_{94} 
\right. \nonumber \\
&& \left. ~~~~ 
+ 746496 T_{94} + 10450944 \zeta_3 T_{95} - 14929920 \zeta_5 T_{95} 
- 7464960 \zeta_3 T_{96} + 7464960 \zeta_5 T_{96} 
\right. \nonumber \\
&& \left. ~~~~ 
+ 5225472 \zeta_3 T_{97} 
- 7464960 \zeta_5 T_{97} \right] \frac{g^9}{1492992} ~+~ O(g^{11}) 
\label{betad}
\end{eqnarray}
and 
\begin{eqnarray}
\gamma_\phi(g) &=&
-~ \frac{T_2}{12} g^2
+ \left[ - 11 T_2 + 24 T_3 \right] \frac{T_2 g^4}{432} \nonumber \\
&& +~ \left[ - 821 T_2^2 + 3222 T_2 T_3 - 3060 T_3^2 + 2592 \zeta_3 T_5 
- 4536 T_5 \right] \frac{T_2 g^6}{62208} \nonumber \\
&& +~ \left[ - 2283 T_2^3 + 576 \zeta_3 T_2^2 T_3 + 12428 T_2^2 T_3
- 6912 \zeta_3 T_2 T_3^2 - 14872 T_2 T_3^2 
+ 14112 \zeta_3 T_2 T_5 
\right. \nonumber \\
&& \left. ~~~~ 
- 2592 \zeta_4 T_2 T_5 - 20928 T_2 T_5 + 41472 \zeta_3 T_3^3 - 43392 T_3^3
- 87552 \zeta_3 T_3 T_5 
\right. \nonumber \\
&& \left. ~~~~ 
- 10368 \zeta_4 T_3 T_5 + 122496 T_3 T_5
+ 13824 \zeta_3 T_{71} - 69120 \zeta_5 T_{71} + 59904 T_{71}
\right. \nonumber \\
&& \left. ~~~~ 
+ 34560 \zeta_3 T_{72} + 31104 \zeta_4 T_{72} 
- 59904 T_{72} \right] \frac{T_2 g^8}{248832} ~+~ O(g^{10}) 
\label{gammad}
\end{eqnarray}
where $\zeta_z$ is the Riemann zeta function. As a check we have reproduced the
three loop results of \cite{23,24} and moreover, the non-simple poles in
$\epsilon$ in the four loop renormalization constants correctly emerge. Their
residues depend on the poles in $\epsilon$ at the previous three loops. A
comment on our conventions is in order at the outset in relation to other
papers. As is usual, \cite{23,24}, we have absorbed the common factor of
$S(d)/(2\pi)^d$, where $S(d)$ is the surface area of the $d$-dimensional unit 
sphere, into $g^2$. This factor plays no role in the values of critical
exponents. Also in comparison with \cite{23,24} our renormalization group
functions are defined with an overall factor of $2$ different. This will be our
convention throughout this and later sections. Finally, to map our results to 
\cite{23,24} and later to \cite{20,21} the sign of $g^2$ needs to be reversed.

For the mass operator
\begin{equation}
{\cal O} ~=~ \frac{1}{2} \phi^i \phi^i
\end{equation}
the anomalous dimension is
\begin{eqnarray}
\gamma_{\cal O}(g) &=&
-~ \frac{T_2}{2} g^2
+ \left[ - T_2 + 24 T_3 \right] \frac{T_2 g^4}{48} \nonumber \\
&& +~ \left[ - 380 T_2^2 + 432 \zeta_3 T_2 T_3 + 711 T_2 T_3
- 864 \zeta_3 T_3^2 - 1170 T_3^2 - 756 T_5 \right] \frac{T_2 g^6}{1728} 
\nonumber \\
&& +~ \left[ - 34560 \zeta_3 T_2^3 + 42635 T_2^3
+ 261792 \zeta_3 T_2^2 T_3 - 69984 \zeta_4 T_2^2 T_3 + 364812 T_2^2 T_3
\right. \nonumber \\
&& \left. ~~~~ 
- 544320 \zeta_3 T_2 T_3^2 + 419904 \zeta_4 T_2 T_3^2 
- 1244160 \zeta_5 T_2 T_3^2 - 200088 T_2 T_3^2 
\right. \nonumber \\
&& \left. ~~~~ 
+ 69984 \zeta_3 T_2 T_5 
+ 23328 \zeta_4 T_2 T_5 - 505872 T_2 T_5 + 3825792 \zeta_3 T_3^3 
- 559872 \zeta_4 T_3^3 
\right. \nonumber \\
&& \left. ~~~~ 
- 2488320 \zeta_5 T_3^3 - 134784 T_3^3
- 2975616 \zeta_3 T_3 T_5 - 466560 \zeta_4 T_3 T_5 + 2488320 \zeta_5 T_3 T_5 
\right. \nonumber \\
&& \left. ~~~~ 
+ 2376000 T_3 T_5 + 870912 \zeta_3 T_{71} - 1866240 \zeta_5 T_{71} 
+ 1638144 T_{71} + 684288 \zeta_3 T_{72} 
\right. \nonumber \\
&& \left. ~~~~ 
+ 559872 \zeta_4 T_{72} + 2177280 \zeta_5 T_{72} - 1638144 T_{72} \right] 
\frac{T_2 g^8}{746496} ~+~ O(g^{10}) ~.
\label{massopd}
\end{eqnarray}
The three loop piece agrees with \cite{23,24} and we note that we have followed
the convention used there to include the wave function renormalization constant
as part of the operator renormalization constant. We have checked that 
$\gamma_{\cal O}(g)$ is equivalent to the $\beta$-function when the expressions 
are reduced to the single coupling theory (\ref{lagphi3}). In the conventions
of \cite{23,24} the relevant relation is 
\begin{equation}
\frac{3}{2} \gamma_\phi(g) - \gamma_{\cal O}(g) ~=~ \beta(g) 
\end{equation}
for (\ref{lagphi3}). For completeness we reproduce the independent 
renormalization group functions for (\ref{lagphi3}) which are 
\begin{eqnarray}
\beta(g) &=& \frac{3}{8} g^3 ~-~ \frac{125}{288} g^5 ~+~ 
5 [ 2592 \zeta_3 + 6617 ] \frac{g^7}{41472} \nonumber \\ 
&& +~ [ -~ 4225824 \zeta_3 + 349920 \zeta_4 + 1244160 \zeta_5 - 3404365 ]
\frac{g^9}{1492992} ~+~ O(g^{11}) \nonumber \\
\gamma_\phi(g) &=& -~ \frac{1}{12} g^2 ~+~ \frac{13}{432} g^4 ~+~ 
[ 2592 \zeta_3 - 5195 ] \frac{g^6}{62208} \nonumber \\
&& +~ [ 10080 \zeta_3 + 18144 \zeta_4 - 69120 \zeta_5 + 53449 ] 
\frac{g^8}{248832} ~+~ O(g^{10}) 
\label{phi3rge}
\end{eqnarray}
which will be relevant to the Lee-Yang edge singularity problem.

As a final application of the renormalization it is interesting to consider the
situation when the $\phi^3$ theory is endowed with $SU(\Nc)$ symmetry. Such
theories were considered at two and three loops in \cite{53,54} as simple 
models of the strong interactions in four dimensions and $\Nc$ denotes the
number of colours. The motivation was partly due to the six dimensional theory 
being asymptotically free and so it could have parallel properties to QCD. 
Indeed this was in part the starting point for the study in \cite{25}. The 
other motivation in \cite{53,54} rested in the idea that the ultraviolet 
properties of one theory could be regarded as being driven by the infrared 
behaviour of another. So the aim in \cite{53,54} was to determine the relevant 
$d$-dimensional critical exponents. In light of the recent development of the 
conformal bootstrap method which aims at examining the fixed point properties 
of various scalar theories, we will therefore extend the results of 
\cite{53,54} to four loops. This is in order to provide complementary data for 
future bootstrap studies. In \cite{53,54} two $\phi^3$ theories with underlying 
$SU(\Nc)$ symmetry were considered. The first scenario was when the tensor 
$d^{ijk}$ is identified as the totally symmetric rank $3$ tensor in $SU(\Nc)$. 
In this case the group invariants become 
\begin{eqnarray}
T_2 &=& \frac{[ \Nc^2 - 4]}{\Nc} ~~,~~ 
T_3 ~=~ \frac{1}{2\Nc} [ \Nc^2 - 12 ] ~~,~~ 
T_5 ~=~ -~ \frac{4}{\Nc^2} [ \Nc^2 - 10 ] \nonumber \\ 
T_{71} &=& \frac{1}{8\Nc^3} [ \Nc^2 - 8 ] [ \Nc^4 - 8 \Nc^2 + 256 ] ~~,~~
T_{72} ~=~ -~ \frac{1}{2\Nc^3} [ \Nc^4 - 68 \Nc^2 + 528 ]
\nonumber \\
T_{91} &=& [ \Nc^8 - 20 \Nc^6 + 352 \Nc^4 - 5120 \Nc^2 + 26880 ] 
\frac{1}{16 \Nc^4}
\nonumber \\
T_{92} &=& T_{96} ~=~ T_{98} ~=~ [ 3 \Nc^6 - 16 \Nc^4 - 896 \Nc^2 + 7296 ] 
\frac{1}{4 \Nc^4} \nonumber \\
T_{93} &=& T_{95} ~=~ T_{97} ~=~ [ \Nc^6 + 16 \Nc^4 - 1024 \Nc^2 + 7104 ] 
\frac{1}{4 \Nc^4} \nonumber \\
T_{94} &=& -~ [ \Nc^6 - 64 \Nc^4 + 1216 \Nc^2 - 6784 ] \frac{1}{4 \Nc^4} 
\nonumber \\
T_{99} &=& [ 5 \Nc^6 - 72 \Nc^4 - 640 \Nc^2 + 7680 ] \frac{1}{4 \Nc^4} 
\end{eqnarray}
where we have made use of the properties of $d^{ijk}$ given in \cite{55} when 
the indices $i$ take values in the adjoint representation. In this instance the
renormalization group functions are
\begin{eqnarray} 
\beta(g) &=& [\Nc^2 - 20] \frac{g^3}{8 \Nc} 
+ [ - 5 \Nc^4 + 496 \Nc^2 - 5360 ] \frac{g^5}{288 \Nc^2} \nonumber \\
&& +~ [ 211 \Nc^6 + 62208 \zeta_3 \Nc^4 - 27132 \Nc^4 - 20736 \zeta_3 \Nc^2 
+ 1220688 \Nc^2 - 4396032 \zeta_3 \nonumber \\
&& ~~~~-~ 9272896 ] \frac{g^7}{41472 \Nc^3} 
\nonumber \\
&& +~ [ - 870048 \zeta_3 \Nc^8 + 1321920 \zeta_5 \Nc^8 - 327893 \Nc^8
+ 14427072 \zeta_3 \Nc^6 + 559872 \zeta_4 \Nc^6 
\nonumber \\
&& ~~~~
-~ 31570560 \zeta_5 \Nc^6 
+ 8142840 \Nc^6 - 155416320 \zeta_3 \Nc^4 - 11384064 \zeta_4 \Nc^4 
\nonumber \\
&& ~~~~
+~ 421770240 \zeta_5 \Nc^4 - 112740480 \Nc^4 + 1477343232 \zeta_3 \Nc^2 
- 35831808 \zeta_4 \Nc^2 
\nonumber \\
&& ~~~~
-~ 1950842880 \zeta_5 \Nc^2 + 1264882304 \Nc^2 
- 7029669888 \zeta_3 + 791285760 \zeta_4 
\nonumber \\
&& ~~~~
-~ 995328000 \zeta_5 - 5761837824] 
\frac{g^9}{1492992 \Nc^4} ~+~ O(g^{11}) \nonumber \\
\gamma_\phi(g) &=& [ - \Nc^2 + 4 ] \frac{g^2}{12 \Nc} 
+ [ \Nc^4 - 104 \Nc^2 + 400 ] \frac{g^4}{432 \Nc^2} \nonumber \\
&& +~ [ 25 \Nc^6 - 10368 \zeta_3 \Nc^4 + 17196 \Nc^4 + 145152 \zeta_3 \Nc^2 
- 296592 \Nc^2 - 414720 \zeta_3 \nonumber \\
&& ~~~~
+~ 909632 ] \frac{g^6}{62208 \Nc^3} \nonumber \\
&& +~ [ 5472 \zeta_3 \Nc^8 - 8640 \zeta_5 \Nc^8 + 2277 \Nc^8 
- 92160 \zeta_3 \Nc^6 + 15552 \zeta_4 \Nc^6 + 172800 \zeta_5 \Nc^6 
\nonumber \\
&& ~~~~
-~ 57760 \Nc^6 + 804096 \zeta_3 \Nc^4 + 393984 \zeta_4 \Nc^4 
- 3317760 \zeta_5 \Nc^4 + 2122784 \Nc^4 
\nonumber \\
&& ~~~~
-~ 4018176 \zeta_3 \Nc^2 
- 7133184 \zeta_4 \Nc^2 + 28753920 \zeta_5 \Nc^2 - 22816768 \Nc^2 
\nonumber \\
&& ~~~~
+~ 7704576 \zeta_3 + 21233664 \zeta_4 - 70778880 \zeta_5 + 60416256 ]
\frac{g^8}{248832 \Nc^4} ~+~ O(g^{10}) \nonumber \\
\gamma_{\cal O}(g) &=& [ - \Nc^2 + 4 ] \frac{g^2}{2 \Nc} 
+ [ 11 \Nc^4 - 184 \Nc^2 + 560 ] \frac{g^4}{48 \Nc^2} \nonumber \\
&& +~ [ - 317 \Nc^6 + 1728 \zeta_3 \Nc^4 + 8664 \Nc^4 - 27648 \zeta_3 \Nc^2 
- 90960 \Nc^2 + 82944 \zeta_3 \nonumber \\
&& ~~~~
+~ 245504 ] \frac{g^6}{1728 \Nc^3} \nonumber \\
&& +~ [ 547344 \zeta_3 \Nc^8 - 855360 \zeta_5 \Nc^8 + 362939 \Nc^8 
- 14211072 \zeta_3 \Nc^6 + 839808 \zeta_4 \Nc^6 
\nonumber \\
&& ~~~~
+~ 20995200 \zeta_5 \Nc^6 
- 8790080 \Nc^6 + 166116096 \zeta_3 \Nc^4 - 12503808 \zeta_4 \Nc^4 
\nonumber \\
&& ~~~~
-~ 170449920 \zeta_5 \Nc^4 + 118589664 \Nc^4 - 943681536 \zeta_3 \Nc^2 
+ 64198656 \zeta_4 \Nc^2 
\nonumber \\
&& ~~~~
+~ 423014400 \zeta_5 \Nc^2 - 812965376 \Nc^2 
+ 1886257152 \zeta_3 - 110481408 \zeta_4 
\nonumber \\
&& ~~~~
-~ 89579520 \zeta_5 + 1824079616 ]
\frac{g^8}{746496 \Nc^4} ~+~ O(g^{10}) 
\end{eqnarray}
where the two loop expressions for $\beta(g)$ and $\gamma_\phi(g)$ are in
agreement with \cite{53} when converted to the conventions used there. For 
large $\Nc$ the $\beta$-function is an alternating series. One interesting case
is when $\Nc$~$=$~$4$ when the $\beta$-function has a Banks-Zaks type fixed 
point, \cite{14}, in strictly six dimensions since
\begin{eqnarray}
\beta^{SU(4)}(g) &=& -~ \frac{1}{8} g^3 ~+~ \frac{9}{32} g^5 ~+~ 
[ 6480 \zeta_3 + 2417 ] \frac{g^7}{1536} \nonumber \\
&& +~ [ - 3053376 \zeta_3 - 58320 \zeta_4 + 4786560 \zeta_5 - 364729 ]
\frac{g^9}{55296} ~+~ O(g^{11}) \nonumber \\
\gamma^{SU(4)}_\phi(g) &=& -~ \frac{1}{4} g^2 ~-~ \frac{7}{48} g^4 ~+~ 
[ - 48 \zeta_3 + 43 ] \frac{g^6}{256} \nonumber \\ 
&& +~ [ 18864 \zeta_3 + 10368 \zeta_4 - 46080 \zeta_5 + 21907 ] 
\frac{g^8}{9216} ~+~ O(g^{10}) \nonumber \\
\gamma^{SU(4)}_{\cal O}(g) &=& -~ \frac{3}{2} g^2 ~+~ \frac{9}{16} g^4 ~+~ 
3 [ 2 \zeta_3 - 7 ] \frac{g^6}{8} \nonumber \\
&& +~ [ 336384 \zeta_3 + 55728 \zeta_4 - 338400 \zeta_5 + 335503 ] 
\frac{g^8}{9216} ~+~ O(g^{10}) ~.
\end{eqnarray}

The second case considered in \cite{53,54} was when the symmetry group is 
$SU(3)$~$\times$~$SU(3)$. We summarize the relevant background to the 
construction of that Lagrangian, as it is more involved than the previous case,
before detailing the extraction of our results. The main ingredient is that 
$\phi^i$ becomes a complex field and the Lagrangian (\ref{lagphi3d}) is 
extended to group to
\begin{equation}
L ~=~ \partial_\mu \bar{\phi}^i \partial^\mu \phi^i ~+~
\frac{g}{6} d^{ijk} \left( \phi^i \phi^j \phi^k ~+~
\bar{\phi}^i \bar{\phi}^j \bar{\phi}^k \right) ~.
\label{lagphi3su3}
\end{equation}
To implement the $SU(3)$~$\times$~$SU(3)$ symmetry the coupling tensor is 
defined by
\begin{equation}
d^{ijk} ~=~ \epsilon_{\alpha \beta \gamma} \epsilon_{\lambda \chi \xi}
T^i_{\alpha\lambda} T^j_{\beta\chi} T^k_{\gamma\xi}
\end{equation}
where the Greek letter subscripts take the values $1$, $2$ or $3$ and
$\epsilon_{\alpha \beta \gamma}$ is the Levi-Civita symbol. The matrices $T^i$ 
are related to the unit matrix and the $SU(3)$ group generators. In other words
$0$~$\leq$~$i$~$\leq$~$8$ and 
\begin{equation}
T^0_{\alpha \beta} ~=~ \sqrt{\frac{3}{2}} \delta_{\alpha \beta}
\end{equation}
with the remaining eight objects corresponding to the $SU(3)$ group generators.
The normalization of $T^0_{\alpha\beta}$ is chosen, \cite{53,54}, so that for 
the nine objects spanning the group we have
\begin{equation}
\mbox{Tr} \left( T^i T^j \right) ~=~ \frac{1}{2} \delta^{ij} ~.
\end{equation} 
With this convention we have the simple rule for the full set of $T^i$ with 
$0$~$\leq$~$i$~$\leq$~$8$ that 
\begin{equation}
T^i_{\alpha\beta} T^i_{\gamma\lambda} ~=~ \frac{1}{2} \delta_{\alpha\lambda}
\delta_{\gamma\beta} ~.
\end{equation}
These relations are crucial in evaluating the products of the coupling tensors
within the renormalization group function determination. We note that unlike
\cite{53,54} our convention is to base the definition of the 
$SU(3)$~$\times$~$SU(3)$ tensor $d^{ijk}$ on the $SU(3)$ group generators 
themselves rather than the Gell-Mann matrices used in \cite{53,54}. With this
construction the fields are in the $[3,\bar{3}]$~$+$~$[\bar{3},3]$
representation of $SU(3)$~$\times$~$SU(3)$. 

To renormalize (\ref{lagphi3su3}) we constructed the relevant set of 
{\sc Qgraf} diagrams to three loops. As noted in \cite{54} the number of 
Feynman diagrams is substantially less than that of the real scalar theory 
(\ref{lagphi3}). This is because not all the basic topologies survive when the 
field is complex. It is simple to see this as each line of a graph must have an
arrow representing the charge. In addition from (\ref{lagphi3su3}) the two 
vertices are represented by the convergence or divergence of three directed 
lines where all lines either have the arrows directed to the point or directed 
away from the point. Therefore, to see which topologies survive from the real 
theory one merely takes the topology and endows its lines with arrows 
consistent with these rules. If this results in a graph with any vertex where
not all arrows converge or diverge then it is absent or excluded. It
transpires, for instance, \cite{54}, that any graph with a subgraph with an odd
number of propagators will immediately vanish. We have checked this by carrying
out the full three loop renormalization explicitly with the Lagrangian 
(\ref{lagphi3su3}). Moreover, the renormalization group functions to this order
are the same as (\ref{betad}) and (\ref{gammad}) but with $T_3$~$=$~$0$ and 
$T_{72}$~$=$~$0$. The first invariant corresponds to the basic one loop 
triangle $3$-point function. So it is clearly absent, \cite{54}. Given this 
observation with regard to the renormalization group functions already 
determined then the four loop expressions will be given by excluding those 
invariants in (\ref{betad}) and (\ref{gammad}) which correspond to absent 
topologies. It is straightforward to check that of the set given in 
(\ref{tensT}) only $T_{94}$ passes the test and the remaining four loop 
invariants are excluded from the renormalization group functions. At this stage
we have only argued for the consequences of the renormalization group functions
when one has a $\phi^3$ theory of the form (\ref{lagphi3su3}). For the 
$SU(3)$~$\times$~$SU(3)$ case we need the explicit values of the non-zero
invariants $T_2$, $T_5$, $T_{71}$ and $T_{94}$. Using the properties for this 
group discussed above we find the values
\begin{equation}
T_2 ~=~ \frac{1}{2} ~~~,~~~ T_5 ~=~ 0 ~~~,~~~ T_{71} ~=~ \frac{1}{32} ~~~,~~~ 
T_{94} ~=~ \frac{1}{256} 
\end{equation}
in our conventions. We note that, \cite{54}, while the two loop nonplanar
topology corresponding to $T_5$ passes the arrow test it actually vanishes for 
the specific $SU(3)$~$\times$~$SU(3)$ group. For other groups $T_5$ may be 
non-zero. These result in 
\begin{eqnarray}
\beta(g) &=& -~ \frac{1}{16} g^3 ~-~ \frac{11}{1152} g^5 ~+~ 
\frac{4363}{331776} g^7 \nonumber \\ 
&& +~ [ - 4032 \zeta_3 - 5760 \zeta_5 + 10279 ] \frac{g^9}{884736} ~+~ 
O(g^{11}) \nonumber \\
\gamma_\phi(g) &=& -~ \frac{1}{24} g^2 ~-~ \frac{11}{1728} g^4 ~-~ 
\frac{821}{497664} g^6 \nonumber \\
&& +~ [ 1152 \zeta_3 - 5760 \zeta_5 + 4231 ] \frac{g^8}{1327104} ~+~ O(g^{10})
\nonumber \\
\gamma_{\cal O}(g) &=& -~ \frac{1}{4} g^2 ~-~ \frac{1}{192} g^4 ~-~ 
\frac{95}{3456} g^6 \nonumber \\
&& +~ [ 183168 \zeta_3 - 466560 \zeta_5 + 452171 ] \frac{g^8}{11943936} ~+~
O(g^{10}) 
\label{rgesu3}
\end{eqnarray}
in the $\MSbar$ scheme.

The {\em two} loop expressions given in (\ref{rgesu3}) are in accord with
\cite{53,54}. That they do not agree with the three loop results is because 
(\ref{lagphi3su3}) was renormalized in a different scheme. The scheme of
\cite{54} is what is now termed the momentum subtraction scheme which was 
developed later in \cite{56} for QCD. Briefly the coupling constant 
renormalization constant is defined, \cite{54}, by requiring that there are no 
corrections beyond the tree term when the vertex function is evaluated at the 
fully symmetric point where the values of the square of the external momentum 
of each leg are equivalent. For (\ref{lagphi3su3}) when the symmetry group is
$SU(3)$~$\times$~$SU(3)$ the constraint from the group theory meant only one
three loop graph needed to be determined at this subtraction point. Moreover,
only the residue of the simple pole was needed for the three loop momentum
subtraction $\beta$-function. The finite part would be necessary for the four
loop correction. That $\gamma_\phi(g)$ agrees at two loop is perhaps somewhat 
surprising given that only the leading term of this is independent of the 
renormalization scheme. In (\ref{lagphi3su3}) the agreement at two loop appears
to be purely coincidental rather as deriving from some property of the two
renormalization schemes. However, we do actually have a check on the three loop
computation which comes via the critical exponents computed in \cite{53,54} at 
the Wilson-Fisher fixed point. While the renormalization group functions are 
central to the evolution of Green's functions for a range of scales, the 
critical exponents at a fixed point are renormalization group invariants. So 
they have the {\em same} value in all schemes. In \cite{53,54} the 
$d$-dimensional momentum subtraction scheme renormalization group functions 
were provided. In other words the finite parts of the underlying momentum 
subtraction scheme renormalization constants manifest themselves as 
$O(\epsilon)$ terms in $\beta(g)$ and $\gamma_\phi(g)$. If one were merely 
interested in purely six dimensions the $O(\epsilon)$ terms would have been
absent in the expressions given in \cite{53,54}. More crucially, if they were 
excluded from a Wilson-Fisher fixed point analysis the derived exponents would 
not be renormalization group invariants. By contrast, in $\MSbar$ there are no 
$O(\epsilon)$ terms in the coefficients of $d$-dimensional renormalization 
group functions aside from the $O(g)$ term of $\beta(g)$ which defines the 
dimension of $g$ in $d$-dimensions. This is because by definition in the 
$\MSbar$ there are no finite parts in the renormalization constants. Therefore,
from our $\MSbar$ expressions (\ref{rgesu3}) we have 
\begin{eqnarray}
\omega &=& 2 \epsilon ~-~ \frac{22}{9} \epsilon^2 ~-~ 
\frac{4847}{81} \epsilon^3 ~+~ [ - 326592 \zeta_3 - 466560 \zeta_5 + 422035 ]
\frac{\epsilon^4}{1458} ~+~ O(\epsilon^5) \nonumber \\
\eta &=& \frac{2}{3} \epsilon ~+~ \frac{32}{3} \epsilon^3 ~+~ 
16 [ 18 \zeta_3 + 1 ] \frac{\epsilon^4}{9} ~+~ O(\epsilon^5) 
\end{eqnarray}
where there is no $O(\epsilon^2)$ term in $\eta$. Comparing the three loop 
exponents with the corresponding expressions given in \cite{53,54} we find 
exact agreement. This is reassuring since the original three loop computation 
and the present one were in different schemes. However, using the $\MSbar$ 
scheme, where we were able to simplify the computation of the divergence 
structure of the $3$-point functions, has allowed us to proceed to a higher 
loop order for these exponents than would be currently possible in a momentum 
subtraction scheme. 

\sect{Lee-Yang edge singularity.}

Equipped with the renormalization group functions we can now study several 
problems at a new order in the perturbative expansion. For the first major 
application of our results we turn to the Lee-Yang edge singularity problem 
which as elucidated in \cite{29} is related to $\phi^3$ theory but with a 
purely imaginary coupling constant. In terms of the results in the previous 
section this requires setting $d^{ijk}$~$=$~$i$ in the initial Lagrangian to 
produce the values 
\begin{equation}
T_2 ~=~ -~ 1 ~~,~~ T_3 ~=~ -~ 1 ~~,~~ T_5 ~=~ 1 ~~,~~ T_{7i} ~=~ -~ 1 ~~,~~ 
T_{9i} ~=~ 1 ~.
\end{equation}
Alternatively one can use the mapping $g$~$\rightarrow$~$ig$ in 
(\ref{phi3rge}). The key quantity of interest for the Lee-Yang problem is the
critical exponent $\sigma$ which is related to the anomalous dimension of 
$\phi$ of (\ref{lagphi3}) through the hyperscaling law
\begin{equation}
\sigma ~=~ \frac{[d-2+\eta]}{[d+2-\eta]}
\end{equation}
in $d$-dimensions. With $d$~$=$~$6$~$-$~$2\epsilon$ determining the critical
coupling constant in $d$-dimensions and expanding the $\phi$ field anomalous
dimension at that point we find
\begin{eqnarray}
\eta &=& -~ \frac{2}{9} \epsilon ~-~ \frac{172}{729} \epsilon^2 ~+~ 
2 [ 15552 \zeta_3 - 8375 ] \frac{\epsilon^3}{59049} \nonumber \\
&& +~ [ -~ 2783808 \zeta_3 + 3779136 \zeta_4 - 2799360 \zeta_5 
- 3883409 ] \frac{\epsilon^4}{4782969} ~+~ O(\epsilon^5) 
\end{eqnarray}
where the terms to $O(\epsilon^3)$ are in agreement with \cite{24} but
expressed in our conventions for $d$.

{\begin{table}[ht]
\begin{center}
\begin{tabular}{|c||c|c||c|c|}
\hline
$d$ & $\eta$ & & $\sigma$ & \\
\hline
    & $3$ loop & $4$ loop & $3$ loop & $4$ loop \\
\hline
$5$ & $-$ $0.1450$ & $-$ $0.1545$ & $0.3996$ & $0.3977$ \\
$4$ & $-$ $0.3173$ & $-$ $0.3824$ & $0.2664$ & $0.2534$ \\ 
$3$ & $-$ $0.4981$ & $-$ $0.6805$ & $0.0913$ & $0.0562$ \\
$2$ & $-$ $0.6826$ & $-$ $1.0484$ & $-$ $0.1458$ & $-$ $0.2077$ \\
$1$ & $-$ $0.8691$ & $-$ $1.4860$ & $-$ $0.4831$ & $-$ $0.5542$ \\
\hline
\end{tabular}
\end{center}
\begin{center}
{Table $1$. Critical exponents $\eta$ and $\sigma$ estimates using Pad\'{e} 
approximants.}
\end{center}
\end{table}}

{\begin{table}[ht]
\begin{center}
\begin{tabular}{|c||c|c||c|c|}
\hline
$d$ & $\eta$ & & $\sigma$ & \\
\hline
    & $3$ loop & $4$ loop & $3$ loop & $4$ loop \\
\hline
5 & $-$ $0.1468$ & $-$ $0.1529$ & $0.3992$ & $0.3980$ \\
4 & $-$ $0.3280$ & $-$ $0.3702$ & $0.2642$ & $0.2558$ \\ 
3 & $-$ $0.5239$ & $-$ $0.6446$ & $0.0862$ & $0.0630$ \\ 
2 & $-$ $0.7281$ & $-$ $0.9742$ & $-$ $0.1540$ & $-$ $0.1958$ \\
1 & $-$ $0.9377$ & $-$ $1.3583$ & $-$ $0.4921$ & $-$ $0.5411$ \\
\hline
\end{tabular}
\end{center}
\begin{center}
{Table $2$. Critical exponents $\eta$ and $\sigma$ estimates using 
Pad\'{e}-Borel method.}
\end{center}
\end{table}}

As the Lee-Yang singularity problem stretches across dimensions to $d$~$=$~$1$
one has to be careful in using the perturbative expansion for large values of
$\epsilon$. Therefore, to gain estimates for $\sigma$ we have used Pad\'{e} and
Pad\'{e}-Borel resummation for $\eta$ and then evaluated $\sigma$ through the 
scaling law. The results for both are given in Tables $1$ and $2$ where we have
used a $[2,1]$ approximant at three loops and $[3,1]$ at four loops. We have 
included the three loop results of \cite{24} for comparison. From Table $1$ 
it is evident that for dimensions close to $6$ convergence appears to be
present from three to four loops for $\sigma$. The large discrepancy in $\eta$ 
estimates down to $d$~$=$~$4$ seems to get washed out in the scaling law. A 
similar feature is apparent in Table $2$ for the Pad\'{e}-Borel application. 
Though the convergence if anything appears marginally improved. The main 
problem is that the exact values of $\sigma$ at $d$~$=$~$1$ and $2$ are not
emerging which are $-$~$\frac{1}{2}$ and $-$~$\frac{1}{6}$ respectively. 
Indeed if anything the four loop estimates in both Tables for these values is
worse than the three loop ones. This might have been expected as naively 
setting a value for a parameter to be of order $2$ in a summed perturbative 
expansion will mean the larger $O(\epsilon^4)$ term will dominate. One way to 
handle this is to use a constrained Pad\'{e} as discussed in \cite{32}. In that
method the two exact values for $\sigma$ are included in the derivation of the
rational polynomial of the Pad\'{e} approximant. We have carried this out for
the four loop estimate of $\sigma$ which is given by
\begin{eqnarray} 
\sigma &=& \frac{1}{2} ~-~ \frac{1}{6} \epsilon ~-~ 
\frac{79}{972} \epsilon^2 ~+~ 
[ 15552 \zeta_3 - 10445 ]\frac{\epsilon^3}{157464} \nonumber \\
&& +~ [ - 2503872 \zeta_3 + 3779136 \zeta_4 - 2799360 \zeta_5 - 4047533 ] 
\frac{\epsilon^4}{25509168} ~+~ O(\epsilon^5) ~.
\end{eqnarray} 
The results are given in Table $3$ where we have reproduced the constrained
three loop $[3,2]$ results of \cite{32} and given our four loop $[4,2]$ 
Pad\'{e} estimates. The constraints are included for completeness. Results from
other methods are included for comparison. These include a strong coupling 
expansion \cite{32}, as well as two Monte Carlo methods which are based on
critical behaviour in problems seemingly unrelated to the Lee-Yang singularity
problem. These are termed (lattice) animals and fluids with the former 
originating in polymers in a solvent and the latter related to pressure in
fluids where there is a repulsive core, \cite{30,31}. The final column in Table
$3$ are recent results from a conformal bootstrap analysis, \cite{33}. With the
inclusion of the exact results for low dimension in the Pad\'{e} approximant
not only is there better convergence for $d$~$=$~$3$ from three to four loops
but there is remarkable agreement with the values from \cite{32}. For the other
methods the four loop estimates lie within error bars except compared to the
$d$~$=$~$3$ value for the fluids method. In light of this it is worth
commenting on why we presented Tables $1$ and $2$ in the first place. This is
partly to make contact with \cite{24}. More crucially given that we will be
using Pad\'{e} approximants later for other problems down to low dimensions it 
is important to be aware of the potential limitations of the technique in the
absence of known exact two dimensional conformal field theory exponents. As 
exact results are known for $d$~$=$~$1$ and $2$ here we can gauge how far off 
estimates may be for these dimensions. The $d$~$=$~$3$ values for $\sigma$ in 
Tables $1$ and $2$ are perhaps not reliable but those for $d$~$=$~$4$ and $5$ 
are in keeping with those of the other methods listed in Table $3$.  

{\begin{table}[ht]
\begin{center}
\begin{tabular}{|c||c|c||c|c|c|c|c|}
\hline
$d$ & $3$ loop & $4$ loop & Ref \cite{32} & Ref \cite{30} & 
Ref \cite{31} & Ref \cite{33} \\
\hline
5 & $0.3989$ & $0.3981$ & $0.401(9)$ & $0.402(5)$ & $0.40(2)$ & $0.4105(5)$ \\
4 & $0.2616$ & $0.2584$ & $0.258(5)$ & $0.2648(15)$ & $0.261(12)$ & 
$0.2685(1)$ \\ 
3 & $0.0785$ & $0.0747$ & $0.076(2)$ & $0.0877(25)$ & $0.080(7)$ & 
$0.085(1)$ \\
2 & $-$ $0.1667$ & $-$ $0.1667$ & $-$ $0.166(5)$ & $-$ $0.161(8)$ & 
$-$ $0.165(6)$ & $-$ $0.1664(5)$ \\
1 & $-$ $0.5000$ & $-$ $0.5000$ & $-$ & $-$ & $-$ & $-$ \\
\hline
\end{tabular}
\end{center}
\begin{center}
{Table $3$. Critical exponent $\sigma$ estimates using constrained Pad\'{e}
approximant and comparison with \cite{30,31,32,33}.}
\end{center}
\end{table}}

\sect{Percolation.}

We now turn to the application of the renormalization to the percolation
problem which requires the evaluation of the tensors (\ref{tensT}) for a
specific configuration. The percolation problem is described by the replica 
limit in the $(N+1)$-state Potts model \cite{34} which in the case of 
(\ref{lagphi3d}) corresponds to a special designation of the coupling tensor 
$d^{ijk}$. A straightforward way of mapping to the Potts model was provided in 
\cite{57} which involves a set of vectors, $e_\alpha^i$. These $(N+1)$ vectors 
describe the vertices of an $N$-dimensional tetrahedron and allow one to 
decompose $d^{ijk}$ as
\begin{equation}
d^{ijk} ~=~ \sum_{\alpha=1}^{N+1} e_\alpha^i e_\alpha^j e_\alpha^k ~.
\end{equation}
In order to represent the tetrahedron the vectors must satisfy the following
relations, 
\begin{equation}
\sum_{\alpha=1}^{N+1} e_\alpha^i ~=~ 0 ~~~,~~~
\sum_{\alpha=1}^{N+1} e_\alpha^i e_\alpha^j ~=~ (N+1) \delta^{ij}
\end{equation}
for sums over the $(N+1)$-dimensional label and
\begin{equation}
\sum_{i=1}^N e_\alpha^i e_\beta^i ~=~ (N+1) \delta_{\alpha\beta} ~-~ 1
\end{equation}
for summations over the original indices denoted by $i$. It is the form of the
final relation which means that the evaluation of the underlying tensors 
(\ref{tensT}) requires special care. In \cite{24} a diagrammatic method was 
outlined to handle the lower rank tensors. However, we have written a 
{\sc Form} routine to reproduce the evaluations given in \cite{24} and then
applied it to the cases $T_{9m}$ for $1$~$\leq$~$m$~$\leq$~$9$. Such a 
systematic path seems more appropriate since the diagrammatic method is tedious
at three loops as it involves seven sums but manageable for only two tensors. 
At four loops the nine summations for nine independent tensors is not 
straightforward. For arbitrary $N$ we have  
\begin{eqnarray}
T_2 &=& (N+1)^2 [ N - 1] ~~,~~ T_3 ~=~ (N+1)^2 [ N - 2] \nonumber \\
T_5 &=& [(N+1)^2 - 6 (N+1) + 10 ] (N+1)^4 \nonumber \\
T_{71} &=& [(N+1)^3 - 9 (N+1)^2 + 29 (N+1) - 32] (N+1)^6 \nonumber \\
T_{72} &=& [ (N+1)^2 - 6 (N+1) + 11] [ N - 2 ] (N+1)^6 \nonumber \\
T_{91} &=& [(N+1)^3 - 9 (N+1)^2 + 30 (N+1) - 35] [ N - 2] (N+1)^8 \nonumber \\
T_{92} &=& T_{96} ~=~ T_{98} ~=~
[(N+1)^3 - 9 (N+1)^2 + 30 (N+1) - 38] [ N - 2] (N+1)^8 \nonumber \\
T_{93} &=& T_{95} ~=~ T_{97} ~=~
[(N+1)^3 - 9 (N+1)^2 + 30 (N+1) - 37] [ N - 2] (N+1)^8 \nonumber \\
T_{94} &=& [(N+1)^4 - 12 (N+1)^3 + 57 (N+1)^2 - 125 (N+1) + 106] (N+1)^8 
\nonumber \\
T_{99} &=& [(N+1)^2 - 5 (N+1) + 10] [ N - 2] [ N - 3] (N+1)^8 
\label{Tpotts}
\end{eqnarray}
where we present the expressions in the same format as \cite{24} and have 
included the known values from \cite{24} for completeness but with $T_{72}$ 
factorized further. Although there are two instances of three tensors giving 
the same value for this tetrahedron configuration, it is clear from the 
underlying Feynman diagram defining the tensors that the graphs themselves are 
topologically distinct. This should not be a surprise since in QCD, for
example, when one examines high loop diagrams different topologies can have the
same combination of colour group Casimirs multiplying them. Taking the 
$N$~$\rightarrow$~$0$ replica limit gives the values we require for the 
evaluation of the critical exponents for the percolation problem. We have  
\begin{eqnarray}
T_2 &=& -~ 1 ~~,~~ T_3 ~=~ -~ 2 ~~,~~ T_5 ~=~ 5 ~~,~~ T_{71} ~=~ -~ 11 ~~,~~ 
T_{72} ~=~ -~ 12 \nonumber \\
T_{91} &=& 26 ~~,~~ T_{92} ~=~ 32 ~~,~~ T_{93} ~=~ 30 ~~,~~ 
T_{94} ~=~ 27 ~~,~~ T_{99} ~=~ 36 
\label{percT0}
\end{eqnarray}
for the independent tensors.

With these values we have computed various critical exponent in powers of
$\epsilon$ to $O(\epsilon^4)$. Using
\begin{eqnarray}
\beta(g) &=& - \frac{\epsilon}{2} g ~-~ \frac{7}{8} g^3 ~-~ 
\frac{671}{288} g^5 ~+~ \left[ -~ \frac{414031}{41472} - \frac{93}{16} \zeta_3
\right] g^7 \nonumber \\
&& +~ \left[ - \frac{121109}{1728} \zeta_3 + \frac{651}{64} \zeta_4 
- \frac{595}{12} \zeta_5 - \frac{84156383}{1492992} \right] g^9 ~+~ O(g^{11}) 
\nonumber \\  
\gamma_\phi(g) &=& \frac{1}{12} g^2 ~+~ \frac{37}{432} g^4 ~+~ 
\left[ \frac{29297}{62208} - \frac{5}{24} \zeta_3 \right] g^6 \nonumber \\
&& +~ \left[ \frac{225455}{82944} + \frac{233}{864} \zeta_3 
+ \frac{33}{32} \zeta_4 - \frac{55}{18} \zeta_5 \right] g^8 ~+~ O(g^{10})
\nonumber \\
\gamma_{\cal O}(g) &=& \frac{1}{2} g^2 ~+~ \frac{47}{48} g^4 ~+~ 
\left[ \frac{3709}{864} + \frac{3}{2} \zeta_3 \right] g^6 \nonumber \\
&& +~ \left[ \frac{18486131}{746496} + \frac{20027}{864} \zeta_3
- \frac{33}{32} \zeta_4 + \frac{15}{2} \zeta_5 \right] g^8 ~+~ O(g^{10})
\end{eqnarray}
they are
\begin{eqnarray}
\eta &=& -~ \frac{2}{21} \epsilon ~-~ \frac{824}{9261} \epsilon^2 
~+~ 4 [290304 \zeta_3 - 93619 ] \frac{\epsilon^3}{4084101} \nonumber \\ 
&& +~ 2 [286336512 \zeta_3 + 384072192 \zeta_4 - 1493614080 \zeta_5 ~-~ 
103309103 ] \frac{\epsilon^4}{1801088541} ~+~ O(\epsilon^5) \nonumber \\
\eta_{\cal O} &=& -~ \frac{4}{7} \epsilon ~-~ \frac{710}{3087} 
\epsilon^2 ~+~ [ 925344 \zeta_3 - 235495 ] \frac{\epsilon^3}{1361367} 
\nonumber \\ 
&& +~ [ 603983520 \zeta_3 + 1224230112 \zeta_4 - 5334336000 \zeta_5 
- 157609181 ] \frac{\epsilon^4}{1200725694} ~+~ O(\epsilon^5) \nonumber \\
\gamma &=& 1 ~+~ \frac{2}{7} \epsilon ~+~ \frac{565}{3087} \epsilon^2 ~+~ 
[ - 925344 \zeta_3 + 408997 ] \frac{\epsilon^3}{2722734} \nonumber \\ 
&& +~ [ - 933950304 \zeta_3 - 1224230112 \zeta_4 + 5334336000 \zeta_5 
+ 302378687 ] \frac{\epsilon^4}{2401451388} ~+~ O(\epsilon^5) \nonumber \\
\nu &=& \frac{1}{2} ~+~ \frac{5}{42} \epsilon ~+~ \frac{589}{9261} 
\epsilon^2 ~+~ [ - 1614816 \zeta_3 + 716519 ] \frac{\epsilon^3}{16336404} 
\nonumber \\
&& +~ [ 344397667 - 1344827232 \zeta_3 - 2136401568 \zeta_4 
+ 10028551680 \zeta_5 ] \frac{\epsilon^4}{14408708328} 
+~ O(\epsilon^5) \nonumber \\
\omega &=& 2 \epsilon ~-~ \frac{1342}{441} \epsilon^2 
+ [ 62496 \zeta_3 + 40639 ] \frac{\epsilon^3}{7203} \nonumber \\ 
&& +~ [ 248046624 \zeta_4 - 702654624 \zeta_3 - 1209116160 \zeta_5 
- 317288185 ] \frac{\epsilon^4}{19059138} ~+~ O(\epsilon^5) ~. 
\end{eqnarray}
The exponents $\eta$, $\eta_{\cal O}$ and $\omega$ are obtained from the 
corresponding renormalization group function and $\gamma$ and $\nu$ are
deduced from the scaling relations
\begin{equation}
\eta_{\cal O} ~=~ \nu^{-1} ~-~ 2 ~+~ \eta ~~~,~~~ 
\gamma ~=~ ( 2 - \eta ) \nu ~.
\end{equation}
With these we have repeated the exercise of the previous section to obtain
Pad\'{e} and Pad\'{e}-Borel estimates of various critical exponents following 
the method of \cite{24}. There estimates were found for $\eta$ and 
$\eta_{\cal O}$ and then values for $\gamma$ and $\nu$ were obtained from the 
scaling laws. Our results are contained in Tables $3$ to $7$ with those at 
three loops agreeing with \cite{24} and included for comparison with the four 
loop estimates. The three loop estimates for $\omega$ are in accord with those 
given in \cite{58}. 

{\begin{table}
\begin{center}
\begin{tabular}{|c||c|c|c|c|c|c|}
\hline
$d$ & $\eta$ & $\eta_{\cal O}$ & $\gamma$ & $\nu$ & $\omega$ \\
\hline
$5$ & $-$ $0.0569$ & $-$ $0.3097$ & $1.1773$ & $0.5723$ & $0.7910$ \\
$4$ & $-$ $0.1186$ & $-$ $0.6319$ & $1.4250$ & $0.6726$ & $1.5155$ \\ 
$3$ & $-$ $0.1812$ & $-$ $0.9566$ & $1.7812$ & $0.8166$ & $2.2326$ \\
$2$ & $-$ $0.2442$ & $-$ $1.2822$ & $2.3328$ & $1.0395$ & $2.9473$ \\
\hline
\end{tabular}
\end{center}
\begin{center}
{Table $4$. Critical exponent estimates using Pad\'{e} approximants to three
loop expressions.}
\end{center}
\end{table}}

{\begin{table}
\begin{center}
\begin{tabular}{|c||c|c|c|c|c|c|}
\hline
$d$ & $\eta$ & $\eta_{\cal O}$ & $\gamma$ & $\nu$ & $\omega$ \\
\hline
$5$ & $-$ $0.0594$ & $-$ $0.3192$ & $1.1834$ & $0.5746$ & $0.7085$ \\
$4$ & $-$ $0.1338$ & $-$ $0.6885$ & $1.4764$ & $0.6919$ & $1.1590$ \\ 
$3$ & $-$ $0.2215$ & $-$ $1.1048$ & $1.9893$ & $0.8955$ & $1.4775$ \\
$2$ & $-$ $0.3222$ & $-$ $1.5678$ & $3.0782$ & $1.3256$ & $1.7150$ \\
\hline
\end{tabular}
\end{center}
\begin{center}
{Table $5$. Critical exponent estimates using Pad\'{e} approximants to four
loop expressions.}
\end{center}
\end{table}}

Overall a similar feature emerges as for the Lee-Yang edge singularity 
estimates in that down to four dimensions there is reasonable convergence but 
below this the results are not as reliable. This situation was ameliorated by 
exploiting known results in two dimensions and using this as a constraint or 
boundary condition on the Pad\'{e} approximant. Therefore, we have followed 
this procedure again and constructed constrained Pad\'{e} approximants to 
$\nu$, $\gamma$ and $\omega$ from their respective exact two dimensional values
of $\frac{4}{3}$, $\frac{43}{18}$ and $2$, \cite{59}. The results of this 
exercise are given in Table $8$. There the two dimensional values of $\eta$ and
the exponent $\beta$ agree with their known exact values. The latter exponent 
as well as $\sigma$ and $\tau$, which are deduced from hyperscaling laws from 
the previous columns in the table, are included for comparison with results 
from other methods. In this respect Table I of \cite{60} gives a comprehensive 
summary of estimates for these exponents. To compare we have included the 
results of the computation of $\nu$, $\gamma$, $\eta$ and $\beta$, \cite{60}, 
in Table $9$ which used a high temperature series method. Though it is worth 
noting that a more recent study, \cite{61}, has obtained estimates for $\gamma$
which are equivalent to the those of the high temperature series of \cite{60}. 
In \cite{61} the series was extended beyond the $15$th order of \cite{60}. 
Examining the results given in Tables $8$ and $9$ on the whole the constrained 
Pad\'{e} estimates are in reasonable consistency with the central values of 
\cite{60}. Perhaps more significantly the three dimensional estimates from the 
perturbative approach are in line not only with \cite{60} but results from 
other methods as is evident from Table I of \cite{60}. There the perturbative 
estimate of $0.34(4)$ was quoted for $\beta$, for example, but that of Table 
$8$ is more in line with other methods now. One exponent not covered by the 
summary table of \cite{60} is the correction to scaling exponent $\omega$. Two 
studies of $\omega$ using Monte Carlo methods are given in \cite{62} and 
\cite{58} which give results in three and four dimensions respectively. These 
are $1.62(13)$ and $1.13(10)$. Other estimates in three dimensions are 
$1.61(5)$, \cite{63}, and $1.77(13)$, \cite{64}. Our estimates in Table $8$ are
in remarkable agreement in three dimensions and within the error of \cite{58}
in four dimensions. With the results in Tables $8$ and $9$ and the close tally 
it gives support to the earlier observation that the usual Pad\'{e} approximant
is only really competitive down to four dimensions. Below that the summation 
fluctuates as is apparent from the four loop estimates and does not capture the
exact two dimensional picture. Hence results down to four dimensions should 
only be considered. For the remaining two exponents $\sigma$ and $\tau$ we note
that the estimates for $\tau$ are in good agreement over all dimensions in 
comparison with Monte Carlo estimates given in Table I of \cite{61}. An 
estimate for $\sigma$ of around $0.452$-$0.454$ is given there too but only in 
three dimensions. So our value is below the central value. The remaining two 
columns in Table $9$ correspond to estimates for $\sigma$ and $\tau$ from 
\cite{63,65,66}. Again the constrained Pad\'{e} estimates are not dissimilar to
the central values.

We close this section by recording the renormalization group functions for the
$N$~$=$~$2$ case. While this is not directly related to the percolation problem
it does correspond to a specific Potts model and so we give the relevant 
expressions for completeness here. The main reason for treating this case 
specially lies in the nature of the group invariants $T_n$ given in 
(\ref{Tpotts}). From the explicit values most vanish at $N$~$=$~$2$. If one 
analyses the underlying Feynman graph which each invariant relates to, then the
non-zero invariants have the same feature as the $SU(3)$~$\times$~$SU(3)$ 
theory analysed earlier. Though they have no other connection aside from the 
graphical one. This similarity is that when $T_n$ is non-zero the corresponding
Feynman graph has no subgraph with an odd number of Feynman propagators. If any
$T_i$~$=$~$0$ then there is at least one subgraph with an odd number of 
propagators. This is clearly the case for $T_3$ which is the one loop triangle.
So in (\ref{betad}) a large number of terms are absent. Moreover, the non-zero 
values of $T_n$ are
\begin{equation}
T_2 ~=~ 9 ~~,~~ T_5 ~=~ 81 ~~,~~ T_{71} ~=~ 729 ~~,~~ T_{94} ~=~ 6561
\end{equation}
which are all powers of $3$ in contrast to (\ref{lagphi3su3}) where the
non-zero invariants were powers of $1/2$ and $T_5$ is non-zero here. With these
values the renormalization group functions are 
\begin{eqnarray}
\beta(g) &=& -~ \frac{\epsilon}{2} g ~-~ \frac{9}{8} g^3 ~-~
\frac{747}{32} g^5 ~+~ 9 [ 2592 \zeta_3 + 10627 ] \frac{g^7}{512} \nonumber \\
&& +~ 243 [ - 25248 \zeta_3 - 864 \zeta_4 - 23040 \zeta_5 + 4607 ]
\frac{g^9}{2048} ~+~ O(g^{11}) \nonumber \\
\gamma_\phi(g) &=& -~ \frac{3}{4} g^2 ~-~ \frac{33}{16} g^4 ~+~
3 [ 2592 \zeta_3 - 5357 ] \frac{g^6}{256} \nonumber \\
&& +~ 243 [ 3104 \zeta_3 - 288 \zeta_4 - 7680 \zeta_5 + 4077 ]
\frac{g^8}{1024} ~+~ O(g^{10}) \nonumber \\
\gamma_{\cal O}(g) &=& -~ \frac{9}{2} g^2 ~-~ \frac{27}{16} g^4 ~-~
\frac{1917}{4} g^6 \nonumber \\
&& +~ 9 [ 906336 \zeta_3 + 23328 \zeta_4 - 1866240 \zeta_5 + 1174907 ]
\frac{g^8}{1024} ~+~ O(g^{10})
\end{eqnarray}
which can be used for a Wilson-Fisher fixed point analysis.

{\begin{table}
\begin{center}
\begin{tabular}{|c||c|c|c|c|c|c|}
\hline
$d$ & $\eta$ & $\eta_{\cal O}$ & $\gamma$ & $\nu$ & $\omega$ \\
\hline
$5$ & $-$ $0.0578$ & $-$ $0.3122$ & $1.1788$ & $0.5729$ & $0.7540$ \\
$4$ & $-$ $0.1229$ & $-$ $0.6431$ & $1.4346$ & $0.6758$ & $1.3880$ \\ 
$3$ & $-$ $0.1903$ & $-$ $0.9800$ & $1.8097$ & $0.8262$ & $1.9937$ \\
$2$ & $-$ $0.2588$ & $-$ $1.3199$ & $2.4057$ & $1.0651$ & $2.5870$ \\
\hline
\end{tabular}
\end{center}
\begin{center}
{Table $6$. Critical exponent estimates using Pad\'{e}-Borel method for three
loop expressions.}
\end{center}
\end{table}}

{\begin{table}
\begin{center}
\begin{tabular}{|c||c|c|c|c|c|c|}
\hline
$d$ & $\eta$ & $\eta_{\cal O}$ & $\gamma$ & $\nu$ & $\omega$ \\
\hline
$5$ & $-$ $0.0582$ & $-$ $0.3160$ & $1.1814$ & $0.5740$ & $0.7539$ \\
$4$ & $-$ $0.1253$ & $-$ $0.6631$ & $1.4517$ & $0.6831$ & $1.3966$ \\ 
$3$ & $-$ $0.1968$ & $-$ $1.0464$ & $1.9096$ & $0.8693$ & $2.0170$ \\
$2$ & $-$ $0.2717$ & $-$ $1.4503$ & $2.7656$ & $1.2174$ & $2.6320$ \\
\hline
\end{tabular}
\end{center}
\begin{center}
{Table $7$. Critical exponent estimates using Pad\'{e}-Borel method for four
loop expressions.}
\end{center}
\end{table}}

{\begin{table}[hb]
\begin{center}
\begin{tabular}{|c||c|c|c|c|c|c|c|}
\hline
$d$ & $\nu$ & $\gamma$ & $\eta$ & $\beta$ & $\omega$ & $\sigma$ & $\tau$ \\
\hline
$5$ & $0.5746$ & $1.1817$ & $-$ $0.0565$ & $0.8457$ & $0.7178$ & $0.4933$ & 
$2.4171$ \\
$4$ & $0.6920$ & $1.4500$ & $-$ $0.0954$ & $0.6590$ & $1.2198$ & $0.4742$ & 
$2.3124$ \\ 
$3$ & $0.8968$ & $1.8357$ & $-$ $0.0470$ & $0.4273$ & $1.6334$ & $0.4419$ & 
$2.1888$ \\
$2$ & $1.3333$ & $2.3888$ & $0.2083$ & $0.1389$ & $2.0000$ & $0.3956$ & 
$2.0549$ \\
\hline
\end{tabular}
\end{center}
\begin{center}
{Table $8$. Critical exponent estimates using constrained Pad\'{e} 
approximant of four loop results.}
\end{center}
\end{table}}

{\begin{table}
\begin{center}
\begin{tabular}{|c||c|c|c|c|c|c|}
\hline
$d$ & $\nu$ & $\gamma$ & $\eta$ & $\beta$ & $\sigma$ & $\tau$ \\
\hline
$5$ & $0.571(3)$ & $1.185(5)$ & $-$ $0.075(20)$ & $0.845(5)$ & $-$ & 
$2.412(4)$ \\
$4$ & $0.678(50)$ & $1.435(15)$ & $-$ $0.12(4)$ & $0.639(20)$ & $-$ & 
$2.313(2)$ \\
$3$ & $0.872(70)$ & $1.805(20)$ & $-$ $0.07(5)$ & $0.405(25)$ & $0.445(10)$ & 
$2.190(2)$ \\
$2$ & $1.3333$ & $2.3888$ & $0.2083$ & $0.1389$ & $-$ & $-$ \\
\hline
\end{tabular}
\end{center}
\begin{center}
{Table $9$. Results for critical exponents from \cite{60,63,65,66}.}
\end{center}
\end{table}}

\sect{$O(N)$ symmetric theory.}

We now turn to another version of $\phi^3$ theory which is the one endowed with
an $O(N)$ symmetry. It has been considered recently in \cite{18,20,26,27,28} in 
the context of the conformal bootstrap programme as a way of accessing five 
dimensional quantum field theories with a conformal symmetry. The basic 
Lagrangian in this case is 
\begin{equation}
L ~=~ \frac{1}{2} \left( \partial_\mu \phi^i \right)^2 ~+~ 
\frac{1}{2} \left( \partial_\mu \sigma \right)^2 ~+~ 
\frac{g_1}{2} \sigma \phi^i \phi^i ~+~ \frac{g_2}{6} \sigma^3 
\label{lagphi3on}
\end{equation}
where there is an $O(N)$ multiplet of fields $\phi^i$ together with a single
scalar field, $\sigma$. To ensure these fields produce a renormalizable
Lagrangian in six dimensions there are two massless coupling constants $g_1$
and $g_2$. For ease of comparison with \cite{18,20} we use the same notation as
those articles. As the main interest in this section is the extension of the
three loop analysis to the next loop order and as the computational techniques
have already been described we will mention only those features which are new
to this calculation. First, with the extra fields we have carried out the full 
computation rather than identify the group structures of section $3$ with those
of the $O(N)$ symmetric Lagrangian as was noted in \cite{20}. While this may 
seem to be inefficient it is actually a necessary first step in the derivation 
of the renormalization of the mass operators of the fields in (\ref{lagphi3on})
which we will discuss later. What is worth noting is the number of $2$-point 
graphs which are required for the renormalization of (\ref{lagphi3on}). For 
$\phi^i$ there are $1$, $5$, $48$ and $637$ one, two, three and four loop 
graphs respectively. The corresponding numbers for the $\sigma$ $2$-point 
function are $2$, $7$, $60$ and $723$. The resultant four loop field anomalous 
dimensions are  
\begin{eqnarray}
\gamma_\phi(g_1,g_2) &=& -~ \frac{g_1^2}{6}
+ \left[ -~ 11 N g_1^2 
+ 26 g_1^2 
+ 48 g_1 g_2 
- 11 g_2^2  \right] \frac{g_1^2}{432} \nonumber \\ 
&& +~ \left[ 13 N^2 g_1^4 
- 232 g_1^4 N 
+ 5184 \zeta_3 g_1^4 
- 9064 g_1^4 + 2646 N g_1^3 g_2
- 3264 g_1^3 g_2 
\right. \nonumber \\
&& \left. ~~~~
- 386 N g_1^2 g_2^2 
+ 5184 \zeta_3 g_1^2 g_2^2 
- 11762 g_1^2 g_2^2 
+ 942 g_1 g_2^3 
+ 327 g_2^4 \right] \frac{g_1^2}{31104} \nonumber \\
&& +~ \left[ 1296 \zeta_3 N^3 g_1^6 
+ 3 N^3 g_1^6 
+ 46656 \zeta_3 N^2 g_1^6 
+ 21412 N^2 g_1^6 
+ 3649536 \zeta_3 N g_1^6 
\right. \nonumber \\
&& \left. ~~~~
+ 1026432 \zeta_4 N g_1^6 
- 3732480 \zeta_5 N g_1^6 
- 1600648 N g_1^6 
- 1275264 \zeta_3 g_1^6 
\right. \nonumber \\
&& \left. ~~~~
+ 1306368 \zeta_4 g_1^6
- 7464960 \zeta_5 g_1^6 
+ 9095944 g_1^6 
- 15552 \zeta_3 N^2 g_1^5 g_2 
\right. \nonumber \\
&& \left. ~~~~
+ 52452 N^2 g_1^5 g_2 
- 2799360 \zeta_3 N g_1^5 g_2 
- 839808 \zeta_4 N g_1^5 g_2 
+ 3945432 N g_1^5 g_2 
\right. \nonumber \\
&& \left. ~~~~
+ 995328 \zeta_3 g_1^5 g_2 
+ 3359232 \zeta_4 g_1^5 g_2 
- 2784240 g_1^5 g_2 
+ 1296 \zeta_3 N^2 g_1^4 g_2^2 
\right. \nonumber \\
&& \left. ~~~~
- 3874 N^2 g_1^4 g_2^2 
+ 1034208 \zeta_3 N g_1^4 g_2^2 
- 116640 \zeta_4 N g_1^4 g_2^2 
- 2864316 N g_1^4 g_2^2 
\right. \nonumber \\
&& \left. ~~~~
+ 3037824 \zeta_3 g_1^4 g_2^2 
+ 233280 \zeta_4 g_1^4 g_2^2 
- 14929920 \zeta_5 g_1^4 g_2^2 
+ 13929064 g_1^4 g_2^2 
\right. \nonumber \\
&& \left. ~~~~
+ 77760 \zeta_3 N g_1^3 g_2^3 
+ 35544 N g_1^3 g_2^3 
- 1772928 \zeta_3 g_1^3 g_2^3 
+ 2239488 \zeta_4 g_1^3 g_2^3 
\right. \nonumber \\
&& \left. ~~~~
+ 1910496 g_1^3 g_2^3 
- 1296 \zeta_3 N g_1^2 g_2^4 
+ 40951 N g_1^2 g_2^4 
+ 1648512 \zeta_3 g_1^2 g_2^4 
\right. \nonumber \\
&& \left. ~~~~
+ 886464 \zeta_4 g_1^2 g_2^4 
- 3732480 \zeta_5 g_1^2 g_2^4 
+ 1056620 g_1^2 g_2^4 
- 342144 \zeta_3 g_1 g_2^5 
\right. \nonumber \\
&& \left. ~~~~
- 279936 \zeta_4 g_1 g_2^5 
+ 459612 g_1 g_2^5 
+ 68688 \zeta_3 g_2^6 
+ 23328 \zeta_4 g_2^6 
\right. \nonumber \\
&& \left. ~~~~
- 204484 g_2^6 \right] \frac{g_1^2}{6718464} ~+~ O(g_i^{10}) 
\end{eqnarray}
and 
\begin{eqnarray}
\gamma_\sigma(g_1,g_2) &=& 
- \left[ N g_1^2 
+ g_2^2 \right] \frac{1}{12} 
+ \left[ 2 N g_1^4 
+ 48 N g_1^3 g_2 
- 11 N g_1^2 g_2^2 
+ 13 g_2^4 \right] \frac{1}{432} \nonumber \\
&& +~ \left[ -~ 2762 N^2 g_1^6 
+ 5184 \zeta_3 N g_1^6 
- 8560 N g_1^6 
+ 1152 N^2 g_1^5 g_2 
+ 1056 N g_1^5 g_2 
\right. \nonumber \\
&& \left. ~~~~
+ 3 N^2 g_1^4 g_2^2 
+ 12960 \zeta_3 N g_1^4 g_2^2 
- 26646 N g_1^4 g_2^2 
- 1560 N g_1^3 g_2^3 
+ 952 N g_1^2 g_2^4 
\right. \nonumber \\
&& \left. ~~~~
+ 2592 \zeta_3 g_2^6 
- 5195 g_2^6 \right] \frac{1}{62208} \nonumber \\
&& +~ \left[ -~ 41472 \zeta_3 N^3 g_1^8 
+ 54266 N^3 g_1^8 
+ 1897344 \zeta_3 N^2 g_1^8 
+ 513216 \zeta_4 N^2 g_1^8 
\right. \nonumber \\
&& \left. ~~~~
- 1866240 \zeta_5 N^2 g_1^8 
- 605816 N^2 g_1^8 
- 238464 \zeta_3 N g_1^8 
+ 653184 \zeta_4 N g_1^8 
\right. \nonumber \\
&& \left. ~~~~
- 3732480 \zeta_5 N g_1^8 
+ 3883280 N g_1^8 
- 8064 N^3 g_1^7 g_2 
- 2488320 \zeta_3 N^2 g_1^7 g_2 
\right. \nonumber \\
&& \left. ~~~~
+ 4171512 N^2 g_1^7 g_2 
+ 1679616 \zeta_4 N g_1^7 g_2 
- 1008768 N g_1^7 g_2 
+ 2592 \zeta_3 N^3 g_1^6 g_2^2 
\right. \nonumber \\
&& \left. ~~~~
- 354 N^3 g_1^6 g_2^2 
+ 1542240 \zeta_3 N^2 g_1^6 g_2^2 
- 233280 \zeta_4 N^2 g_1^6 g_2^2 
- 2324552 N^2 g_1^6 g_2^2 
\right. \nonumber \\
&& \left. ~~~~
+ 3535488 \zeta_3 N g_1^6 g_2^2 
+ 746496 \zeta_4 N g_1^6 g_2^2 
- 14929920 \zeta_5 N g_1^6 g_2^2 
\right. \nonumber \\
&& \left. ~~~~
+ 10883728 N g_1^6 g_2^2 
- 233280 \zeta_3 N^2 g_1^5 g_2^3 
+ 416016 N^2 g_1^5 g_2^3 
- 1026432 \zeta_3 N g_1^5 g_2^3 
\right. \nonumber \\
&& \left. ~~~~
+ 2799360 \zeta_4 N g_1^5 g_2^3 
- 240816 N g_1^5 g_2^3 
+ 12960 \zeta_3 N^2 g_1^4 g_2^4 
- 19101 N^2 g_1^4 g_2^4 
\right. \nonumber \\
&& \left. ~~~~
+ 1031616 \zeta_3 N g_1^4 g_2^4 
+ 1283040 \zeta_4 N g_1^4 g_2^4 
- 7464960 \zeta_5 N g_1^4 g_2^4 
\right. \nonumber \\
&& \left. ~~~~
+ 6462626 N g_1^4 g_2^4 
- 108864 \zeta_3 N g_1^3 g_2^5 
+ 289416 N g_1^3 g_2^5 
+ 196992 \zeta_3 N g_1^2 g_2^6 
\right. \nonumber \\
&& \left. ~~~~
- 93312 \zeta_4 N g_1^2 g_2^6 
- 306528 N g_1^2 g_2^6 
+ 272160 \zeta_3 g_2^8 
+ 489888 \zeta_4 g_2^8 
\right. \nonumber \\
&& \left. ~~~~
- 1866240 \zeta_5 g_2^8 
+ 1443123 g_2^8 \right] \frac{1}{6718464} ~+~ O(g_i^{10})
\end{eqnarray}
where the order symbol refers to any combination of the two couplings. As we
do not use the method of subtractions but follow the automatic renormalization
algorithm of \cite{41} the cancellation of the double, triple and quadruple 
poles in $\epsilon$ act as a useful computational check. Moreover, we correctly
reproduced the three loop results given in \cite{20}.

For the two $\beta$-functions, $\beta_1(g_1,g_2)$ and $\beta_2(g_1,g_2)$, we
again do not need to compute any nullified $3$-point vertex graphs but instead
use the method outlined in section $2$ where the propagators of each field had 
an extra term. In terms of graphs this means we avoid computing $6455$ graphs 
at four loops for $g_1$ and $6681$ for $g_2$. However, in expanding the 
propagator within a graph to reproduce the corresponding nullified $3$-point 
vertices, one has to label the extra term of the $\phi^i$ propagator with the 
coupling constant $g_1$. This is because this is the only vertex with two 
$\phi^i$ fields. Equally the additional term for the $\sigma$ propagator is 
labelled with $g_2$. Once this step has been achieved the renormalization 
process outlined for the basic $\phi^3$ Lagrangian is followed. The result is 
the two $\beta$-functions  
\begin{eqnarray}
\beta_1(g_1,g_2) &=& 
-~ \frac{\epsilon g_1 }{2}
+ \left[ -~ N g_1^2 
+ 8 g_1^2 
+ 12 g_1 g_2 
- g_2^2 \right] \frac{g_1}{24} \nonumber \\
&& +~ \left[ -~ 86 N g_1^4 
- 536 g_1^4 
+ 132 N g_1^3 g_2 
- 360 g_1^3 g_2 
- 11 N g_1^2 g_2^2 
- 628 g_1^2 g_2^2 
\right. \nonumber \\
&& \left. ~~~~
- 24 g_1 g_2^3 
+ 13 g_2^4 \right] \frac{g_1}{864} \nonumber \\
&& +~ \left[ 3662 N^2 g_1^6 
+ 129600 \zeta_3 N g_1^6 
- 40688 N g_1^6 
+ 20736 \zeta_3 g_1^6 
+ 251360 g_1^6 
\right. \nonumber \\
&& \left. ~~~~
- 36 N^2 g_1^5 g_2 
- 155520 \zeta_3 N g_1^5 g_2 
+ 124704 N g_1^5 g_2 
+ 186624 \zeta_3 g_1^5 g_2 
+ 18000 g_1^5 g_2 
\right. \nonumber \\
&& \left. ~~~~
+ 3 N^2 g_1^4 g_2^2 
+ 12960 \zeta_3 N g_1^4 g_2^2 
- 53990 N g_1^4 g_2^2 
- 41472 \zeta_3 g_1^4 g_2^2 
+ 358480 g_1^4 g_2^2 
\right. \nonumber \\
&& \left. ~~~~
- 4560 N g_1^3 g_2^3 
+ 124416 \zeta_3 g_1^3 g_2^3 
+ 97776 g_1^3 g_2^3 
+ 952 N g_1^2 g_2^4 
+ 62208 \zeta_3 g_1^2 g_2^4 
\right. \nonumber \\
&& \left. ~~~~
+ 9960 g_1^2 g_2^4 
- 31104 \zeta_3 g_1 g_2^5 
+ 33612 g_1 g_2^5 
+ 2592 \zeta_3 g_2^6 
- 5195 g_2^6 \right] \frac{g_1}{124416} \nonumber \\
&& +~ \left[ 93312 \zeta_3 N^3 g_1^8 
- 12310 N^3 g_1^8 
+ 21959424 \zeta_3 N^2 g_1^8 
- 5365440 \zeta_4 N^2 g_1^8 
\right. \nonumber \\
&& \left. ~~~~
- 1866240 \zeta_5 N^2 g_1^8 
- 11535384 N^2 g_1^8 
- 172969344 \zeta_3 N g_1^8 
+ 14183424 \zeta_4 N g_1^8 
\right. \nonumber \\
&& \left. ~~~~
+ 111974400 \zeta_5 N g_1^8 
+ 12401088 N g_1^8 
+ 31290624 \zeta_3 g_1^8 
+ 1492992 \zeta_4 g_1^8 
\right. \nonumber \\
&& \left. ~~~~
- 82114560 \zeta_5 g_1^8 
- 104680384 g_1^8 
- 31104 \zeta_3 N^3 g_1^7 g_2 
+ 4248 N^3 g_1^7 g_2 
\right. \nonumber \\
&& \left. ~~~~
- 17262720 \zeta_3 N^2 g_1^7 g_2 
+ 2799360 \zeta_4 N^2 g_1^7 g_2 
+ 11998152 N^2 g_1^7 g_2 
\right. \nonumber \\
&& \left. ~~~~
- 29673216 \zeta_3 N g_1^7 g_2 
+ 3919104 \zeta_4 N g_1^7 g_2 
+ 89579520 \zeta_5 N g_1^7 g_2 
\right. \nonumber \\
&& \left. ~~~~
- 93820080 N g_1^7 g_2 
- 89268480 \zeta_3 g_1^7 g_2 
+ 12877056 \zeta_4 g_1^7 g_2 
- 78382080 \zeta_5 g_1^7 g_2 
\right. \nonumber \\
&& \left. ~~~~
+ 5902944 g_1^7 g_2 
+ 2592 \zeta_3 N^3 g_1^6 g_2^2 
- 354 N^3 g_1^6 g_2^2 
+ 3403296 \zeta_3 N^2 g_1^6 g_2^2 
\right. \nonumber \\
&& \left. ~~~~
- 233280 \zeta_4 N^2 g_1^6 g_2^2 
- 4985248 N^2 g_1^6 g_2^2 
+ 27454464 \zeta_3 N g_1^6 g_2^2 
\right. \nonumber \\
&& \left. ~~~~
- 12223872 \zeta_4 N g_1^6 g_2^2 
- 59719680 \zeta_5 N g_1^6 g_2^2 
+ 25092064 N g_1^6 g_2^2 
\right. \nonumber \\
&& \left. ~~~~
- 176380416 \zeta_3 g_1^6 g_2^2 
+ 18382464 \zeta_4 g_1^6 g_2^2 
+ 82114560 \zeta_5 g_1^6 g_2^2 
\right. \nonumber \\
&& \left. ~~~~
- 109678192 g_1^6 g_2^2 
- 342144 \zeta_3 N^2 g_1^5 g_2^3 
+ 576648 N^2 g_1^5 g_2^3 
- 17324928 \zeta_3 N g_1^5 g_2^3 
\right. \nonumber \\
&& \left. ~~~~
+ 839808 \zeta_4 N g_1^5 g_2^3 
+ 111974400 \zeta_5 N g_1^5 g_2^3 
- 41588256 N g_1^5 g_2^3 
\right. \nonumber \\
&& \left. ~~~~
- 70232832 \zeta_3 g_1^5 g_2^3 
+ 559872 \zeta_4 g_1^5 g_2^3 
+ 44789760 \zeta_5 g_1^5 g_2^3 
- 146215152 g_1^5 g_2^3 
\right. \nonumber \\
&& \left. ~~~~
+ 12960 \zeta_3 N^2 g_1^4 g_2^4 
- 19101 N^2 g_1^4 g_2^4 
+ 3478464 \zeta_3 N g_1^4 g_2^4 
- 3195936 \zeta_4 N g_1^4 g_2^4 
\right. \nonumber \\
&& \left. ~~~~
- 7464960 \zeta_5 N g_1^4 g_2^4 
+ 7579786 N g_1^4 g_2^4 
- 85784832 \zeta_3 g_1^4 g_2^4 
+ 5412096 \zeta_4 g_1^4 g_2^4 
\right. \nonumber \\
&& \left. ~~~~
+ 3732480 \zeta_5 g_1^4 g_2^4 
- 34030688 g_1^4 g_2^4 
- 2255040 \zeta_3 N g_1^3 g_2^5 
+ 1119744 \zeta_4 N g_1^3 g_2^5 
\right. \nonumber \\
&& \left. ~~~~
+ 1918896 N g_1^3 g_2^5 
- 24292224 \zeta_3 g_1^3 g_2^5 
+ 8957952 \zeta_4 g_1^3 g_2^5 
- 55987200 \zeta_5 g_1^3 g_2^5 
\right. \nonumber \\
&& \left. ~~~~
+ 17096616 g_1^3 g_2^5 
+ 196992 \zeta_3 N g_1^2 g_2^6 
- 93312 \zeta_4 N g_1^2 g_2^6 
- 306528 N g_1^2 g_2^6 
\right. \nonumber \\
&& \left. ~~~~
- 6371136 \zeta_3 g_1^2 g_2^6 
+ 4665600 \zeta_4 g_1^2 g_2^6 
- 17426144 g_1^2 g_2^6 
- 4494528 \zeta_3 g_1 g_2^7 
\right. \nonumber \\
&& \left. ~~~~
- 4199040 \zeta_4 g_1 g_2^7 
+ 22394880 \zeta_5 g_1 g_2^7 
- 9944304 g_1 g_2^7 
+ 272160 \zeta_3 g_2^8 
\right. \nonumber \\
&& \left. ~~~~
+ 489888 \zeta_4 g_2^8 
- 1866240 \zeta_5 g_2^8 
+ 1443123 g_2^8 \right] \frac{g_1}{13436928} ~+~ O(g_i^{11}) 
\label{betaon1}
\end{eqnarray}
and
\begin{eqnarray}
\beta_2(g_1,g_2) &=& 
-~ \frac{\epsilon g_2}{2} 
+ \left[4 N g_1^3
- N g_1^2 g_2 
+ 3 g_2^3 \right] \frac{1}{8} \nonumber \\
&& +~ \left[ -~ 24 N g_1^5 
- 322 N g_1^4 g_2 
- 60 N g_1^3 g_2^2 
+ 31 N g_1^2 g_2^3 
- 125 g_2^5 \right] \frac{1}{288} \nonumber \\
&& +~ \left[ 27696 N^2 g_1^7 
+ 34224 N g_1^7 
- 38474 N^2 g_1^6 g_2 
+ 5184 \zeta_3 N g_1^6 g_2 
+ 59408 N g_1^6 g_2 
\right. \nonumber \\
&& \left. ~~~~
+ 11304 N^2 g_1^5 g_2^2 
+ 62208 \zeta_3 N g_1^5 g_2^2 
+ 25296 N g_1^5 g_2^2 
- 789 N^2 g_1^4 g_2^3 
\right. \nonumber \\
&& \left. ~~~~
+ 44064 \zeta_3 N g_1^4 g_2^3 
+ 127890 N g_1^4 g_2^3 
- 20736 \zeta_3 N g_1^3 g_2^4 
- 8688 N g_1^3 g_2^4 
\right. \nonumber \\
&& \left. ~~~~
- 6272 N g_1^2 g_2^5 
+ 12960 \zeta_3 g_2^7 
+ 33085 g_2^7 \right] \frac{1}{41472} \nonumber \\
&& +~ \left[ 1088640 \zeta_3 N^3 g_1^9 
- 1031208 N^3 g_1^9 
- 8771328 \zeta_3 N^2 g_1^9 
- 3359232 \zeta_4 N^2 g_1^9 
\right. \nonumber \\
&& \left. ~~~~
+ 6915984 N^2 g_1^9 
- 6117120 \zeta_3 N g_1^9 
- 559872 \zeta_4 N g_1^9 
+ 11197440 \zeta_5 N g_1^9 
\right. \nonumber \\
&& \left. ~~~~
- 20404128 N g_1^9 
- 1021248 \zeta_3 N^3 g_1^8 g_2 
+ 706478 N^3 g_1^8 g_2 
+ 3856896 \zeta_3 N^2 g_1^8 g_2 
\right. \nonumber \\
&& \left. ~~~~
+ 8071488 \zeta_4 N^2 g_1^8 g_2 
- 13063680 \zeta_5 N^2 g_1^8 g_2 
- 26286776 N^2 g_1^8 g_2 
\right. \nonumber \\
&& \left. ~~~~
- 34763904 \zeta_3 N g_1^8 g_2 
- 4385664 \zeta_4 N g_1^8 g_2 
+ 1866240 \zeta_5 N g_1^8 g_2 
\right. \nonumber \\
&& \left. ~~~~
- 401008 N g_1^8 g_2 
+ 279936 \zeta_3 N^3 g_1^7 g_2^2 
- 147384 N^3 g_1^7 g_2^2 
\right. \nonumber \\
&& \left. ~~~~
- 12192768 \zeta_3 N^2 g_1^7 g_2^2 
+ 1399680 \zeta_4 N^2 g_1^7 g_2^2 
+ 22394880 \zeta_5 N^2 g_1^7 g_2^2 
\right. \nonumber \\
&& \left. ~~~~
+ 5773632 N^2 g_1^7 g_2^2 
- 39688704 \zeta_3 N g_1^7 g_2^2 
+ 8398080 \zeta_4 N g_1^7 g_2^2 
\right. \nonumber \\
&& \left. ~~~~
+ 44789760 \zeta_5 N g_1^7 g_2^2 
- 67219056 N g_1^7 g_2^2 
- 23328 \zeta_3 N^3 g_1^6 g_2^3 
\right. \nonumber \\
&& \left. ~~~~
+ 9906 N^3 g_1^6 g_2^3 
+ 4388256 \zeta_3 N^2 g_1^6 g_2^3 
- 3172608 \zeta_4 N^2 g_1^6 g_2^3 
\right. \nonumber \\
&& \left. ~~~~
+ 10267192 N^2 g_1^6 g_2^3 
- 56619648 \zeta_3 N g_1^6 g_2^3 
+ 7744896 \zeta_4 N g_1^6 g_2^3 
\right. \nonumber \\
&& \left. ~~~~
+ 7464960 \zeta_5 N g_1^6 g_2^3 
- 9887792 N g_1^6 g_2^3 
- 1477440 \zeta_3 N^2 g_1^5 g_2^4 
\right. \nonumber \\
&& \left. ~~~~
+ 559872 \zeta_4 N^2 g_1^5 g_2^4 
- 3730536 N^2 g_1^5 g_2^4 
- 8398080 \zeta_3 N g_1^5 g_2^4 
+ 559872 \zeta_4 N g_1^5 g_2^4 
\right. \nonumber \\
&& \left. ~~~~
- 55987200 \zeta_5 N g_1^5 g_2^4 
- 8554440 N g_1^5 g_2^4 
+ 44064 \zeta_3 N^2 g_1^4 g_2^5 
+ 224817 N^2 g_1^4 g_2^5 
\right. \nonumber \\
&& \left. ~~~~
- 22187520 \zeta_3 N g_1^4 g_2^5 
- 956448 \zeta_4 N g_1^4 g_2^5 
+ 14929920 \zeta_5 N g_1^4 g_2^5 
\right. \nonumber \\
&& \left. ~~~~
- 44490442 N g_1^4 g_2^5 
- 9782208 \zeta_3 N g_1^3 g_2^6 
+ 1959552 \zeta_4 N g_1^3 g_2^6 
\right. \nonumber \\
&& \left. ~~~~
+ 22394880 \zeta_5 N g_1^3 g_2^6 
+ 3707040 N g_1^3 g_2^6 
+ 1223424 \zeta_3 N g_1^2 g_2^7 
\right. \nonumber \\
&& \left. ~~~~
- 513216 \zeta_4 N g_1^2 g_2^7 
+ 1351296 N g_1^2 g_2^7 
- 12677472 \zeta_3 g_2^9 
+ 1049760 \zeta_4 g_2^9 
\right. \nonumber \\
&& \left. ~~~~
+ 3732480 \zeta_5 g_2^9 
- 10213095 g_2^9 \right] \frac{1}{4478976} ~+~ O(g_i^{11}) ~. 
\label{betaon2}
\end{eqnarray}
As with the two field anomalous dimensions we have reproduced the three loop
expressions given in \cite{20} and equally recovered the expressions given
earlier for (\ref{lagphi3}) when $N$~$=$~$0$. This comparison is made with
reference to the comments on our conventions. In particular the leading term of 
each $\beta$-function derives from our choice of $d$~$=$~$6$~$-$~$2\epsilon$ 
and the factor of $2$ in our definition of the renormalization group functions 
in comparison to \cite{20,21}.

To complete the four loop renormalization of (\ref{lagphi3on}) we compute the
renormalization group functions associated with the two mass operators which
we will denote by
\begin{equation}
{\cal O}_1 ~=~ \frac{1}{2} \phi^i \phi^i ~~~,~~~
{\cal O}_2 ~=~ \frac{1}{2} \sigma^2 ~. 
\end{equation}
These two operators have the same canonical dimension of $2$ and therefore mix 
under renormalization. Here we note that our canonical dimension convention 
derives from the dimensionality of the associated coupling constant of the 
operators in a Lagrangian. This is an important distinction and is motivated in
part for later discussion in relation to checks with the large $N$ expansion of
critical exponents. In other words denoting the bare operators with a subscript
${}_{\mbox{\footnotesize{o}}}$
\begin{equation}
{\cal O}_{i\,\mbox{\footnotesize{o}}} ~=~ Z_{ij} {\cal O}_j
\end{equation}
where $Z_{ij}$ is the mixing matrix of renormalization constants. These produce
a mixing matrix of mass anomalous dimensions denoted by $\gamma_{ij}(g_1,g_2)$.
To extract the renormalization constants one ordinarily inserts the operators
into separate $\phi^i$ and $\sigma$ $2$-point functions of all possible 
$1$-particle irreducible Feynman graphs and follows the normal procedure. 
However, as the operators do not involve derivatives there is no complication 
with mixing into total derivative operators. Moreover, this means that for the 
determination of the $\MSbar$ renormalization constants the operators are 
inserted with no momentum flowing in or out of the vertex itself. This is a 
standard method and reduces the problem to a simple $2$-point function 
computation. In the context of the underlying $\phi^3$ interaction this leads 
to a similar computational simplification which we exploited before. Again one 
need not generate any more Feynman graphs than those already used for the wave 
function renormalization. Similar to the coupling constant we expand each 
$\phi^i$ and $\sigma$ propagator as if there was a respective mass present 
using the mapping
\begin{equation}
\frac{1}{k^2} ~\mapsto~ \frac{1}{k^2} ~+~ \frac{m^2}{(k^2)^2} ~.
\end{equation}
The additional complication here is that one has to to label the $O(m^2)$ term 
to indicate whether that insertion is from a $\phi^i$ or $\sigma$ field mass 
term. By contrast if one instead evaluated the $3$-point functions with a 
nullified operator insertion then there would be $4$ one loop, $38$ two loop, 
$722$ three loop and $13136$ four loop graphs to determine for the mixing 
matrix in total at each loop order. Following the procedure for renormalizing a
mixing matrix we find the elements of $\gamma_{ij}(g_1,g_2)$ to four loops are 
\begin{eqnarray}
\gamma_{11}(g_1,g_2) &=& 
\frac{g_1^2}{3} ~+~ 
\left[ - 22 N g_1^2 - 134 g_1^2 - 30 g_1 g_2 + 5 g_2^2 \right] 
\frac{g_1^2}{216} 
\nonumber \\
&& 
+~ \left[ 803 N^2 g_1^4 
+ 15552 \zeta_3 N g_1^4 
- 4016 N g_1^4 
+ 2592 \zeta_3 g_1^4 
+ 31420 g_1^4 
- 7776 \zeta_3 N g_1^3 g_2 
\right. \nonumber \\
&& \left. ~~~~
+ 2259 N g_1^3 g_2 
+ 15552 \zeta_3 g_1^3 g_2 
- 2964 g_1^3 g_2 
+ 3926 N g_1^2 g_2^2 
- 5184 \zeta_3 g_1^2 g_2^2 
\right. \nonumber \\
&& \left. ~~~~
+ 18512 g_1^2 g_2^2 
- 2859 g_1 g_2^3 
- 51 g_2^4 \right] \frac{g_1^2}{15552} 
\nonumber \\
&& 
+~ \left[ 16848 \zeta_3 N^3 g_1^6 
- 8322 N^3 g_1^6 
+ 2507760 \zeta_3 N^2 g_1^6 
- 734832 \zeta_4 N^2 g_1^6 
\right. \nonumber \\
&& \left. ~~~~
- 1366196 N^2 g_1^6 
- 21591360 \zeta_3 N g_1^6 
+ 1691280 \zeta_4 N g_1^6 
+ 14463360 \zeta_5 N g_1^6 
\right. \nonumber \\
&& \left. ~~~~
+ 1064726 N g_1^6 
+ 3911328 \zeta_3 g_1^6 
+ 186624 \zeta_4 g_1^6 
- 10264320 \zeta_5 g_1^6 
\right. \nonumber \\
&& \left. ~~~~
- 13085048 g_1^6 
- 909792 \zeta_3 N^2 g_1^5 g_2 
+ 139968 \zeta_4 N^2 g_1^5 g_2 
+ 370827 N^2 g_1^5 g_2 
\right. \nonumber \\
&& \left. ~~~~
- 225504 \zeta_3 N g_1^5 g_2 
+ 1959552 \zeta_4 N g_1^5 g_2 
+ 2799360 \zeta_5 N g_1^5 g_2 
\right. \nonumber \\
&& \left. ~~~~
- 7312056 N g_1^5 g_2 
- 8040384 \zeta_3 g_1^5 g_2 
+ 979776 \zeta_4 g_1^5 g_2 
- 8398080 \zeta_5 g_1^5 g_2 
\right. \nonumber \\
&& \left. ~~~~
+ 3962952 g_1^5 g_2 
+ 104328 \zeta_3 N^2 g_1^4 g_2^2 
- 31177 N^2 g_1^4 g_2^2 
- 1489104 \zeta_3 N g_1^4 g_2^2 
\right. \nonumber \\
&& \left. ~~~~
- 991440 \zeta_4 N g_1^4 g_2^2 
- 2799360 \zeta_5 N g_1^4 g_2^2 
+ 3771432 N g_1^4 g_2^2 
- 9916992 \zeta_3 g_1^4 g_2^2 
\right. \nonumber \\
&& \left. ~~~~
+ 1877904 \zeta_4 g_1^4 g_2^2 
+ 8864640 \zeta_5 g_1^4 g_2^2 
- 11767142 g_1^4 g_2^2 
- 524880 \zeta_3 N g_1^3 g_2^3 
\right. \nonumber \\
&& \left. ~~~~
- 69984 \zeta_4 N g_1^3 g_2^3 
+ 2799360 \zeta_5 N g_1^3 g_2^3 
- 2267862 N g_1^3 g_2^3 
+ 1741824 \zeta_3 g_1^3 g_2^3 
\right. \nonumber \\
&& \left. ~~~~
- 979776 \zeta_4 g_1^3 g_2^3 
- 5598720 \zeta_5 g_1^3 g_2^3 
+ 32664 g_1^3 g_2^3 
+ 76464 \zeta_3 N g_1^2 g_2^4 
\right. \nonumber \\
&& \left. ~~~~
- 104976 \zeta_4 N g_1^2 g_2^4 
- 303299 N g_1^2 g_2^4 
- 3335904 \zeta_3 g_1^2 g_2^4 
- 443232 \zeta_4 g_1^2 g_2^4 
\right. \nonumber \\
&& \left. ~~~~
+ 7464960 \zeta_5 g_1^2 g_2^4 
- 2428708 g_1^2 g_2^4 
- 73872 \zeta_3 g_1 g_2^5 
+ 139968 \zeta_4 g_1 g_2^5 
\right. \nonumber \\
&& \left. ~~~~
+ 134451 g_1 g_2^5 
- 11016 \zeta_3 g_2^6 
- 11664 \zeta_4 g_2^6 
+ 36596 g_2^6 \right] \frac{g_1^2}{1679616} ~+~ O(g_i^{10}) \nonumber \\ 
\gamma_{12}(g_1,g_2) &=& 
\frac{N g_1^2}{2} ~+~ \left[ - 2 g_1^2 - 18 g_1 g_2 - 3 g_2^2 \right] 
\frac{N g_1^2}{24} 
\nonumber \\
&& 
+~ \left[ 1154 N g_1^4 
+ 1426 g_1^4 
- 992 N g_1^3 g_2 
+ 1822 g_1^3 g_2 
+ 141 N g_1^2 g_2^2 
+ 864 \zeta_3 g_1^2 g_2^2 
\right. \nonumber \\
&& \left. ~~~~
+ 1430 g_1^2 g_2^2 
+ 864 \zeta_3 g_1 g_2^3 
+ 1420 g_1 g_2^3 
- 21 g_2^4 \right] \frac{N g_1^2}{1728}
\nonumber \\
&&
+~ \left[ 45360 \zeta_3 N^2 g_1^6 
- 42967 N^2 g_1^6 
- 365472 \zeta_3 N g_1^6 
- 139968 \zeta_4 N g_1^6 
+ 288166 N g_1^6 
\right. \nonumber \\
&& \left. ~~~~
- 254880 \zeta_3 g_1^6 
- 23328 \zeta_4 g_1^6 
+ 466560 \zeta_5 g_1^6 
- 850172 g_1^6 
- 27216 \zeta_3 N^2 g_1^5 g_2 
\right. \nonumber \\
&& \left. ~~~~
+ 18117 N^2 g_1^5 g_2 
+ 111456 \zeta_3 N g_1^5 g_2 
+ 233280 \zeta_4 N g_1^5 g_2 
- 311040 \zeta_5 N g_1^5 g_2 
\right. \nonumber \\
&& \left. ~~~~
- 856464 N g_1^5 g_2 
- 1161216 \zeta_3 g_1^5 g_2 
- 186624 \zeta_4 g_1^5 g_2 
+ 155520 \zeta_5 g_1^5 g_2 
\right. \nonumber \\
&& \left. ~~~~
+ 88128 g_1^5 g_2 
+ 3888 \zeta_3 N^2 g_1^4 g_2^2 
- 1935 N^2 g_1^4 g_2^2 
- 260928 \zeta_3 N g_1^4 g_2^2 
\right. \nonumber \\
&& \left. ~~~~
+ 66096 \zeta_4 N g_1^4 g_2^2 
+ 311040 \zeta_5 N g_1^4 g_2^2 
+ 130289 N g_1^4 g_2^2 
- 1468800 \zeta_3 g_1^4 g_2^2 
\right. \nonumber \\
&& \left. ~~~~
+ 124416 \zeta_4 g_1^4 g_2^2 
+ 2488320 \zeta_5 g_1^4 g_2^2 
- 2360254 g_1^4 g_2^2 
+ 111888 \zeta_3 N g_1^3 g_2^3 
\right. \nonumber \\
&& \left. ~~~~
- 42768 \zeta_4 N g_1^3 g_2^3 
+ 135612 N g_1^3 g_2^3 
- 506304 \zeta_3 g_1^3 g_2^3 
+ 62208 \zeta_4 g_1^3 g_2^3 
\right. \nonumber \\
&& \left. ~~~~
- 777600 \zeta_5 g_1^3 g_2^3 
+ 107418 g_1^3 g_2^3 
- 7776 \zeta_3 N g_1^2 g_2^4 
- 19521 N g_1^2 g_2^4 
\right. \nonumber \\
&& \left. ~~~~
- 251856 \zeta_3 g_1^2 g_2^4 
+ 31104 \zeta_4 g_1^2 g_2^4 
- 466560 \zeta_5 g_1^2 g_2^4 
- 209457 g_1^2 g_2^4 
\right. \nonumber \\
&& \left. ~~~~
- 247104 \zeta_3 g_1 g_2^5 
+ 50544 \zeta_4 g_1 g_2^5 
- 363531 g_1 g_2^5 
- 37584 \zeta_3 g_2^6 
+ 11664 \zeta_4 g_2^6 
\right. \nonumber \\
&& \left. ~~~~
+ 51165 g_2^6 \right] \frac{N g_1^2}{186624} ~+~ O(g_i^{10}) \nonumber \\
\gamma_{21}(g_1,g_2) &=& 
\frac{g_1^2}{2} ~~+ \left[ 7 N g_1^2 - 20 g_1^2 - 54 g_1 g_2 - 2 g_2^2 \right] 
\frac{g_1^2}{72} 
\nonumber \\
&&
+~ \left[ - 99 N^2 g_1^4 
- 7776 \zeta_3 N g_1^4 
+ 8798 N g_1^4 
+ 5184 \zeta_3 g_1^4 
+ 3476 g_1^4 
- 4896 N g_1^3 g_2 
\right. \nonumber \\
&& \left. ~~~~
+ 17532 g_1^3 g_2 
- 250 N g_1^2 g_2^2 
+ 10368 \zeta_3 g_1^2 g_2^2 
+ 10054 g_1^2 g_2^2 
+ 5184 \zeta_3 g_1 g_2^3 
\right. \nonumber \\
&& \left. ~~~~
+ 864 g_1 g_2^3 
- 2592 \zeta_3 g_2^4 
+ 2801 g_2^4 \right] \frac{g_1^2}{10368} 
\nonumber \\
&& 
+~ \left[ - 1296 \zeta_3 N^3 g_1^6 
+ 513 N^3 g_1^6 
- 312336 \zeta_3 N^2 g_1^6 
+ 69984 \zeta_4 N^2 g_1^6 
+ 202501 N^2 g_1^6 
\right. \nonumber \\
&& \left. ~~~~
- 1161216 \zeta_3 N g_1^6 
- 559872 \zeta_4 N g_1^6 
+ 2799360 \zeta_5 N g_1^6 
- 1429786 N g_1^6 
\right. \nonumber \\
&& \left. ~~~~
- 1039392 \zeta_3 g_1^6 
+ 209952 \zeta_4 g_1^6 
- 466560 \zeta_5 g_1^6 
- 1075028 g_1^6 
\right. \nonumber \\
&& \left. ~~~~
+ 42768 \zeta_3 N^2 g_1^5 g_2 
- 100470 N^2 g_1^5 g_2 
+ 1492992 \zeta_3 N g_1^5 g_2 
- 209952 \zeta_4 N g_1^5 g_2 
\right. \nonumber \\
&& \left. ~~~~
- 933120 \zeta_5 N g_1^5 g_2 
- 665130 N g_1^5 g_2 
- 4043520 \zeta_3 g_1^5 g_2 
+ 139968 \zeta_4 g_1^5 g_2 
\right. \nonumber \\
&& \left. ~~~~
+ 466560 \zeta_5 g_1^5 g_2 
- 647544 g_1^5 g_2 
- 4536 \zeta_3 N^2 g_1^4 g_2^2 
+ 6693 N^2 g_1^4 g_2^2 
\right. \nonumber \\
&& \left. ~~~~
- 504144 \zeta_3 N g_1^4 g_2^2 
- 58320 \zeta_4 N g_1^4 g_2^2 
+ 3732480 \zeta_5 N g_1^4 g_2^2 
- 966856 N g_1^4 g_2^2 
\right. \nonumber \\
&& \left. ~~~~
- 3506976 \zeta_3 g_1^4 g_2^2 
+ 349920 \zeta_4 g_1^4 g_2^2 
+ 3732480 \zeta_5 g_1^4 g_2^2 
- 6103186 g_1^4 g_2^2 
\right. \nonumber \\
&& \left. ~~~~
+ 76464 \zeta_3 N g_1^3 g_2^3 
- 151632 \zeta_4 N g_1^3 g_2^3 
+ 147648 N g_1^3 g_2^3 
- 2462400 \zeta_3 g_1^3 g_2^3 
\right. \nonumber \\
&& \left. ~~~~
+ 373248 \zeta_4 g_1^3 g_2^3 
- 2332800 \zeta_5 g_1^3 g_2^3 
- 608376 g_1^3 g_2^3 
- 89424 \zeta_3 N g_1^2 g_2^4 
\right. \nonumber \\
&& \left. ~~~~
+ 46656 \zeta_4 N g_1^2 g_2^4 
+ 67895 N g_1^2 g_2^4 
- 987552 \zeta_3 g_1^2 g_2^4 
+ 326592 \zeta_4 g_1^2 g_2^4 
\right. \nonumber \\
&& \left. ~~~~
- 2332800 \zeta_5 g_1^2 g_2^4 
+ 667542 g_1^2 g_2^4 
- 261792 \zeta_3 g_1 g_2^5 
+ 198288 \zeta_4 g_1 g_2^5 
\right. \nonumber \\
&& \left. ~~~~
- 738288 g_1 g_2^5 
- 187272 \zeta_3 g_2^6 
- 174960 \zeta_4 g_2^6 
+ 933120 \zeta_5 g_2^6 
\right. \nonumber \\
&& \left. ~~~~
- 414346 g_2^6 \right] \frac{g_1^2}{559872} ~+~ O(g_i^{10})
\end{eqnarray}
and 
\begin{eqnarray}
\gamma_{22}(g_1,g_2) &=& 
\left[ - N g_1^2  + 5 g_2^2 \right] \frac{1}{12} ~+~ 
\left[ - 80 N g_1^4 - 30 N g_1^3 g_2 
+ 26 N g_1^2 g_2^2 - 97 g_2^4 \right] \frac{1}{216} 
\nonumber \\
&& 
+~ \left[ - 20618 N^2 g_1^6 
+ 5184 \zeta_3 N g_1^6 
+ 27800 N g_1^6 
+ 11304 N^2 g_1^5 g_2 
+ 62208 \zeta_3 N g_1^5 g_2 
\right. \nonumber \\
&& \left. ~~~~
- 14064 N g_1^5 g_2 
- 1185 N^2 g_1^4 g_2^2 
+ 28512 \zeta_3 N g_1^4 g_2^2 
+ 154038 N g_1^4 g_2^2 
\right. \nonumber \\
&& \left. ~~~~
- 31104 \zeta_3 N g_1^3 g_2^3 
- 11496 N g_1^3 g_2^3 
- 9884 N g_1^2 g_2^4 
+ 18144 \zeta_3 g_2^6 
\right. \nonumber \\
&& \left. ~~~~
+ 52225 g_2^6 \right] \frac{1}{62208} 
\nonumber \\
&& 
+~ \left[ - 132840 \zeta_3 N^3 g_1^8 
+ 95093 N^3 g_1^8 
+ 206064 \zeta_3 N^2 g_1^8 
+ 863136 \zeta_4 N^2 g_1^8 
\right. \nonumber \\
&& \left. ~~~~
- 1866240 \zeta_5 N^2 g_1^8 
- 2073638 N^2 g_1^8 
- 2555712 \zeta_3 N g_1^8 
- 46656 \zeta_4 N g_1^8 
\right. \nonumber \\
&& \left. ~~~~
- 233280 \zeta_5 N g_1^8 
- 1428940 N g_1^8 
+ 69984 \zeta_3 N^3 g_1^7 g_2 
- 36846 N^3 g_1^7 g_2 
\right. \nonumber \\
&& \left. ~~~~
- 1912896 \zeta_3 N^2 g_1^7 g_2 
- 69984 \zeta_4 N^2 g_1^7 g_2 
+ 5598720 \zeta_5 N^2 g_1^7 g_2 
+ 471072 N^2 g_1^7 g_2 
\right. \nonumber \\
&& \left. ~~~~
- 1664064 \zeta_3 N g_1^7 g_2 
+ 1819584 \zeta_4 N g_1^7 g_2 
- 5598720 \zeta_5 N g_1^7 g_2 
\right. \nonumber \\
&& \left. ~~~~
- 3838764 N g_1^7 g_2 
- 9072 \zeta_3 N^3 g_1^6 g_2^2 
+ 3759 N^3 g_1^6 g_2^2 
+ 445824 \zeta_3 N^2 g_1^6 g_2^2 
\right. \nonumber \\
&& \left. ~~~~
- 775656 \zeta_4 N^2 g_1^6 g_2^2 
+ 2920258 N^2 g_1^6 g_2^2 
- 17117568 \zeta_3 N g_1^6 g_2^2 
\right. \nonumber \\
&& \left. ~~~~
+ 2251152 \zeta_4 N g_1^6 g_2^2 
+ 11664000 \zeta_5 N g_1^6 g_2^2 
- 6035150 N g_1^6 g_2^2 
\right. \nonumber \\
&& \left. ~~~~
- 454896 \zeta_3 N^2 g_1^5 g_2^3 
+ 209952 \zeta_4 N^2 g_1^5 g_2^3 
- 1275264 N^2 g_1^5 g_2^3 
- 754272 \zeta_3 N g_1^5 g_2^3 
\right. \nonumber \\
&& \left. ~~~~
- 419904 \zeta_4 N g_1^5 g_2^3 
- 16796160 \zeta_5 N g_1^5 g_2^3 
- 1292700 N g_1^5 g_2^3 
+ 14904 \zeta_3 N^2 g_1^4 g_2^4 
\right. \nonumber \\
&& \left. ~~~~
+ 86694 N^2 g_1^4 g_2^4 
- 6225336 \zeta_3 N g_1^4 g_2^4 
- 973944 \zeta_4 N g_1^4 g_2^4 
+ 6531840 \zeta_5 N g_1^4 g_2^4 
\right. \nonumber \\
&& \left. ~~~~
- 14219965 N g_1^4 g_2^4 
- 3316464 \zeta_3 N g_1^3 g_2^5 
+ 629856 \zeta_4 N g_1^3 g_2^5 
\right. \nonumber \\
&& \left. ~~~~
+ 8398080 \zeta_5 N g_1^3 g_2^5 
+ 893478 N g_1^3 g_2^5 
+ 434160 \zeta_3 N g_1^2 g_2^6 
- 180792 \zeta_4 N g_1^2 g_2^6 
\right. \nonumber \\
&& \left. ~~~~
+ 545052 N g_1^2 g_2^6 
- 4788072 \zeta_3 g_2^8 
+ 332424 \zeta_4 g_2^8 
+ 1632960 \zeta_5 g_2^8 
\right. \nonumber \\
&& \left. ~~~~
- 4010301 g_2^8 \right] \frac{1}{1679616} ~+~ O(g_i^{10}) ~. 
\end{eqnarray}
The immediate checks on these expressions are the internal ones similar to
those alluded to for the wave function and coupling constant renormalizations.
Specifically the higher order poles in $\epsilon$ in the renormalization
constants are not independent but depend on the lower order simple poles. That
these higher order poles correctly emerge indicate that the procedure is not
inconsistent. We note that unlike the expression given in (\ref{massopd}) we
have not included the wave function renormalization constants in the 
determination of the renormalization constants of the mixing matrix as is the
usual procedure for a set of operators. See, for instance, \cite{67} for a 
similar renormalization in QCD.

A more appropriate check on our results rests in comparing with results from
another quantum field theory. This is possible through the presence of the
$O(N)$ symmetry which means that the anomalous dimensions can be extracted
using another expansion method which is the large $N$ expansion. While this
was noted in \cite{20} it is worth summarizing the background to the large $N$
technique because of the overlap with the renormalization group at a critical
point. If the $\beta$-function has a non-trivial fixed point at the value
$g_c$, where this could also represent a vector of coupling constants such as 
we have here, then the renormalization group functions evaluated at $g_c$ are
termed critical exponents which are renormalization group invariants. Moreover
at a fixed point the critical exponents from field theories which are
invariably different in nature can be the same. This universality is the key
to our large $N$ checks. Here the relevant scalar field theories which lie in
the same universality class as the $O(N)$ nonlinear $\sigma$ model, $O(N)$
$\phi^4$ theory and (\ref{lagphi3on}). The former two ordinarily reside in
dimensions less than six and are respectively perturbatively renormalizable in
two and four dimensions. In dimensions differing from their canonical
dimension they may cease to be perturbatively renormalizable. Instead they are
renormalizable above their canonical dimension in the sense that at their
Wilson-Fisher fixed point in $d$-dimensions the critical exponents can be 
determined in the large $N$ expansion. As $N$ is a dimensionless parameter,
which remains dimensionless in $d$-dimensions unlike the coupling constant in
dimensionally regularized perturbation theory, then the quantity $1/N$ 
becomes a valid parameter for a perturbative expansion when $N$ is large.
Since the critical exponents can be computed to several orders in $1/N$ and
in $d$-dimensions and because they are related to the critical renormalization 
group functions, they contain information on the anomalous dimensions of
{\em all} theories in the same universality class at that fixed point. For the
case we consider here, (\ref{lagphi3on}), the critical exponents corresponding
to the two wave function renormalization constants and masses are known to
$O(1/N^2)$ in $d$-dimensions and for $\phi^i$ to $O(1/N^3)$, 
\cite{35,36,37,38}. 

To be more specific and make contact with earlier work, \cite{35,36}, it is 
worth recalling the situation for $O(N)$ $\phi^4$ theory and define the 
relevant critical exponents. The form of the Lagrangian which is most 
appropriate is
\begin{equation}
L ~=~ \frac{1}{2} \left( \partial_\mu \phi^i \right)^2 ~+~ 
\frac{1}{2} \sigma \phi^i \phi^i ~-~ \frac{1}{2g} \sigma^2 
\label{lagphi4on}
\end{equation}
where $\sigma$ is regarded as an auxiliary field which if eliminated produces 
the usual $\phi^4$ Lagrangian. The single coupling constant $g$ here appears
with the quadratic term in $\sigma$ as it is in this particular form that the
large $N$ evaluation of the critical exponents is developed at high order in
$1/N$, \cite{35,36,37}. Moreover, this formulation allows one to observe which 
theories lie in the same universality class. For instance, the $O(N)$ nonlinear
$\sigma$ model has a similar formulation but the final term is linear rather 
than quadratic in $\sigma$ and has a different coupling constant which is
dimensionless in two dimensions. Moreover, there is a similarity to 
(\ref{lagphi3on}) from the point of view of the interaction but the six 
dimensional theory has an additional coupling. However, only one of the 
critical points of (\ref{lagphi3on}) is in the same universality class as 
(\ref{lagphi4on}). As a point of reference we use similar notation to 
\cite{35,36} to define the full scaling dimensions of the fields. In 
\cite{35,36} the critical exponents of $\phi^i$ and $\sigma$ were
$\tilde{\alpha}$ and $\tilde{\beta}$ respectively. It should be noted that 
these are not related to the physical exponents of earlier sections and are 
distinct to this section. We have modified the notation of \cite{35,36} here 
for the full dimension of the fields to avoid confusion with the usual use of 
$\alpha$ and $\beta$ as critical exponents. In terms of the anomalous 
contributions they are defined by
\begin{equation}
\tilde{\alpha} ~=~ \half d ~-~ 1 ~+~ \half \eta ~~~,~~~
\tilde{\beta} ~=~ 2 ~-~ \eta ~-~ \chi
\end{equation}
where $\eta$ is the anomalous dimension of $\phi^i$ and $\chi$ is the anomalous
dimension of the interaction of (\ref{lagphi4on}). The canonical dimension of 
$\tilde{\beta}$ is in keeping with our mass operator canonical dimension 
convention. In terms of the renormalization group functions of 
(\ref{lagphi3on})
\begin{equation}
\gamma_\phi(g_{1c},g_{2c}) ~=~ \half \eta ~~~,~~~
\gamma_\sigma(g_{1c},g_{2c}) ~=~ -~ \eta ~-~ \chi ~.
\end{equation}
Using (\ref{betaon1}) and (\ref{betaon2}) we have determined the location of
fixed point in the large $N$ expansion, $(g_{1c},g_{2c})$, and evaluated the 
field anomalous dimensions to the orders in $1/N$ to which they are known. It 
is satisfying to record that we find total agreement with our expressions. 

It is possible to repeat this check for $\gamma_{ij}(g_1,g_2)$ which requires 
some care. In the $\phi^4$ formulation of the universal theory at the fixed
point, (\ref{lagphi4on}), the critical exponents of the two masses are 
straightforward to deduce. That for ${\cal O}_1$ was discussed in early 
sections and in fact is equivalent to $\tilde{\beta}$ above. The critical 
exponent for ${\cal O}_2$ can be deduced from the final term of 
(\ref{lagphi4on}). In relation to extracting information on the 
$\beta$-function of $O(N)$ $\phi^4$ theory in the neighbourhood of four 
dimensions the relevant critical exponent is denoted by $\omega$ and is 
proportional to the critical slope of the $\beta$-function. However, in 
relation to the equivalent theory defined in six dimensions that exponent would
give the mass anomalous dimension of $\sigma$ as is evident from the form of 
the final term of (\ref{lagphi4on}). Therefore, in terms of the exponents the 
critical point mass dimensions of the two mass operators are
\begin{equation}
\Delta_1 ~=~ 2 ~-~ \eta ~-~ \chi ~~~,~~~ 
\Delta_2 ~=~ 2 \omega ~. 
\end{equation} 
The remaining matter is to reconcile this argument with $\gamma_{ij}(g_1,g_2)$
at criticality and check if the exponents computed from it at the large $N$
fixed point agree with the above known exponents at $O(1/N^2)$. The key to this
is to compute the eigen-anomalous dimensions of $\gamma_{ij}(g_1,g_2)$. These
are given by  
\begin{equation}
\gamma_\pm(g_1,g_2) ~=~ 2 ~-~ \left[ \gamma_{11}(g_1,g_2) 
+ \gamma_{22}(g_1,g_2) \pm \Delta_d(g_1,g_2) \right]
\end{equation}
where the discriminant is given by
\begin{equation}
\Delta_d(g_1,g_2) ~=~ \sqrt{ \left[ [ \gamma_{11}(g_1,g_2) 
- \gamma_{22}(g_1,g_2) ]^2 + 4 \gamma_{12}(g_1,g_2) \gamma_{21}(g_1,g_2) 
\right] } 
\end{equation}
and we have included the canonical dimension of $2$ here. In \cite{20} it 
appears that the canonical dimension of $(d$~$-$~$2)$ was used for the mass
dimensions based on the dimensions of the constituent fields. However, here we
retain the value of $2$ as that convention is essential in getting consistency 
with this particular check between different theories. Evaluating 
$\gamma_\pm(g_1,g_2)$ at the point $(g_{1c},g_{2c})$ and expanding to 
$O(1/N^2)$ we find that $\gamma_-(g_{1c},g_{2c})$ agrees exactly to four loops 
with $\Delta_1$. Similarly $\gamma_+(g_{1c},g_{2c})$ is in precise agreement 
with $\Delta_2$ to the same accuracy. In particular the canonical dimension of 
$2$ is crucial for ensuring the consistency of the latter as it derives from 
the exponent $\omega$ in the $O(N)$ $\phi^4$ theory and that exponent near four
dimensions corresponds to corrections to scaling rather than a mass operator. 
We regard these large $N$ comparisons and in particular the second on the mass 
operators as non-trivial checks on our perturbative computations. It is worth 
mentioning that a one loop evaluation of a mass mixing matrix was given in 
\cite{18} as well as a two loop version for the related $Sp(N)$ version of 
(\ref{lagphi3on}) in \cite{21}. The renormalization group functions of the 
$Sp(N)$ symmetric version of (\ref{lagphi3on}) are related to those of the 
$O(N)$ theory by replacing $N$ in each expression by $(-N)$. However, as far as
we can see both of the mass mixing matrix computations appear to have included 
$1$-particle {\em reducible} diagrams. Low order in $\epsilon$ checks of the 
critical exponents with respect to the large $N$ mass operator exponents were 
discussed in \cite{18,20}.

One of the motivations of extending the three loop results of \cite{20} to
the next loop is to examine the critical point structure of the $O(N)$ two 
coupling theory. In \cite{20} it was noted that there are various critical 
value of $N$ for which the fixed point structure has different properties. For
instance for $N$~$>$~$N_{cr}$ there are three distinct critical points at real 
values of $g_1$ and $g_2$ which are perturbatively unitary. In addition for 
values of $N$ in the range $(N_{cr}^\prime,N_{cr}^{\prime\prime})$ there are 
non-unitary fixed points. In \cite{20} the values of these critical values of 
$N$ were computed in an $\epsilon$ expansion from the three loop 
$\beta$-functions building on the one loop work of \cite{25,18}. Therefore, we 
extend those estimates here using (\ref{betaon1}) and (\ref{betaon2}) and use 
the same notation and method but in our conventions. First, we introduce the 
new scaled coupling variables $x$ and $y$ by
\begin{equation}
g_1 ~=~ i \sqrt{\frac{12\epsilon}{N}} x ~~~,~~~
g_2 ~=~ i \sqrt{\frac{12\epsilon}{N}} y 
\end{equation}
where we recall $d$~$=$~$6$~$-$~$2\epsilon$. The presence of $i$ here is to be 
consistent with \cite{20,21} given our coupling constant conventions. Then to 
find the critical couplings and $N_{cr}$ one solves the set of equations
\begin{equation}
\beta_1(g_1,g_2) ~=~ \beta_2(g_1,g_2) ~=~ 
\frac{\partial \beta_1}{\partial g_1} \frac{\partial \beta_2}{\partial g_2} 
~-~ \frac{\partial \beta_1}{\partial g_2} 
\frac{\partial \beta_2}{\partial g_1} ~=~ 0 
\end{equation}
where the final equation is the condition for at least one zero eigenvalue of
the Hessian of the matrix of derivatives of the $\beta$-functions. Solving the
resultant three equations perturbatively in $d$-dimensions there are three
solutions which we designate $A$, $B$ and $C$. The critical values in each of
the three cases are 
\begin{eqnarray}
N^A_{cr} &=& 1038.26605 ~-~ 1219.67959 \epsilon ~-~ 1456.69332 \epsilon^2 ~+~ 
3621.68482 \epsilon^3 ~+~ O(\epsilon^4) \nonumber \\
x^A_{cr} &=& 1.018036 ~-~ 0.01879 \epsilon ~+~ 0.027606 \epsilon^2 ~-~ 
0.02587 \epsilon^3 ~+~ O(\epsilon^4)\nonumber \\
y^A_{cr} &=& 8.90305 ~-~ 0.42045 \epsilon ~+~ 4.06719 \epsilon^2 ~-~ 
2.00941 \epsilon^3 ~+~ O(\epsilon^4) 
\end{eqnarray}
\begin{eqnarray}
N^B_{cr} &=& 1.02145 ~+~ 0.06506 \epsilon ~-~ 0.00652 \epsilon^2 ~+~ 
0.20347 \epsilon^3 ~+~ O(\epsilon^4) \nonumber \\
x^B_{cr} &=& i [ 0.23185 ~+~ 0.17773 \epsilon ~-~ 0.15822 \epsilon^2 ~+~
0.61640 \epsilon^3 ~+~ O(\epsilon^4) ] \nonumber \\
y^B_{cr} &=& i [ 0.25582 ~+~ 0.22746 \epsilon ~-~ 0.17106 \epsilon^2 ~+~
0.77176 \epsilon^3 ~+~ O(\epsilon^4) ] 
\end{eqnarray}
and
\begin{eqnarray}
N^C_{cr} &=& -~ 0.08750 ~+~ 0.69453 \epsilon ~-~ 3.53076 \epsilon^2 ~+~
22.49021 \epsilon^3 ~+~ O(\epsilon^4) \nonumber \\
x^C_{cr} &=& 0.13175 ~-~ 0.33427 \epsilon ~+~ 0.48270 \epsilon^2 ~-~ 
3.84349 \epsilon^3 ~+~ O(\epsilon^4) \nonumber \\
y^C_{cr} &=& -~ 0.03277 ~+~ 0.26911 \epsilon ~-~ 1.43791 \epsilon^2 ~+~
10.20700 \epsilon^3 ~+~ O(\epsilon^4) ~. 
\end{eqnarray}
We have checked that the results to $O(\epsilon^3)$ agree exactly with those
given in \cite{20}. With the $O(\epsilon^3)$ terms now present we can revisit
the analysis of the location of the boundaries for the unitary versus
non-unitary theories for five dimensions. In \cite{20} a value of 
$N_{cr}$~$=$~$64.253$ emerged by setting $\epsilon$~$=$~$\half$ in $N^A_{cr}$. 
This is a significant distance from the leading value of $1038.266$. As the 
$O(\epsilon^3)$ correction is both large and has a positive sign this means 
that the estimate for $N_{cr}$ will increase. Using the simple substitution 
approach then $N_{cr}$~$=$~$516.963$ which is larger than expectations from 
other methods. However, this large increase is perhaps more indicative of a 
lack of convergence and it seems more appropriate to estimate $N_{cr}$ by using 
Pad\'{e} approximants. Using a $[0,L-1]$ Pad\'{e} approximant for the $L$th 
loop we find the values given in Table $10$. It appears that the value of 
$N_{cr}$ settles to around $400$ which while larger than the three loop 
estimate given in \cite{20} is significantly smaller than the leading value. 
The corresponding values for solutions $B$ and $C$ are given respectively in 
Tables $11$ and $12$. In both instances the values of $N_{cr}$ appear to 
converge to $1.05$ and $0$ respectively. Indeed for $B$ the $O(\epsilon^3)$ 
correction confirms the observation of \cite{20} that $N_{cr}^\prime$~$>$~$1$. 

{\begin{table}
\begin{center}
\begin{tabular}{|c||c|c|c|c|}
\hline
   & $O(1)$ & $O(\epsilon)$ & $O(\epsilon^2)$ & $O(\epsilon^3)$ \\
\hline
$N^A_{cr}$ & $1038.2660$ & $654.0820$ & $454.7593$ & $421.7574$ \\
$x^A_{cr}$ & $1.0180$ & $1.0087$ & $1.0126$ & $1.0133$ \\ 
$y^A_{cr}$ & $8.9031$ & $8.6977$ & $8.8670$ & $9.5082$ \\
\hline
\end{tabular}
\end{center}
\begin{center}
{Table $10$. Estimates of critical value of $N$ and location of fixed points
for solution $A$ in five dimensions using Pad\'{e} approximants.}
\end{center}
\end{table}}
{\begin{table}
\begin{center}
\begin{tabular}{|c||c|c|c|c|}
\hline
   & $O(1)$ & $O(\epsilon)$ & $O(\epsilon^2)$ & $O(\epsilon^3)$ \\
\hline
$N^B_{cr}$ & $1.0215$ & $1.0551$ & $1.0524$ & $1.0539$ \\
$x^B_{cr}$ & $0.2318i$ & $-$ & $0.2933i$ & $0.3073i$ \\ 
$y^B_{cr}$ & $0.2558i$ & $-$ & $0.3385i$ & $0.3564i$ \\
\hline
\end{tabular}
\end{center}
\begin{center}
{Table $11$. Estimates of critical value of $N$ and location of fixed points
for solution $B$ in five dimensions using Pad\'{e} approximants.}
\end{center}
\end{table}}
{\begin{table}
\begin{center}
\begin{tabular}{|c||c|c|c|c|}
\hline
   & $O(1)$ & $O(\epsilon)$ & $O(\epsilon^2)$ & $O(\epsilon^3)$ \\
\hline
$N^C_{cr}$ & $-$ $0.0875$ & $-$ $0.0176$ & $-$ $0.0082$ & $-$ $0.0035$ \\
$x^C_{cr}$ & $0.1317$ & $0.0581$ & $0.0347$ & $-$ $0.0112$ \\ 
$y^C_{cr}$ & $-$ $0.0328$ & $-$ $0.0064$ & $0.0039$ & $0.0228$ \\
\hline
\end{tabular}
\end{center}
\begin{center}
{Table $12$. Estimates of critical value of $N$ and location of fixed points
for solution $C$ in five dimensions using Pad\'{e} approximants.}
\end{center}
\end{table}}

One interesting application of our analysis is to compare estimates for the
dimension of $\phi$ with recent estimates using the conformal bootstrap method 
of \cite{28}. There the $N$~$=$~$500$ theory was considered directly in 
strictly five dimensions and the estimate of $\Delta_\phi$~$=$~$1.500409$ was
given. This is remarkably close to the estimate obtained using the exponent 
evaluated to $O(1/N^3)$ in the large $N$ method which was 
$\Delta_\phi$~$=$~$1.500414$. From the location of the $\beta$-function zeroes 
for $N$~$=$~$500$ we have
\begin{eqnarray}
x &=& 0.805458 ~+~ 0.276177 \epsilon ~+~ 0.939812 \epsilon^2 ~+~ 
8.067242 \epsilon^3 ~+~ O(\epsilon^4) \nonumber \\
y &=& -~ 9.455850 ~-~ 4.724456 \epsilon ~+~ 3.073550  \epsilon^2 ~+~ 
140.713929 \epsilon^3 ~+~ O(\epsilon^4) ~. 
\end{eqnarray}
Using the four loop term and a Pad\'{e} approximant we find the estimate of
$\Delta_\phi$~$=$~$1.500537$ compared to a value of $1.500976$ using the three
loop expression. While not precisely on top of the other two methods there is
a hint that a five loop computation may bridge the difference. 

As our final part of the analysis we briefly consider several low values of $N$
in order to extend the results of \cite{25,18,20} where $N$~$=$~$0$ and $1$ 
were considered. The former value has been already considered in section $3$ as
it is the case of purely one field. For $N$~$=$~$1$ there are two non-trivial 
fixed points one of which is stable at $g_2$~$=$~$\frac{6}{5}g_1$ to leading 
order in $\epsilon$ and the other is at $g_1$~$=$~$g_2$ exactly, \cite{25,20}. 
For this unstable one the Lagrangian can be rewritten in such a way as to have 
a double copy of the basic theory (\ref{lagphi3}) and it is believed that the
flow is away from the unstable case to the infrared stable fixed point,
\cite{20}. Given our four loop analysis it is a simple exercise to extend the
expressions for the critical exponents at each fixed point to the next order
in $\epsilon$. At the $g_2$~$=$~$g_1$ fixed point we have 
\begin{equation}
\Delta_\phi ~=~ \Delta_\sigma ~=~ 2 ~-~ 1.111111 \epsilon ~-~ 
0.117970 \epsilon^2 ~+~ 0.174760 \epsilon^3 ~-~ 0.631636 \epsilon^4 ~+~ 
O(\epsilon^5) 
\end{equation} 
for the dimensions of the fields and
\begin{eqnarray}
\omega_+ &=& 2.000000 \epsilon ~-~ 3.086420 \epsilon^2 ~+~ 
12.725343 \epsilon^3 ~-~ 72.522012 \epsilon^4 ~+~ O(\epsilon^5) \nonumber \\
\omega_- &=& -~ 0.222222 \epsilon ~-~ 0.235940 \epsilon^2 ~+~ 
0.349520 \epsilon^2 ~-~ 1.263272 \epsilon^4 ~+~ O(\epsilon^5) 
\label{omegsig1}
\end{eqnarray}
for the eigen-critical exponents of the matrix
\begin{equation}
\omega_{ij} ~=~ \left. \frac{\partial \beta_i}{\partial g_j}
\right|_{g_1=g_{1c}\,g_2=g_{2c}}
\end{equation}
at the fixed point. In (\ref{omegsig1}) we use the convention that $d$ is not
added to the eigenvalues of $\omega_{ij}$ and also note that the signs of the
leading terms confirms the fixed point is unstable. In addition 
\begin{eqnarray}
\Delta_+ &=& 2 ~-~ 1.111111 \epsilon ~-~ 0.117970 \epsilon^2 ~+~ 
0.174760 \epsilon^3 ~-~ 0.631636 \epsilon^4 ~+~ O(\epsilon^5) \nonumber \\ 
\Delta_- &=& 2 ~+~ 0.222222 \epsilon ~+~ 0.235940 \epsilon^2 ~-~ 
0.349520 \epsilon^3 + 1.263272 \epsilon^4 ~+~ O(\epsilon^5) 
\end{eqnarray}
for the mass matrix eigen-exponents which shows $\Delta_+$~$=$~$\Delta_\sigma$.
By contrast at the infrared stable point we have
\begin{eqnarray}
\Delta_\phi &=& 2 ~-~ 1.100200 \epsilon ~-~ 0.093791 \epsilon^2 ~+~ 
0.160519 \epsilon^3 ~-~ 0.545803 \epsilon^4 ~+~ O(\epsilon^5) \nonumber \\
\Delta_\sigma &=& 2 ~-~ 1.122244 \epsilon ~-~ 0.143537 \epsilon^2 ~+~ 
0.188846 \epsilon^3 ~-~ 0.721707 \epsilon^4 ~+~ O(\epsilon^5) \nonumber \\ 
\omega_+ &=& 2.000000 \epsilon ~-~ 3.092766 \epsilon^2 ~+~ 
12.776556 \epsilon^3 ~-~ 72.867332 \epsilon^4 ~+~ O(\epsilon^5) \nonumber \\
\omega_- &=& 0.220441 \epsilon ~+~ 0.175093 \epsilon^2 ~-~ 
0.316680 \epsilon^3 ~+~ 1.200781 \epsilon^4 ~+~ O(\epsilon^5) \nonumber \\ 
\Delta_+ &=& 2 ~-~ 1.122244 \epsilon ~-~ 0.143537 \epsilon^2 ~+~ 
0.188846 \epsilon^3 ~-~ 0.721707 \epsilon^4 ~+~ O(\epsilon^5) \nonumber \\
\Delta_- &=& 2 ~+~ 0.100200 \epsilon ~-~ 0.046403 \epsilon^2 ~-~ 
0.156191 \epsilon^3 ~+~ 0.336654 \epsilon^4 ~+~ O(\epsilon^5) ~. 
\end{eqnarray}
We note that our expressions for $\omega_i$ agree with \cite{20} at three loops
when the different conventions are accommodated. Clearly 
$\Delta_-$~$=$~$\Delta_\sigma$ as expected, \cite{20}, in contrast to the
situation at the unstable fixed point. There the other eigen-exponent of the
mass matrix was equivalent to the field dimension for $N$~$=$~$1$. In 
Appendix B we have recorded the field and mass eigen-exponents for a variety
of values of $N$ for the three fixed points that occur in the solution of
$\beta_i(g_1,g_2)$~$=$~$0$. The specific cases we considered are those chosen 
in \cite{28} for a conformal bootstrap analysis. The same feature which has 
just been noted is apparent there. In other words at certain fixed points one 
of the mass eigen-exponents is equivalent to the anomalous dimension of the 
$\sigma$ field similar to the large $N$ results. However, which mass exponent 
is identified depends on the nature of the underlying fixed point. 

\sect{Discussion.}

We close with various observations. First, we have carried out the four loop
renormalization of $\phi^3$ theory in six dimensions. Clearly this has been a
nontrivial exercise since we had to calculate the full set of master four loop
massless $2$-point functions to the requisite orders in $\epsilon$. Though
tedious due to the large amount of integration by parts required, the master
evaluation rested on the corresponding known masters in four dimensions,
\cite{46}. That one can relate them was possible through the methods of 
\cite{51,52} and using general properties such as Weinberg's theorem, 
\cite{47}. One of the original aims was to refine the $\epsilon$ expansion 
estimates of critical exponents for several physical problems. Overall the 
exponents we computed at the next order of precision are in reasonable accord 
with numerical approaches. Though the accuracy for low dimensions was driven by
a deeper underlying property and that was the use of two dimensional conformal 
symmetry, \cite{32}. There since critical exponents are known exactly one could
constrain the exponent estimates and allow us to extract the behaviour across 
several dimensions. Such techniques now make the $\epsilon$-expansion 
reasonably competitive, but it is perhaps in the application to the more recent
studies of $O(N)$ symmetric theories that will be beneficial in future. One 
issue examined in \cite{20} was the range of $N$ defining the conformal window 
which was suggested in \cite{20} to drop from the one loop value of 
$N_{cr}$~$=$~$1038$, \cite{25,18}, to around $N_{cr}$~$=$~$64$. Here we took a 
more conservative approach in applying summation techniques to suggest that 
while the bounding value drops, it maybe does not reduce so far as this. While 
successive three and four loop estimates suggest a value settling to around 
$N_{cr}$~$=$~$400$ it would be premature to regard this problem as having been 
resolved. It may be that the perturbative approach is not as fully equipped to 
give a definitive answer in comparison to, say, the conformal bootstrap 
machinery. Also as noted in \cite{20} the $N$~$=$~$1$ theory may be related to 
the deformed $(3,10)$ minimal conformal field theory in two dimensions. 
Therefore, if true then one could perform a similar analysis across the gap to 
six dimensions. It would be interesting then to try and match predictions with 
physical systems. 

One intriguing possibility is the potential application of the universality 
between $2$ and $6$ dimensions to non-scalar theories such as those with gauge 
symmetry or supersymmetry. Indeed the conformal window of QCD is of interest,
\cite{14}. Akin to the connections of four dimensional $\phi^4$ theory and six 
dimensional $\phi^3$ theory one question which would be worth considering in 
future is what if any is the higher dimensional theory which is in the same 
universality class as the Banks-Zaks fixed point in QCD. To an extent there is 
already a parallel in QCD with what has been discussed here and in \cite{20} 
for the large $N$ connection in the scalar field theories. At the Wilson-Fisher
fixed point in $2$~$<$~$d$~$<$~$4$ it is known that QCD is in the same 
universality class as the non-abelian Thirring model in the large $\Nf$ 
expansion, \cite{68}. Here $\Nf$ is the number of (massless) quarks and it is 
important to appreciate that the connection is with respect to this particular 
parameter being large as opposed to $\Nc$ being large where $\Nc$ is the number
of colours in $SU(\Nc)$ QCD. Like $\phi^4$ theory QCD is perturbatively 
renormalizable in four dimensions and the non-abelian Thirring model is 
renormalizable in two dimensions. Indeed this fact has been used to compute 
large $\Nf$ critical exponents to varying orders in $1/\Nf$ in the Thirring 
model and demonstrate that their $\epsilon$ expansion with respect to four 
dimensions agrees with the analogous renormalization group functions of QCD. 
See, for example, \cite{69,70}. While parallel to the lower end of the chain of
nonlinear $\sigma$ model, $\phi^4$ and $\phi^3$ theories which are 
renormalizable in two, four and six dimensions respectively, what is missing in
the QCD instance is the corresponding six dimensional theory. To extend the QCD
chain would require an understanding of the critical point properties of a 
spin-$1$ field in contrast to a spin-$0$ field. We have referred to this field 
not as a gauge field or gluon because in the two dimensional non-abelian 
Thirring model the spin-$1$ field plays the role of an auxiliary field rather 
than a gauge field much in the same way that one replaces the interaction in 
$\phi^4$ theory by a $3$-point vertex at the expense of introducing an 
auxiliary field. In (\ref{lagphi3on}) this appears as $\sigma$ and is the 
connecting field for the chain in $4$~$<$~$d$~$<$~$6$ as well as being the 
Lagrange multipler field in the two dimensional $O(N)$ nonlinear $\sigma$ 
model. There it imposes the constraint that the fields lie on a 
multidimensional sphere. So the $\sigma$ field is effectively the lynchpin 
field which also underlies the chain of theories at the Wilson-Fisher fixed 
point across the dimensions.

\vspace{1cm}
\noindent
{\bf Acknowledgements.} This work was carried out with the support of STFC
Consolidated Grant ST/L000431/1. We thank R.M. Simms for discussions and Dr 
R.N. Lee for pointing out an error in ${\cal M}_{22}$ of equation 
(\ref{mastint}) in an earlier version of the article.

\appendix

\sect{Master Integrals.}

In this appendix we record the explicit values of the master integrals required
for the four loop renormalization to various orders in $\epsilon$ near six 
dimensions. Terms beyond the $O(1/\epsilon)$ are needed due to the presence of 
spurious poles in $\epsilon$ from the integration by parts algorithm of 
Laporta, \cite{42}. We have not included all the basic integrals used but only 
those which are not products of lower loop integrals nor which contain only
simple self-energy subgraphs in order to save space. These are straightforward 
to construct directly by expanding products and ratios of Euler 
$\Gamma$-functions. We use the same notation used in \cite{46} for the 
definition of the topology. The explicit graphs of the integrals are given in
\cite{46}. We have 
\begin{eqnarray}
M_{21} &=& 
\frac{1}{373248} \frac{1}{\epsilon^3} 
+ \frac{4093}{261273600} \frac{1}{\epsilon^2} 
+ \frac{17541299}{329204736000} \frac{1}{\epsilon} 
+ \left[
\frac{12061889939}{138265989120000}
+ \frac{19}{466560} \zeta_3
\right] 
\nonumber \\
&&
+ \left[
- \frac{26183347978621}{58071715430400000}
+ \frac{19}{311040} \zeta_4
+ \frac{15131}{65318400} \zeta_3
\right]
\epsilon 
\nonumber \\
&&
+ \left[
- \frac{151264019628699781}{24390120480768000000}
+ \frac{341}{155520} \zeta_5
+ \frac{15131}{43545600} \zeta_4
+ \frac{124760533}{82301184000} \zeta_3
\right]
\epsilon^2
\nonumber \\
&&
+ \left[
- \frac{490045281662428129541}{10243850601922560000000}
+ \frac{251}{46656} \zeta_6
+ \frac{275209}{21772800} \zeta_5
+ \frac{124760533}{54867456000} \zeta_4
\right. \nonumber \\
&& \left. ~~~
+ \frac{388425339013}{34566497280000} \zeta_3
- \frac{493}{233280} \zeta_3^2
\right]
\epsilon^3
\nonumber \\
&&
+ \left[
- \frac{1331159757988712929353901}{4302417252807475200000000}
+ \frac{16619}{311040} \zeta_7
+ \frac{1583}{51030} \zeta_6
+ \frac{1700403287}{27433728000} \zeta_5
\right. \nonumber \\
&& \left. ~~~
+ \frac{388425339013}{23044331520000} \zeta_4
+ \frac{1138231977555493}{14517928857600000} \zeta_3
- \frac{493}{77760} \zeta_3 \zeta_4
- \frac{397643}{32659200} \zeta_3^2
\right]
\epsilon^4
\nonumber \\
&&
~+~ O(\epsilon^5) \nonumber \\
M_{22} &=& 
-~ \frac{1}{544320} \frac{1}{\epsilon^3} 
- \frac{517}{36578304} \frac{1}{\epsilon^2} 
- \frac{3058789}{38407219200} \frac{1}{\epsilon} 
+ \left[ - \frac{13379432663}{32262064128000}
+ \frac{1}{13608} \zeta_3
\right]
\nonumber \\
&&
+~ \left[
- \frac{398545304569}{188195374080000}
+ \frac{1}{9072} \zeta_4
+ \frac{1609}{3048192} \zeta_3
\right]
\epsilon
\nonumber \\
&&
+~ \left[
- \frac{20429737763758291}{1897009370726400000}
+ \frac{37}{27216} \zeta_5
+ \frac{1609}{2032128} \zeta_4
+ \frac{27941057}{9601804800} \zeta_3
\right]
\epsilon^2
\nonumber \\
&&
+~ \left[
- \frac{65407910058930860539}{1195115903557632000000}
+ \frac{25}{7776} \zeta_6
+ \frac{29203}{3048192} \zeta_5
+ \frac{27941057}{6401203200} \zeta_4
\right. \nonumber \\
&& \left. ~~~~
+ \frac{13768662841}{896168448000} \zeta_3
- \frac{17}{11340} \zeta_3^2
\right]
\epsilon^3
\nonumber \\
&&
+~ \left[
- \frac{279112798944169319679083}{1003897358988410880000000}
+ \frac{13157}{544320} \zeta_7
+ \frac{365}{16128} \zeta_6
+ \frac{500709689}{9601804800} \zeta_5
\right. \nonumber \\
&& \left. ~~~~
+ \frac{13768662841}{597445632000} \zeta_4
+ \frac{22798640898757}{282293061120000} \zeta_3
- \frac{17}{3780} \zeta_3 \zeta_4
- \frac{5353}{508032} \zeta_3^2
\right]
\epsilon^4 
~+~ O(\epsilon^5) \nonumber \\
M_{27} &=& 
\frac{1}{1555200} \frac{1}{\epsilon^3} 
+ \frac{43}{11664000} \frac{1}{\epsilon^2} 
+ \frac{88957}{9797760000} \frac{1}{\epsilon} 
+ \left[
- \frac{86009717}{2743372800000}
+ \frac{7}{194400} \zeta_3
\right]
\nonumber \\
&&
+~ \left[
- \frac{34572378221}{57610828800000}
+ \frac{7}{129600} \zeta_4
+ \frac{301}{1458000} \zeta_3
\right]
\epsilon
\nonumber \\
&&
+~ \left[
- \frac{2548180213275329}{483930961920000000}
+ \frac{221}{129600} \zeta_5
+ \frac{301}{972000} \zeta_4
+ \frac{2924447}{2449440000} \zeta_3
\right]
\epsilon^2
\nonumber \\
&&
+~ \left[
- \frac{1134533046211608059}{30487650600960000000}
+ \frac{649}{155520} \zeta_6
+ \frac{9503}{972000} \zeta_5
+ \frac{2924447}{1632960000} \zeta_4
\right. \nonumber \\
&& \left. ~~~~
+ \frac{5613786161}{685843200000} \zeta_3
- \frac{29}{19440} \zeta_3^2
\right]
\epsilon^3
\nonumber \\
&&
+~ \left[
- \frac{12113190120183510866647}{51219253009612800000000}
+ \frac{3763}{86400} \zeta_7
+ \frac{27907}{1166400} \zeta_6
+ \frac{37569353}{816480000} \zeta_5
\right. \nonumber \\
&& \left. ~~~~
+ \frac{5613786161}{457228800000} \zeta_4
+ \frac{4104046049389}{72013536000000} \zeta_3
- \frac{29}{6480} \zeta_3 \zeta_4
- \frac{1247}{145800} \zeta_3^2
\right]
\epsilon^4 
~+~ O(\epsilon^5) \nonumber \\
M_{32} &=& 
-~ \frac{1}{7776} \frac{1}{\epsilon^4} 
- \frac{13}{31104} \frac{1}{\epsilon^3} 
- \frac{3281}{5598720} \frac{1}{\epsilon^2} 
+ \left[
\frac{188299}{111974400}
- \frac{7}{3888} \zeta_3
\right]
\frac{1}{\epsilon}
\nonumber \\
&&
+~ \left[
\frac{397898731}{20155392000}
- \frac{7}{2592} \zeta_4
- \frac{91}{15552} \zeta_3
\right]
\nonumber \\
&&
+~ \left[
\frac{16226423357}{134369280000}
- \frac{7}{144} \zeta_5
- \frac{91}{10368} \zeta_4
- \frac{65699}{2799360} \zeta_3
\right]
\epsilon
\nonumber \\
&&
+~ \left[
\frac{44885504009599}{72559411200000}
- \frac{455}{3888} \zeta_6
- \frac{91}{576} \zeta_5
- \frac{65699}{1866240} \zeta_4
- \frac{7076939}{55987200} \zeta_3
+ \frac{113}{3888} \zeta_3^2
\right]
\epsilon^2 \nonumber \\
&& +~ O(\epsilon^3) \nonumber \\
M_{33} &=& 
\frac{1}{25920} \frac{1}{\epsilon^4} 
+ \frac{109}{777600} \frac{1}{\epsilon^3} 
+ \frac{71}{1866240} \frac{1}{\epsilon^2}
\nonumber \\
&&
+~ \left[
- \frac{1037519}{311040000}
+ \frac{31}{12960} \zeta_3
\right]
\frac{1}{\epsilon} 
+ \left[
- \frac{1856641247}{55987200000}
+ \frac{31}{8640} \zeta_4
+ \frac{3379}{388800} \zeta_3
\right]
\nonumber \\
&&
+~ \left[
- \frac{89611835501}{373248000000}
+ \frac{449}{4320} \zeta_5
+ \frac{3379}{259200} \zeta_4
+ \frac{1013903}{23328000} \zeta_3
\right]
\epsilon
\nonumber \\
&&
+~ \left[
- \frac{924742645942453}{604661760000000}
+ \frac{329}{1296} \zeta_6
+ \frac{48941}{129600} \zeta_5
+ \frac{1013903}{15552000} \zeta_4
+ \frac{47128153}{155520000} \zeta_3
\right. \nonumber \\
&& \left. ~~~~
- \frac{983}{12960} \zeta_3^2
\right]
\epsilon^2
\nonumber \\
&&
+~ \left[
- \frac{65933071345592731}{7255941120000000}
+ \frac{5669}{2160} \zeta_7
+ \frac{35861}{38880} \zeta_6
+ \frac{11025007}{7776000} \zeta_5
+ \frac{47128153}{103680000} \zeta_4
\right. \nonumber \\
&& \left. ~~~~
+ \frac{11647067957}{5598720000} \zeta_3
- \frac{983}{4320} \zeta_3 \zeta_4
- \frac{107147}{388800} \zeta_3^2
\right]
\epsilon^3 ~+~ O(\epsilon^4) \nonumber \\
M_{34} &=& 
-~ \frac{1}{103680} \frac{1}{\epsilon^4} 
- \frac{253}{6220800} \frac{1}{\epsilon^3} 
- \frac{6031}{74649600} \frac{1}{\epsilon^2}
+ \left[
\frac{10551067}{22394880000}
- \frac{5}{10368} \zeta_3
\right]
\frac{1}{\epsilon}
\nonumber \\
&&
+~ \left[
\frac{3403463359}{447897600000}
- \frac{5}{6912} \zeta_4
- \frac{253}{124416} \zeta_3
\right]
\nonumber \\
&&
+~ \left[
\frac{588761983211}{8957952000000}
- \frac{31}{1152} \zeta_5
- \frac{253}{82944} \zeta_4
- \frac{2444147}{186624000} \zeta_3
\right]
\epsilon
\nonumber \\
&&
+~ \left[
\frac{2243912853260021}{4837294080000000}
- \frac{685}{10368} \zeta_6
- \frac{7843}{69120} \zeta_5
- \frac{2444147}{124416000} \zeta_4
- \frac{1133018717}{11197440000} \zeta_3
\right. \nonumber \\
&& \left. ~~~~
+ \frac{1247}{51840} \zeta_3^2
\right]
\epsilon^2
\nonumber \\
&&
+~ \left[
\frac{171045558222119119}{58047528960000000}
- \frac{12503}{17280} \zeta_7
- \frac{34661}{124416} \zeta_6
- \frac{131641}{256000} \zeta_5
- \frac{1133018717}{7464960000} \zeta_4
\right. \nonumber \\
&& \left. ~~~~
- \frac{161255734073}{223948800000} \zeta_3
+ \frac{1247}{17280} \zeta_3 \zeta_4
+ \frac{315491}{3110400} \zeta_3^2
\right]
\epsilon^3 ~+~ O(\epsilon^4) \nonumber \\
M_{35} &=& 
\frac{1}{41472} \frac{1}{\epsilon^3} 
+ \frac{173}{1382400} \frac{1}{\epsilon^2}
+ \left[
\frac{245651}{746496000}
+ \frac{1}{28800} \zeta_3
\right]
\frac{1}{\epsilon}
\nonumber \\
&&
+~ \left[
- \frac{46303}{4976640000}
+ \frac{1}{19200} \zeta_4
+ \frac{2989}{5184000} \zeta_3
\right]
\nonumber \\
&&
+~ \left[
- \frac{18489907021}{2687385600000}
- \frac{23}{28800} \zeta_5
+ \frac{2989}{3456000} \zeta_4
+ \frac{13087}{3840000} \zeta_3
\right]
\epsilon
\nonumber \\
&&
+~ \left[
- \frac{1087166778487}{17915904000000}
- \frac{1}{480} \zeta_6
+ \frac{25751}{1728000} \zeta_5
+ \frac{13087}{2560000} \zeta_4
+ \frac{318793381}{18662400000} \zeta_3
\right. \nonumber \\
&& \left. ~~~~
+ \frac{29}{28800} \zeta_3^2
\right]
\epsilon^2
\nonumber \\
&&
+~ \left[
- \frac{3908335481909509}{9674588160000000}
- \frac{221}{11520} \zeta_7
+ \frac{9283}{259200} \zeta_6
+ \frac{404999}{3840000} \zeta_5
+ \frac{318793381}{12441600000} \zeta_4
\right. \nonumber \\
&& \left. ~~~~
+ \frac{33530592221}{373248000000} \zeta_3
+ \frac{29}{9600} \zeta_3 \zeta_4
- \frac{67019}{5184000} \zeta_3^2
\right]
\epsilon^3 ~+~ O(\epsilon^4) \nonumber \\
M_{36} &=& 
\frac{1}{69120} \frac{1}{\epsilon^2} 
+ \frac{47}{497664} \frac{1}{\epsilon} 
+ \frac{9241}{29859840}
+ \left[
\frac{383903}{358318080}
- \frac{1}{288} \zeta_5
+ \frac{1}{405} \zeta_3
\right]
\epsilon
\nonumber \\
&&
+~ \left[
\frac{49253441}{4299816960}
- \frac{5}{576} \zeta_6
- \frac{65}{3456} \zeta_5
+ \frac{1}{270} \zeta_4
+ \frac{107}{38880} \zeta_3
+ \frac{7}{1440} \zeta_3^2
\right]
\epsilon^2
\nonumber \\
&&
+~ \left[
\frac{160435019}{1146617856}
- \frac{127}{2880} \zeta_7
- \frac{325}{6912} \zeta_6
+ \frac{10889}{207360} \zeta_5
+ \frac{107}{25920} \zeta_4
- \frac{36259}{233280} \zeta_3
+ \frac{7}{480} \zeta_3 \zeta_4
\right. \nonumber \\
&& \left. ~~~~
+ \frac{91}{3456} \zeta_3^2
\right]
\epsilon^3 ~+~ O(\epsilon^4) \nonumber \\
M_{41} &=& 
-~ \frac{1}{20736} \frac{1}{\epsilon^3} 
- \frac{383}{746496} \frac{1}{\epsilon^2} 
+ \left[
- \frac{128657}{44789760}
+ \frac{1}{2880} \zeta_3
\right]
\frac{1}{\epsilon}
\nonumber \\
&&
+~ \left[
- \frac{1111315}{107495424}
+ \frac{1}{1920} \zeta_4
+ \frac{253}{103680} \zeta_3
\right]
\nonumber \\
&&
+~ \left[
- \frac{93732961}{6449725440}
- \frac{49}{8640} \zeta_5
+ \frac{253}{69120} \zeta_4
- \frac{19}{746496} \zeta_3
\right]
\epsilon
\nonumber \\
&&
+~ \left[
\frac{82848167}{573308928}
- \frac{13}{864} \zeta_6
- \frac{199}{103680} \zeta_5
- \frac{19}{497664} \zeta_4
- \frac{677957}{8957952} \zeta_3
+ \frac{31}{1728} \zeta_3^2
\right]
\epsilon^2 ~+~ O(\epsilon^3) \nonumber \\
M_{42} &=& 
\frac{1}{62208} \frac{1}{\epsilon^3} 
+ \frac{79}{746496} \frac{1}{\epsilon^2}
+ \left[
\frac{5689}{4976640}
+ \frac{1}{2880} \zeta_3
\right]
\frac{1}{\epsilon}
+ \left[
\frac{1277555}{107495424}
+ \frac{1}{1920} \zeta_4
- \frac{11}{311040} \zeta_3
\right]
\nonumber \\
&&
+~ \left[
\frac{635773633}{6449725440}
- \frac{13}{2880} \zeta_5
- \frac{11}{207360} \zeta_4
- \frac{67793}{3732480} \zeta_3
\right]
\epsilon
\nonumber \\
&&
+~ \left[
\frac{17878937029}{25798901760}
- \frac{7}{576} \zeta_6
- \frac{6019}{103680} \zeta_5
- \frac{67793}{2488320} \zeta_4
- \frac{903631}{4976640} \zeta_3
+ \frac{19}{2880} \zeta_3^2
\right]
\epsilon^2 \nonumber \\
&& +~ O(\epsilon^3) \nonumber \\
M_{43} &=& 
-~ \frac{1}{5760} \frac{1}{\epsilon^2} 
+ \left[
- \frac{131}{69120}
+ \frac{1}{2880} \zeta_3
\right]
\frac{1}{\epsilon}
+ \left[
- \frac{10553}{829440}
+ \frac{1}{1920} \zeta_4
+ \frac{101}{34560} \zeta_3
\right]
\nonumber \\
&&
+~ \left[
- \frac{684343}{9953280}
- \frac{1}{360} \zeta_5
+ \frac{101}{23040} \zeta_4
+ \frac{8687}{414720} \zeta_3
\right]
\epsilon
\nonumber \\
&&
+~ \left[
- \frac{13206019}{39813120}
- \frac{1}{128} \zeta_6
- \frac{101}{4320} \zeta_5
+ \frac{8687}{276480} \zeta_4
+ \frac{642169}{4976640} \zeta_3
- \frac{1}{180} \zeta_3^2
\right]
\epsilon^2
\nonumber \\
&&
+~ \left[
- \frac{719292013}{477757440}
- \frac{67}{960} \zeta_7
- \frac{101}{1536} \zeta_6
- \frac{4367}{51840} \zeta_5
+ \frac{642169}{3317760} \zeta_4
+ \frac{13908749}{19906560} \zeta_3
\right. \nonumber \\
&& \left. ~~~~
- \frac{1}{60} \zeta_3 \zeta_4
- \frac{101}{2160} \zeta_3^2
\right]
\epsilon^3 ~+~ O(\epsilon^4) \nonumber \\
M_{44} &=& 
\frac{7}{103680} \frac{1}{\epsilon^3} 
+ \frac{61}{77760} \frac{1}{\epsilon^2} 
+ \frac{32939}{7464960} \frac{1}{\epsilon}
+ \left[
\frac{277411}{17915904}
+ \frac{1}{1296} \zeta_3
\right]
\nonumber \\
&&
+~ \left[
\frac{19619333}{1074954240}
+ \frac{1}{864} \zeta_4
+ \frac{781}{155520} \zeta_3
\right]
\epsilon
\nonumber \\
&&
+~ \left[
- \frac{4976176237}{12899450880}
- \frac{147}{64} \zeta_7
+ \frac{1999}{1728} \zeta_5
+ \frac{781}{103680} \zeta_4
+ \frac{113243}{93312} \zeta_3
\right]
\epsilon^2 ~+~ O(\epsilon^3) \nonumber \\
M_{45} &=& 
\frac{7}{20736} \frac{1}{\epsilon^3} 
+ \frac{277}{138240} \frac{1}{\epsilon^2}
+ \left[
\frac{88751}{14929920}
+ \frac{1}{2880} \zeta_3
\right]
\frac{1}{\epsilon}
\nonumber \\
&&
+~ \left[
- \frac{30421}{59719680}
+ \frac{1}{1920} \zeta_4
+ \frac{1183}{103680} \zeta_3
\right]
\nonumber \\
&&
+~ \left[
- \frac{333199777}{2149908480}
- \frac{23}{2880} \zeta_5
+ \frac{1183}{69120} \zeta_4
+ \frac{5915}{82944} \zeta_3
\right]
\epsilon
\nonumber \\
&&
+~ \left[
- \frac{88173683267}{60197437440}
- \frac{1}{48} \zeta_6
+ \frac{511}{1280} \zeta_5
+ \frac{5915}{55296} \zeta_4
+ \frac{39770197}{104509440} \zeta_3
+ \frac{55}{4032} \zeta_3^2
\right]
\epsilon^2 \nonumber \\
&& +~ O(\epsilon^3) \nonumber \\
M_{51} &=& 
\frac{1}{864} \frac{1}{\epsilon^3} 
+ \left[
\frac{85}{10368}
- \frac{1}{216} \zeta_3
\right]
\frac{1}{\epsilon^2}
+ \left[
\frac{1171}{41472}
- \frac{1}{144} \zeta_4
- \frac{7}{324} \zeta_3
\right]
\frac{1}{\epsilon}
\nonumber \\
&&
+~ \left[
\frac{575}{1492992}
- \frac{1}{108} \zeta_5
- \frac{7}{216} \zeta_4
+ \frac{43}{7776} \zeta_3
\right]
\nonumber \\
&&
+~ \left[
- \frac{15492679}{17915904}
- \frac{5}{432} \zeta_6
- \frac{191}{1296} \zeta_5
+ \frac{43}{5184} \zeta_4
+ \frac{45173}{93312} \zeta_3
- \frac{5}{36} \zeta_3^2
\right]
\epsilon ~+~ O(\epsilon^2) \nonumber \\
M_{52} &=& 
\frac{1}{216} \frac{1}{\epsilon^3} 
+ \left[
\frac{37}{1296}
- \frac{1}{108} \zeta_3
\right]
\frac{1}{\epsilon^2}
+ \left[
\frac{779}{7776}
- \frac{1}{72} \zeta_4
- \frac{11}{324} \zeta_3
\right]
\frac{1}{\epsilon}
\nonumber \\
&&
+~ \left[
\frac{10289}{46656}
+ \frac{2}{27} \zeta_5
- \frac{11}{216} \zeta_4
- \frac{25}{486} \zeta_3
\right]
\nonumber \\
&&
+~ \left[
\frac{3109}{93312}
+ \frac{5}{24} \zeta_6
+ \frac{22}{81} \zeta_5
- \frac{25}{324} \zeta_4
+ \frac{445}{5832} \zeta_3
- \frac{7}{54} \zeta_3^2
\right]
\epsilon ~+~ O(\epsilon^2) \nonumber \\
M_{61} &=& 
-~ \frac{1}{72} \frac{1}{\epsilon^2} 
+ \left[
- \frac{17}{108}
+ \frac{5}{36} \zeta_5
- \frac{1}{36} \zeta_3
\right]
\frac{1}{\epsilon}
\nonumber \\
&&
+~ \left[
- \frac{407}{432}
+ \frac{25}{72} \zeta_6
+ \frac{85}{216} \zeta_5
- \frac{1}{24} \zeta_4
+ \frac{25}{216} \zeta_3
- \frac{1}{36} \zeta_3^2
\right] ~+~ O(\epsilon) \nonumber \\
M_{62} &=& 
\left[
\frac{1}{216}
- \frac{1}{72} \zeta_3
\right]
\frac{1}{\epsilon^2}
+ \left[
\frac{37}{648}
- \frac{1}{48} \zeta_4
- \frac{13}{216} \zeta_3
\right]
\frac{1}{\epsilon}
\nonumber \\
&&
+~ \left[
\frac{505}{1296}
- \frac{1}{16} \zeta_5
- \frac{13}{144} \zeta_4
- \frac{41}{144} \zeta_3
\right] ~+~ O(\epsilon) \nonumber \\
M_{63} &=& 
\left[
\frac{1}{216}
- \frac{1}{72} \zeta_3
\right]
\frac{1}{\epsilon^2}
+ \left[
\frac{65}{1296}
- \frac{1}{48} \zeta_4
- \frac{13}{216} \zeta_3
\right]
\frac{1}{\epsilon}
+ \left[
\frac{179}{648}
- \frac{1}{36} \zeta_5
- \frac{13}{144} \zeta_4
- \frac{2}{9} \zeta_3
\right]
\nonumber \\
&&
+ \left[
\frac{4109}{5832}
- \frac{5}{144} \zeta_6
- \frac{97}{432} \zeta_5
- \frac{1}{3} \zeta_4
- \frac{457}{972} \zeta_3
- \frac{5}{12} \zeta_3^2
\right]
\epsilon
~+~ O(\epsilon^2) ~. 
\label{mastint}
\end{eqnarray}
In several cases we have given terms up to the level of $\zeta_7$ despite the
fact that this number does not appear in the final renormalization group 
functions. It would be present in the finite parts of the various Green's 
functions. For compactness we have expressed the master integrals in 
(\ref{mastint}) in the corresponding $G$-scheme in six dimensions which is the 
standard way to present them, \cite{46}. In other words for {\em each} loop 
order the factor
\begin{equation}
G ~=~ \frac{\Gamma(1+\epsilon) (\Gamma(1-\epsilon))^2}{\Gamma(2-2\epsilon)}
\end{equation}
is included. This circumvents the need to include, for instance, the
Euler-Mascheroni constant $\gamma$ which would otherwise appear throughout and
which is ordinarily included in part in the change from the minimal subtraction
to modified minimal subtraction scheme, \cite{71}.
 
\sect{Field dimensions at various fixed points.}

In this appendix we give the various critical exponents for (\ref{lagphi3on})
to $O(\epsilon^4)$ at the same fixed points value of $N$ as given in the 
appendix of \cite{28}. This is partly for completeness but also to note 
interesting structure. For the various values of $N$ considered there are three
distinct fixed points which are labelled in the same way as \cite{28}. These 
were termed `critical', `theory $2$' and `theory 3' and we retain that
nomenclature here but label the exponents with the respective superscripts $c$,
$1$ and $2$. The first case considered in \cite{28} was $N$~$=$~$600$ and the 
three sets of exponents are    
\begin{eqnarray}
\Delta^c_\phi &=& 2 + (0.000036 i - 0.996357) \epsilon 
+ (0.000142 i - 0.008332) \epsilon^2 \nonumber \\
&& +~ ( - 0.000852 i + 0.001801) \epsilon^3 
+ ( - 0.002328 i + 0.014101) \epsilon^4 ~+~ O(\epsilon^5) \nonumber \\ 
\Delta^{c}_\sigma &=& 2 + ( - 0.060308 i + 0.167456) \epsilon 
+ (0.172529 i - 0.263035) \epsilon^2 \nonumber \\
&& +~ (0.274055 i + 0.000784) \epsilon^3 
+ ( - 0.846566 i + 0.464339) \epsilon^4 ~+~ O(\epsilon^5) \nonumber \\ 
\Delta^c_+ &=& 2 + ( - 1.594272 - 0.673433 i) \epsilon 
+ ( - 0.231855 + 1.280485 i) \epsilon^2 \nonumber \\
&& +~ (0.890531 + 3.353446 i) \epsilon^3 
+ ( 1.853548 - 1.733472 i) \epsilon^4 ~+~ O(\epsilon^5) \nonumber \\ 
\Delta^c_- &=& 2 + (0.167456 - 0.060308 i) \epsilon 
+ ( - 0.263035 + 0.172529 i) \epsilon^2 \nonumber \\
&& +~ (0.000784 + 0.274055 i) \epsilon^3 
+ ( 0.464339 - 0.846567 i) \epsilon^4 ~+~ O(\epsilon^5) 
\end{eqnarray}
\begin{eqnarray}
\Delta^2_\phi &=& 2 + ( - 0.000036 i - 0.996357) \epsilon 
+ ( - 0.000142 i - 0.008332) \epsilon^2 \nonumber \\
&& +~ (0.000853 i + 0.001801) \epsilon^3 
+ (0.002328 i + 0.014101) \epsilon^4 ~+~ O(\epsilon^5) \nonumber \\ 
\Delta^2_\sigma &=& 2 + (0.060308 i + 0.167456) \epsilon 
+ ( - 0.172529 i - 0.263035) \epsilon^2 \nonumber \\
&& +~ ( - 0.274055 i + 0.000784) \epsilon^3
+ ( 0.846566 i + 0.464339) \epsilon^4 ~+~ O(\epsilon^5) \nonumber \\
\Delta^2_+ &=& 2 + ( - 1.594272 + 0.673433 i) \epsilon 
+ ( - 0.231855 - 1.280485 i) \epsilon^2 \nonumber \\
&& +~ (0.890531 - 3.353446 i) \epsilon^3 
+ ( 1.853548 + 1.733472 i) \epsilon^4 ~+~ O(\epsilon^5) \nonumber \\ 
\Delta^2_- &=& 2 + (0.167457 + 0.060308 i) \epsilon 
+ ( - 0.263035 - 0.172530 i) \epsilon^2 \nonumber \\
&& +~ (0.000784 - 0.274055 i) \epsilon^3 
+ ( 0.464339 + 0.846566 i) \epsilon^4 ~+~ O(\epsilon^5) 
\end{eqnarray}
and 
\begin{eqnarray}
\Delta^3_\phi &=& 2 - 0.997795 \epsilon - 0.00219 \epsilon^2 
+ 0.003555 \epsilon^3 + 0.034922 \epsilon^4 ~+~ O(\epsilon^5) \nonumber \\ 
\Delta^3_\sigma &=& 2 - 0.159643 \epsilon + 0.167857 \epsilon^2 
+ 0.763443 \epsilon^3 + 4.639369 \epsilon^4 ~+~ O(\epsilon^5) \nonumber \\
\Delta^3_+ &=& 2 - 0.159642 \epsilon + 0.167857 \epsilon^2 
+ 0.763443 \epsilon^3 + 4.639369 \epsilon^4 ~+~ O(\epsilon^5) \nonumber \\ 
\Delta^3_- &=& 2 + 0.632464 \epsilon + 0.506339 \epsilon^2 
+ 4.963931 \epsilon^3 + 35.781014 \epsilon^4 ~+~ O(\epsilon^5) ~. 
\end{eqnarray}
Here and for the expressions we give for the other values of $N$ we note that
\begin{equation}
\Delta^i_\pm ~=~ \gamma_\pm(g_{1c},g_{2c}) 
\end{equation}
includes the canonical dimension and these mass exponents definitions are the
same as those used for the large $N$ comparison with the large $N$ critical
exponents. For this value of $N$ and others two of the points have critical 
exponents which are complex conjugates. Next the parallel values for 
$N$~$=$~$1000$ are
\begin{eqnarray}
\Delta^c_\phi &=& 2 + ( - 0.000010 i - 0.997921) \epsilon 
+ (0.000167 i - 0.004378) \epsilon^2 \nonumber \\
&& +~ (0.001518 i + 0.000406) \epsilon^3 
+ ( 0.021951 i + 0.007247) \epsilon^4 ~+~ O(\epsilon^5) \nonumber \\ 
\Delta^c_\sigma &=& 2 + (0.010791 i + 0.115748) \epsilon 
+ ( - 0.186373 i - 0.172907) \epsilon^2 \nonumber \\
&& +~ ( - 1.392417 i - 0.116672) \epsilon^3
+ ( - 21.819202 i - 0.017897) \epsilon^4 ~+~ O(\epsilon^5) \nonumber \\
\Delta^c_+ &=& 2 + ( - 1.424029 + 0.158680 i) \epsilon 
+ ( - 0.198690 - 2.582892 i) \epsilon^2 \nonumber \\
&& +~ ( - 0.142920 - 22.664631 i) \epsilon^3 
+ ( - 2.551351 - 349.159818 i) \epsilon^4 ~+~ O(\epsilon^5) \nonumber \\ 
\Delta^c_- &=& 2 + (0.115748 + 0.010791 i) \epsilon 
+ ( - 0.172907 - 0.186373 i) \epsilon^2 \nonumber \\
&& +~ ( - 0.116672 - 1.392417 i) \epsilon^3 
+ ( - 0.017897 - 21.819201 i) \epsilon^4 ~+~ O(\epsilon^5) 
\end{eqnarray}
\begin{eqnarray}
\Delta^2_\phi &=& 2 + (0.000010 i - 0.997921) \epsilon 
+ ( - 0.000167 i - 0.0043778) \epsilon^2 \nonumber \\ 
&& +~ ( - 0.001518 i + 0.000406) \epsilon^3 
+ ( - 0.021951 i + 0.007247) \epsilon^4 ~+~ O(\epsilon^5) \nonumber \\ 
\Delta^2_\sigma &=& 2 + ( - 0.010791 i + 0.115748) \epsilon 
+ (0.186373 i - 0.172907) \epsilon^2 \nonumber \\
&& +~ (1.392417 i - 0.116671) \epsilon^3 
+ ( 21.819201 i - 0.017897) \epsilon^4 ~+~ O(\epsilon^5) \nonumber \\
\Delta^2_+ &=& 2 + ( - 1.424029 - 0.158680i) \epsilon 
+ ( - 0.198690 + 2.582892 i) \epsilon^2 \nonumber \\ 
&& +~ ( - 0.142920 + 22.664631 i) \epsilon^3 
+ ( - 2.551351 + 349.159818 i) \epsilon^4 ~+~ O(\epsilon^5) \nonumber \\ 
\Delta^2_- &=& 2 + (0.115748 - 0.010791 i) \epsilon 
+ ( - 0.172907 + 0.186373 i) \epsilon^2 \nonumber \\
&& +~ ( - 0.116671 + 1.392417 i) \epsilon^3 
+ ( - 0.017897 + 21.819202 i) \epsilon^4 ~+~ O(\epsilon^5)
\end{eqnarray}
and
\begin{eqnarray}
\Delta^3_\phi &=& 2 - 0.998612 \epsilon - 0.001535 \epsilon^2 
+ 0.001864 \epsilon^3 + 0.018909 \epsilon^4 ~+~ O(\epsilon^5) \nonumber \\ 
\Delta^3_\sigma &=& 2 - 0.127848 \epsilon + 0.138289 \epsilon^2 
+ 0.598452 \epsilon^3 + 3.499963 \epsilon^4 ~+~ O(\epsilon^5) \nonumber \\
\Delta^3_+ &=& 2 - 0.127848 \epsilon + 0.138289 \epsilon^2 
+ 0.598452 \epsilon^3 + 3.499963 \epsilon^4 ~+~ O(\epsilon^5) \nonumber \\
\Delta^3_- &=& 2 + 0.525620 \epsilon + 0.386259 \epsilon^2 
+ 3.806034 \epsilon^3 + 27.088909 \epsilon^4 ~+~ O(\epsilon^5) ~. 
\end{eqnarray}
Finally, for $N$~$=$~$1400$ we find 
\begin{eqnarray}
\Delta^c_\phi &=& 2 - 0.998526 \epsilon - 0.002924 \epsilon^2 
- 0.000056 \epsilon^3 + 0.004014 \epsilon^4 ~+~ O(\epsilon^5) \nonumber \\ 
\Delta^c_\sigma &=& 2 + 0.067715 \epsilon - 0.143131 \epsilon^2 
+ 0.005374 \epsilon^3 + 0.348439 \epsilon^4 ~+~ O(\epsilon^5) \nonumber \\
\Delta^c_+ &=& 2 - 1.765085 \epsilon - 0.767483 \epsilon^2 
+ 0.291437 \epsilon^3 + 2.728177 \epsilon^4 ~+~ O(\epsilon^5) \nonumber \\ 
\Delta^c_- &=& 2 + 0.067715 \epsilon - 0.143131 \epsilon^2 
+ 0.005374 \epsilon^3 + 0.348439 \epsilon^4 ~+~ O(\epsilon^5)
\end{eqnarray}
\begin{eqnarray}
\Delta^2_\phi &=& 2 - 0.998575 \epsilon - 0.002939 \epsilon^2 
+ 0.000342 \epsilon^ 3 + 0.005877 \epsilon^4 ~+~ O(\epsilon^5) \nonumber \\ 
\Delta^2_\sigma &=& 2 + 0.115801 \epsilon - 0.123622 \epsilon^2 
- 0.2790432 \epsilon^3 - 0.776356 \epsilon^4 ~+~ O(\epsilon^5) \nonumber \\
\Delta^2_+ &=& 2 - 0.924175 \epsilon + 0.427891 \epsilon^2 
- 1.113372 \epsilon^3 - 10.101218 \epsilon^4 ~+~ O(\epsilon^5) \nonumber \\
\Delta^2_- &=& 2 + 0.115801 \epsilon - 0.1236220 \epsilon^2 
- 0.279043 \epsilon^3 - 0.776356 \epsilon^4 ~+~ O(\epsilon^5) 
\end{eqnarray}
and
\begin{eqnarray}
\Delta^3_\phi &=& 2 - 0.998982 \epsilon - 0.001191 \epsilon^2 
+ 0.001224 \epsilon^3 + 0.012696 \epsilon^4 ~+~ O(\epsilon^5) \nonumber \\ 
\Delta^3_\sigma &=& 2 - 0.109951 \epsilon + 0.120821 \epsilon^2 
+ 0.507233 \epsilon^3 + 2.895587 \epsilon^4 ~+~ O(\epsilon^5) \nonumber \\
\Delta^3_+ &=& 2 - 0.109951 \epsilon + 0.120821 \epsilon^2 
+ 0.507233 \epsilon^3 + 2.895587 \epsilon^4 ~+~ O(\epsilon^5) \nonumber \\ 
\Delta^3_- &=& 2 + 0.462254 \epsilon + 0.321631 \epsilon^2 
+ 3.184495 \epsilon^3 + 22.495823 \epsilon^4 ~+~ O(\epsilon^5) ~. 
\end{eqnarray}
We note that the expression for $\Delta^2_\sigma$ corrects an obvious
typographical error in equation (A.23) of \cite{28}. There the $O(\epsilon)$
term is not recorded although its actual coefficient appears as the coefficient
of the $O(\epsilon^2)$ term. For the $N$~$=$~$1400$ set the three fixed points
again produces real critical exponents. Unfortunately in each set improving the 
series convergence using Pad\'{e} approximants only applies to $\Delta_\phi$ 
along the lines discussed in the main text for $N$~$=$~$500$. In each case we 
have recorded the mass mixing matrix eigen-critical exponents in our 
conventions as there is an interesting feature which extends the observation in
the large $N$ comparison in section $6$. In each set of exponents and values of
$N$ $\Delta_\sigma$ is equivalent to one of the mass eigen-critical exponents. 
In other words for finite values of $N$ the field critical exponent and its 
mass exponent are equivalent. However, the particular eigen-exponent the field 
dimension equates to depends on the specific fixed point. In each of the cases 
presented here $\Delta_\sigma$ corresponds to the minus exponent for the points
designated critical and theory $2$ but to the plus exponent for theory $3$. Of 
the three only theory $3$ has real exponents and this picture tallies with the 
large $N$ checks discussed earlier.

\end{document}